\documentclass[a4paper,12pt]{article}
\usepackage{epsfig}
\usepackage{multirow}

\def\beq{\begin{equation}}
\def\eeq{\end{equation}}
\def\bea{\begin{eqnarray}}
\def\eea{\end{eqnarray}}
\def\ksl{\hbox{\hbox{${k}$}}\kern-1.9mm{\hbox{${/}$}}}

\def\lsim{\raise0.3ex\hbox{$\;<$\kern-0.75em\raise-1.1ex\hbox{$\sim\;$}}}
\def\gsim{\raise0.3ex\hbox{$\;>$\kern-0.75em\raise-1.1ex\hbox{$\sim\;$}}}

\def\dosixfigs#1#2#3#4#5#6#7{\centerline{
\epsfxsize=#1\epsfig{file=#2, width=6cm,height=7.5cm, angle=-90}
\hspace{0cm}
\hfil
\epsfxsize=#1\epsfig{file=#3,  width=6cm, height=7.5cm, angle=-90}}

\vspace{0.2cm}
\centerline{
\epsfxsize=#1\epsfig{file=#4, width=6cm,height=7.5cm, angle=-90}
\hspace{0cm}
\hfil
\epsfxsize=#1\epsfig{file=#5,  width=6cm, height=7.5cm, angle=-90}}

\vspace{0.2cm}
\centerline{
\epsfxsize=#1\epsfig{file=#6, width=6cm,height=7.5cm, angle=-90}
\hspace{0cm}
\hfil
\epsfxsize=#1\epsfig{file=#7,  width=6cm, height=7.5cm, angle=-90}}
}

\def\dofourfigs#1#2#3#4#5{\centerline{
\epsfxsize=#1\epsfig{file=#2, width=6cm,height=7.5cm, angle=-90}
\hspace{0cm}
\hfil
\epsfxsize=#1\epsfig{file=#3,  width=6cm, height=7.5cm, angle=-90}}

\vspace{0.5cm}
\centerline{
\epsfxsize=#1\epsfig{file=#4, width=6cm,height=7.5cm, angle=-90}
\hspace{0cm}
\hfil
\epsfxsize=#1\epsfig{file=#5,  width=6cm, height=7.5cm, angle=-90}}
}

\def\dofigs#1#2#3{\centerline{\epsfxsize=#1\epsfig{file=#2, width=7cm, 
height=8cm, angle=-90}
\hfil\epsfxsize=#1\epsfig{file=#3,  width=7cm, height=8cm, angle=-90}}}

\def\dofigsV#1#2#3{\centerline{\epsfxsize=#1\epsfig{file=#2, width=9cm, 
height=12cm, angle=-90}}

\vspace{0.5cm}
\noindent
\centerline{
\epsfxsize=#1\epsfig{file=#3,  width=9cm, height=12cm, angle=-90}}}

\def\dofig#1#2{\centerline{\epsfxsize=#1\epsfig{file=#2, width=15cm, 
height=10cm, angle=0}}}

\newcommand{\text}{\rm}

\def\lsim{\raise0.3ex\hbox{$\;<$\kern-0.75em\raise-1.1ex\hbox{$\sim\;$}}} 

\def\gsim{\raise0.3ex\hbox{$\;>$\kern-0.75em\raise-1.1ex\hbox{$\sim\;$}}}

\begin{document}
\begin{titlepage}

{\flushright{
        \begin{minipage}{5cm}
          ROME1-1469/2010\\
        \end{minipage}        }

}

\vskip 2cm
\begin{center} 

\boldmath
{\Large\bf 
Testing Effective Yukawa Couplings

\vspace{0.2cm}
in Higgs Searches at the Tevatron and  LHC}

\vspace{1cm}

E. Gabrielli$^a$ and B. Mele$^b$
\vskip 1.cm

\emph{
$^a$ CERN, PH-TH, CH-1211 Geneva 23, Switzerland\\
$^b$ INFN, Sezione di Roma, c/o Dip. di Fisica,
 Universit\`a di Roma  ``La Sapienza'',\\ Piazzale A. Moro 2, I-00185 Rome, Italy
}
\end{center}

\vskip 0.7cm
\begin{abstract}
We explore  the possibility that, while the Higgs mechanism provides masses to the weak-gauge bosons at the 
electroweak scale as  in the standard model, fermion masses are generated by an unknown mechanism
at a higher energy scale.
At low energies, the standard model
can then be regarded as an effective field theory, where fermion masses explicitly break the electroweak 
$\rm SU(2)_{\rm L}\times U(1)_{\rm Y}$ gauge symmetry.
If $\Lambda$ is the renormalization scale where 
the renormalized Yukawa couplings  vanish,  then at energies lower than $\Lambda$, effective Yukawa couplings  will be radiatively induced by nonzero fermion masses. In this scenario,  Higgs-boson  
decays into photons and weak gauge-bosons pairs are in general quite enhanced for a light Higgs.
However, depending on  $\Lambda$,
a substantial decay rate into $b \bar{b}$  can arise, 
 that can be of the same order as, or 
larger than, the  {\it enhanced} $H\to\gamma \gamma$ rate.
A new framework for Higgs searches at hadron colliders is
outlined,   vector-boson fusion becoming the dominant  production mechanism at the CERN LHC, with an important role also played 
by the $WH/ZH$ associated production. 
A detailed analysis of the Higgs branching fractions and their implications 
in Higgs searches is provided, versus  the energy scale $\Lambda$.

\end{abstract}
\vfill
\end{titlepage}

\section{Introduction}
Arguably, one of the most intriguing aspects of particle physics
is the origin of fermion masses.
Despite the impressive phenomenological success of the standard model (SM) \cite{SMref}, 
a clear understanding of the fermion mass spectrum is still missing.
The Higgs mechanism alone does not provide any explanation 
for the observed huge hierarchy  in the fermion masses, that, 
not including the neutrino sector, ranges over 6 orders of magnitude.
The Higgs Yukawa couplings, responsible for 
fermion-mass generation in the SM, are put in by hand,   
and tuned with the corresponding fermion masses 
times the inverse of the vacuum expectation value of the Higgs field.
It is a matter of fact that, out of 19 free parameters 
of the SM,  13 belong to the Yukawa sector.
This 
suggests that some new physics  beyond
the SM might be responsible for  fermion masses and/or 
 the flavor structure in the Yukawa couplings. 
Many extensions of the SM  have been proposed to solve this puzzle,
but none of them can be considered as conclusive.

A nonzero fermion mass explicitly breaks
the chiral symmetry, and therefore the electroweak (EW) gauge symmetry 
$\rm SU(2)_{\rm L}\times U(1)_{\rm Y}$.
However, while the EW gauge symmetry breaking is needed for giving
masses to the weak-gauge bosons, its mechanism could be different from the one responsible for the generation of fermion masses.
In the SM there is only one mass scale: the
vacuum expectation value $v$ of the Higgs fields
($v \simeq 246$ GeV), which sets masses for both fermions and 
gauge bosons. 
On the other hand, 
the fermion mass generation scale could in principle be different
from the EW symmetry breaking scale  \cite{quigg}.

Once the Higgs boson is discovered
at the CERN Large Hadron Collider (LHC), the analysis of its decay modes and production processes 
will help to unravel  the mechanisms for both EW symmetry breaking 
and fermion mass generation (see, e.g., \cite{djouadi}).
Direct Higgs boson searches at CERN LEP 
have excluded Higgs masses up to
114.4 GeV at 95\% C.L.\cite{LEP} , while the Tevatron at Fermilab has recently ruled out the mass range 
163  GeV$ < m_H < 166$  GeV at 95\%C.L. \cite{Tevatron}.
Moreover, indirect 
constraints on the Higgs boson mass and couplings 
come from the analysis of its virtual contributions to the EW processes.
These contributions are not extremely 
sensitive to the Higgs boson mass, since, due to the decoupling theorem, 
one-loop radiative corrections depend only logarithmically on the Higgs mass.
Nevertheless,  electroweak precision tests (EWPT) point to
a Higgs mass quite close to the LEP2 direct bound of 
114.4 GeV. In particular, one gets an upper limit  
$m_H \le $ 157 GeV at 95\% C.L., that can be relaxed up to 
$m_H \le $ 186 GeV, if combined with the direct  limit $m_H > 114.4$ GeV  \cite{HboundsEW}.

On the contrary, present EWPT do not  really constrain
 Yukawa couplings. In fact,
radiative corrections induced by Yukawa couplings of light fermions are tiny in the SM. Also, the SM Yukawa coupling contribution to the $Zb\bar b$ vertex is by far too small to be tested.
The effects of the top-quark Yukawa corrections could be
potentially large. On the other hand, in processes with 
external light quarks or leptons, they 
enter only at two loops, which is beyond the present experimental sensitivity.

Flavor-changing neutral current (FCNC) processes generated at 
one loop in the SM 
are an excellent probe of the fermion mass-matrix structure,
but they are  weakly affected by the Yukawa corrections, too.  The latter enter at the two-loop level, since no Higgs-boson exchange is needed at one loop to cancel the ultraviolet (UV) 
divergencies\footnote{This is not true in the case  of 
the so-called flavor-changing Higgs penguin diagrams,
where a Higgs-boson is exchanged between a flavor-conserving current
and a flavor-changing one induced at one-loop. However, 
this contribution to  low-energy FCNC processes 
is tiny, being suppressed by the Yukawa couplings to the external 
light quarks.}. Even though  fermion masses were put in by hand in the SM 
Lagrangian, due to the Glashow-Iliopoulos-Maiani mechanism \cite{GIM},
the FCNC processes would turn out to be finite at one loop with 
no need of  Higgs-boson exchange between fermions\footnote{This is not  true at the two-loop level,
where Higgs boson exchange is necessary  to 
cancel the UV divergencies.}.

By the way, even gauge-couplings 
unification in grand unified theories (GUT) 
is weakly affected by  Yukawa radiative corrections, since their contribution enters at  the two-loop 
level in the renormalization group (RG) equations  of gauge-couplings
\cite{arason}.

In conclusion, the fact that present experimental data do not   significantly constrain Yukawa couplings  still leaves  room
to speculations on the Higgs couplings to fermions, and eventually
on the origin of fermion masses.

Following these considerations, we will explore  the possibility 
that there is indeed a light Higgs boson 
that is mainly responsible for the EW symmetry breaking,  
but  the chiral-symmetry breaking (ChSB) has a different origin.
In particular, the conjecture explored in the present paper is the one where 
the (unavoidable) contribution of ChSB to the gauge boson masses 
is negligible with respect to the  contribution provided 
by the  Higgs boson.
We will also assume that all the new physics effects 
responsible for  the fermion mass generation 
can be reabsorbed in the nonvanishing fermion masses
plus the renormalization conditions of the Yukawa couplings 
at a large  energy scale $\Lambda$. 

In order to implement the decoupling of the Higgs boson 
to standard fermions, we define $\Lambda$ as the renormalization
scale where the renormalized Yukawa couplings  vanish.
In general, $\Lambda$ might not necessarily coincide with 
the scale of fermion mass generation. 
\\
We will also conservatively assume that potential 
new heavy degrees of freedom will not significantly affect 
the running of gauge and (effective) Higgs Yukawa couplings at high energies.
Hence,
from the EW scale up to the $\Lambda$ scale, the SM can be
regarded as an effective field theory, where the fermion mass terms
explicitly break the SU(2)$_{\rm L}$ gauge symmetry.
Because of the presence of fermion mass terms,
the decoupling of the Higgs boson 
from fermions is spoilt at  energy scales different from $\Lambda$.
Because of the breaking of chiral symmetry, Yukawa couplings are not 
protected  against radiative corrections, and 
effective Yukawa couplings will be radiatively induced at low energies.

In this article, we will evaluate the effective Yukawa couplings 
at  energies lower than $\Lambda$, by  the techniques of RG equations. This will allow us 
 to resum  the 
leading logarithmic terms $g_i^{2n }\log^n{(\Lambda/m_H)}$
(where $g_i$ are the SM gauge couplings), at any order in perturbation theory.

We will explicitly check that contributions 
from  higher-dimensional operators, that could spoil
the validity of perturbation theory in the Yukawa sector, are well under 
control for any value of $\Lambda$ up to the GUT scale ($10^{16}$GeV), 
provided the Higgs mass is not much larger than the 
EW scale~$v$.
A critical discussion on the 
validity of our approach for large values of $\Lambda$ 
will be provided in the next sections. 
In our analysis we will set the upper value of $\Lambda$ at the GUT scale,
which guarantees the perturbativity of the Yukawa sector in all the 
$\Lambda$ range for $m_H < v$. Nevertheless, the existing models that could
provide a specific UV completion to our effective theory are in general 
characterized by a $\Lambda$ scale much lower than $10^{16}$ GeV.

The present model  differs 
substantially from the fermiophobic Higgs boson scenario, appearing in several
extensions of the SM \cite{HiggsFP}. Indeed, 
we will see that the Higgs boson decay rate into $b \bar{b}$ 
can be strongly enhanced at low energy, depending on the size of 
$\Lambda$. In the fermiophobic Higgs scenario, 
the Higgs couplings to fermions are
set to zero at the EW scale $\Lambda_{\rm EW}\sim m_H$.
This condition can  naturally arise 
within new physics models where the scale $\Lambda$
is not  far from the EW scale.
In our approach, when $\Lambda\to m_H$,  all the leading-log  terms
$\log^n{(\Lambda/m_H)}$, 
as well as the renormalization effects on the Yukawa couplings at the 
EW scale, vanish. One then smoothly recovers  the fermiophobic Higgs model.

Note that the suppression of the top Yukawa coupling in the present approach somewhat relieves the SM hierarchical problem in the one-loop 
radiative corrections to the Higgs mass (see, e.g., \cite{quigg}), where the 
dominant contribution to the  quadratic cut-off dependence is now 
the one of order ${\cal O}(M_W^2)$ instead of ${\cal  O}(m_t^2)$. 

A straightforward way to test our scenario  for $m_H \lsim 150$ GeV  is through the
measurement of  Higgs boson decay and production rates at colliders. 
The vanishing of the top-quark Yukawa coupling at  tree-level
has a dramatic impact on the Higgs boson production at 
hadron colliders.   The vector boson fusion (VBF)
\cite{VBF,VBFNLO,VBFNNLO} will replace
the  gluon fusion via the top-quark loop \cite{Hgg,HggNLO} as the main 
production mechanism at the LHC.
If the Higgs boson is quite light, as suggested by EWPT, the dominant  Higgs decays channels will be 
into $WW$, $ZZ$  (where one of the weak gauge bosons, or both, 
can be off-shell), and $\gamma\gamma$. The rate for $H\to b\bar b$ will turn out to be comparable to the $H\to\gamma\gamma$ rate at 
$\Lambda \sim$10 TeV, while it  increases at larger $\Lambda$'s. Furthermore, the enhanced branching ratio BR($H\to\gamma\gamma$) makes the $WH/ZH$ associated production with $H\to\gamma\gamma$ a remarkably interesting channel that could be studied for inclusive $W/Z$ decays.

Present direct experimental limits on $m_H$ crucially depend on the SM Yukawa coupling assumption.
Bounds on $m_H$ in the purely fermiophobic Higgs scenario 
have been obtained at LEP, where one finds $m_H > $ 109.7 GeV at 95\% C.L. \cite{LEPFP}, and the Tevatron, where the D0 experiment 
provides a limit $m_H > $ 101 GeV at 95\% C.L.\cite{D0} and the CDF experiment gives a bound $m_H > $ 106 GeV at 95\% C.L.\cite{CDF}.
In the framework we are going to discuss, non trivial variations 
in the Higgs decay pattern are expected with respect to the fermiophobic scenario that would require a new analysis of experimental data in order to get  dedicated bounds on 
$m_H$, depending on the scale $\Lambda$.

The paper is organized as follows.
In Sec.  2, we present the theoretical framework, derive
the RG equations for the effective 
Yukawa couplings, and give some numerical results for the latter.
After a few comments on the theoretical consistency of the present approach, discussed in Sec. 3,
 we provide, in Sec. 4, the numerical results for the Higgs branching
ratios, as a function of the energy scale $\Lambda$.
In Sec. 5, we study  Higgs production rates corresponding to different decay signatures at hadronic colliders.
Our conclusions will be presented in  Sec. 6.
In the Appendix, we supply relevant formulas for the Higgs boson decays.
\section{RG equations for effective Yukawa couplings}

In this section, we will derive the RG equations for the effective Yukawa couplings arising from the 
 conjecture that the standard 
Higgs mechanism provides masses to the weak gauge-bosons at the EW scale, while
  fermion masses are not generated by Yukawa couplings as in the SM framework.
We will assume that from the EW 
scale up to some larger energy scale,
 only SM  degrees of freedom are relevant, and fermion masses are put in by hand.
Since by switching off the SM 
tree-level couplings of the Higgs-boson to fermions, 
the SM becomes nonrenormalizable,
 we should consider it as an 
effective field theory valid up to some high-energy scale.\footnote{
Notice that one can always introduce in the Lagrangian
 fermion mass terms without Yukawa couplings  
in a ${\rm SU(2)\times U(1)}_{\rm Y}$
gauge-invariant way by considering the nonlinear realization of the EW gauge symmetry. Also in this case, the theory 
is manifestly nonrenormalizable.}

Let us  define the relevant EW Lagrangian in the unitary gauge,
with only physical degrees of freedom  propagating
\bea
{\cal L}={\cal L}_{G,H} + {\cal L}_{F} .
\label{LAG}
\eea
${\cal L}_{G,H}$ is the bosonic sector of the SM EW Lagrangian 
in the unitary gauge, containing
the gauge fields $G=\left\{A_{\mu},W_{\mu}^{\pm},Z_{\mu}
\right\}$, and $A$ and $H$ are the photon and Higgs fields respectively. 
${\cal L}_{F}$ regards
the fermionic sector of the Lagrangian
\bea
{\cal L}_{F}= {\cal L}_{{ Kin}} + {\cal L}_{CC} + {\cal L}_{NC}-
\frac{1}{\sqrt 2}
\sum_f 
{\rm Y}_f \left(\bar{\psi}_{f} \psi_{f} H\right)\,  -\,  \sum_f m_f \bar{\psi}_f \psi_f  \, ,
\label{LF}
\eea
where  ${\cal L}_{{ Kin}}$ is the kinetic term, 
${\cal L}_{CC}$ and ${\cal L}_{NC}$ are the SM Lagrangian for 
the charged- and neutral-currents interactions of quarks and leptons, 
and ${\rm Y}_f$ are the Yukawa couplings, in the basis of fermion mass 
eigenstates. In eq.~(\ref{LF}), we neglect the CKM mixing, since we are interested in flavor-conserving Higgs transitions. Hence, all parameters in eq.~(\ref{LF}) are real.

The Lagrangian in eq.~(\ref{LF}) is basically the SM one, 
where tree-level relations between  fermion masses 
and  Yukawa interactions have been relaxed.  Yukawa  couplings
will be introduced in any case, since, after breaking the chiral 
 symmetry explicitly through the fermion mass terms in eq.~(\ref{LF}), 
they are no more protected against radiative corrections.
Even  assuming  vanishing Yukawa couplings 
at tree level, 
 divergent terms, proportional to 
fermion masses, will arise radiatively at loop level. 
The latter  can only be canceled by counterterms proportional 
to the Yukawa-Higgs-fermion interactions. 
Therefore, the operators 
$({Y_f}\bar{\psi}_{f} \psi_{f} H)$, even if not present in the classical
Lagrangian, will reappear at the quantum level, 
due to the nonvanishing Dirac mass terms.

Starting from the effective Lagrangian in eqs.(\ref{LAG})-(\ref{LF}),
new contributions  to the $\beta$ functions of Yukawa couplings are expected with respect to the SM results,
due to the fact that $m_f$ terms do not arise from spontaneous chiral symmetry breaking.
Since
the theory is not  renormalizable,
the off shell renormalization of the Yukawa operator
$O_{Y_f}=\bar{\psi_f} \psi_f H$
 implies in general higher-dimensional  
counterterms, like for instance $\sim (\partial_{\mu} \bar{\psi_f}) 
(\partial^{\mu}\psi_f) H$,  introducing  new coupling constants.
However, since here we are interested in {\it on shell}  Higgs coupling to fermions  (that are relevant, e.g., in Higgs boson decays), the Yukawa operator 
$O_{Y_f}$ will be renormalized by evaluating one-loop 
matrix elements with all external legs {\it on shell}.
Then, as can be checked by explicit calculation, all the divergencies 
can be reabsorbed in  Yukawa couplings
and field renormalization constants. 
This is no more true when the Higgs field is off shell, 
and the contribution from the mixing of  {\it off shell} 
higher-dimensional operators is to be taken into account in the renormalization of 
Yukawa couplings. Notice that this discussion 
does not apply to the SM, which is a renormalizable theory.

The relevant $\beta$ function will be worked out through the usual steps.
First, one has to derive the relation between the bare (${\rm Y}_f^0$) and the 
renormalized (${\rm Y}_f$) Yukawa couplings (where $f$ refers to a generic 
fermion field).  
In general, one has
\bea
{\rm Y}^0_f=Z_{{\rm Y}_f} Z_{{\psi_f}}^{-1} Z_H^{-1/2} {\rm Y}_f \,.
\label{Y}
\eea
As usual, the symbol $Z_{{\rm Y}_f}$ is the renormalization constant 
canceling  the divergencies
associated to the {\it one-particle irreducible}  matrix elements
of the Yukawa operator $O_{\rm Y_f}$, while the terms 
$Z_{{\psi_f}}$ 
and $Z_{H}$ are the  wave-function renormalization constants 
of the $f$ fermion and Higgs fields, respectively. 
Neglecting the contributions to the $\beta$ function 
induced by the CKM mixing in the
one-loop calculation, the renormalization
constants, as well as the bare and renormalized Yukawa couplings, 
will be diagonal matrices in the flavor space.

We perform the one-loop integrals in  $D$-dimensions, with  $D=4-\epsilon$, and subtract the divergent part 
by using the $\overline{{\rm MS}}$ renormalization scheme.
In dimensional regularization, the bare Yukawa coupling ${\rm Y}_f^0$
can be expressed as a Laurent series in $\epsilon$ 
\bea
{\rm Y}_f^0=\mu^{\epsilon/2}\left({\rm Y}_f+
\sum_{n=1}^{\infty}\frac{a_{n f}
({\rm Y}_f,g_i)} {\epsilon^n}\right)\,,
\label{Yb}
\eea
where $\mu$ is the  renormalization scale as defined 
in the $\overline{{\rm MS}}$ scheme, and $g_i$, with $i={ 1,2,3}$, indicate 
the SM gauge coupling constants associated to the ${\rm SU({3})_c
\times SU({2})_{\rm L}\times U({1})}_Y$ gauge group.
In eq.(\ref{Yb}) we have omitted the dependence on the renormalization scale 
$\mu$ in the (renormalized) Yukawa couplings and gauge couplings $g_i$.
Notice that the coefficients $a_{n f}({\rm Y}_f,g_i)$ 
in eq.(\ref{Yb}) depend on $\mu$ through the $\mu$ dependence of
both the renormalized Yukawa and gauge couplings.

As usual, the Yukawa $\beta$ function is defined as 
$\beta_{{\rm Y}_f}=d {\rm Y}_f/dt$, with $t=\log{\mu}$, and 
is connected to the residue of the simple pole $1/\epsilon$ 
in eq.(\ref{Yb}), that is to $a_{1f}({\rm Y}_f,g_i)$.
Then, according to the notation in eq.(\ref{Yb}), one obtains
\bea
\beta_{{\rm Y}_f}=\frac{1}{2}\left(-1+
\sum_{f^{\prime}} {\rm Y}_{f^{\prime}}\frac{\partial}{\partial {\rm Y}_{f^{\prime}}}
+\sum_{i=1}^3 
g_i\frac{\partial}{\partial g_i}\right) a_{1f}({\rm Y}_f,g_i)
\, ,
\eea
where the sum on $f^{\prime}$ runs 
over all the nonvanishing fermion contributions.

\begin{figure}[!htb]
\begin{center}
\dofig{3.1in}{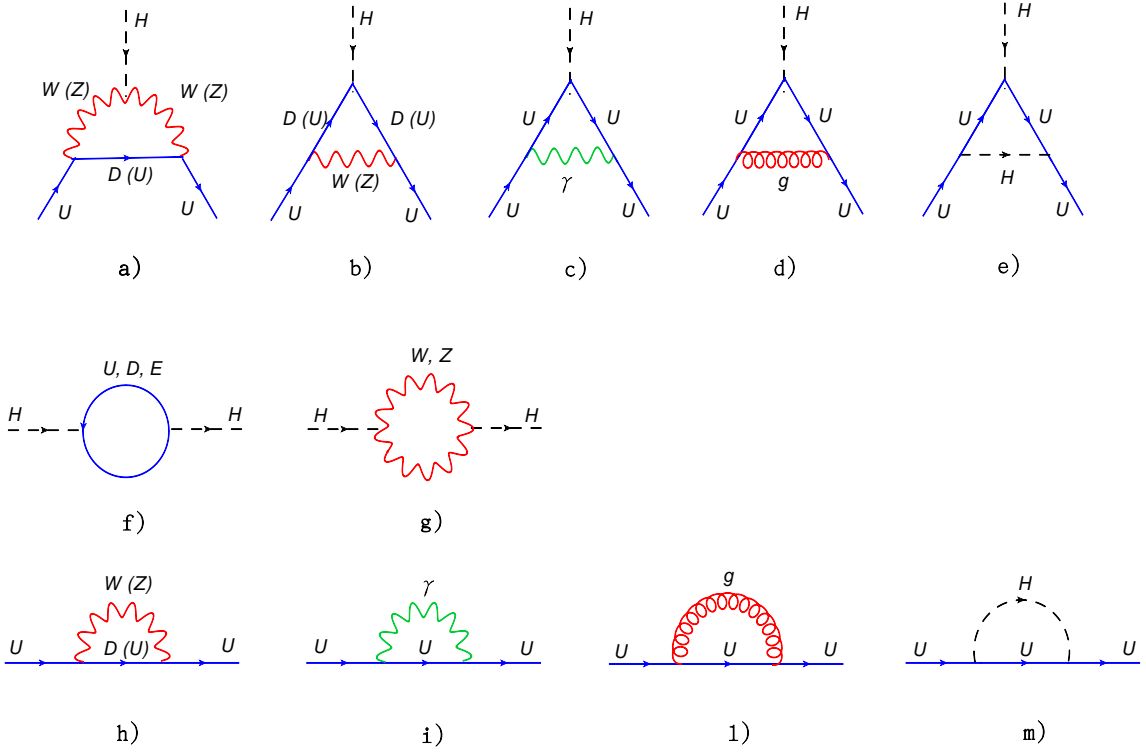}
\end{center}
\caption{\small One-loop Feynman diagrams  
 contributing, in the unitary gauge, to the $\beta$ function of the Yukawa couplings of
up-type quarks $U=\{u,c,t\}$. Labels $\gamma$ and $g$
 correspond to the photon and gluon propagators, respectively.
Diagrams in Figs. 1(a)-1(e) refer to the vertex corrections, while
1(f)-1(g) and 1(h)-1(k) refer to the Higgs boson $H$ and the up-type quarks' self-energy corrections,
respectively. The labels $D=\{d,s,b\}$ and $E=\{e,\mu,\tau\}$ 
stand for the down-type quarks and charged leptons, respectively.}
\label{Diag}
\end{figure}

The full set of the
one-loop diagrams contributing (in the unitary gauge for the $W$ and $Z$ propagators, and with the CKM matrix set to unity) to the term $a_{1f}$ 
in eq.(\ref{Yb}) 
 is given in Fig.\ref{Diag} for the up-type quarks $U=\{u,c,t\}$. (The case of down-type quarks $D=\{d,s,b\}$ and charged leptons $E=\{e,\mu,\tau\}$ can be obtained from Fig.\ref{Diag} in a straightforward way.)
Diagrams in Figs. 1(a)-1(e) correspond to
the vertex corrections, while Figs. 1(f)-1(g) and 1(h)-1(k) are the 
self-energy
corrections to the Higgs boson and up-type quark fields, respectively.
The vertex correction with two internal Higgs boson lines, and 
the Higgs self-energy diagram with a  Higgs boson loop 
have not been included in Fig.\ref{Diag}. Indeed, this vertex correction 
and the contribution to the Higgs wave-function renormalization 
arising from the Higgs-boson loop are finite, 
and so do not contribute to the $\beta$ function.

The RG equations for the effective Yukawa couplings  at one-loop are then given by
\bea
\frac{d {\bf Y_U}}{d t}&=&\frac{1}{16 \pi^2}\left\{
3\,\xi_H^2 \left( {\bf Y_U}- {\bf Y^{\scriptscriptstyle SM}_U}\right) - 
3{\bf Y^{\scriptscriptstyle SM}_U}{\bf Y^{\scriptscriptstyle SM}_D}\left({\bf Y_D}-{\bf Y^{\scriptscriptstyle SM}_D}\right)
+\frac{3}{2} {\bf Y_U}\left({\bf Y_U}{\bf Y_U} -
{\bf Y^{\scriptscriptstyle SM}_D} {\bf Y^{\scriptscriptstyle SM}_D}\right)
\right.
\nonumber\\
&-& \left. 
{\bf Y_U}\left(\frac{17}{20} g_1^2 +\frac{9}{4}g_2^2+8g_3^2-
{\bf Tr(Y)} \right)
\right\}\, ,
\label{RGE1}\\
\nonumber\\
\frac{d {\bf Y_D}}{d t}&=&\frac{1}{16 \pi^2}\left\{
3\, \xi_H^2 \left( {\bf Y_D}- {\bf Y^{\scriptscriptstyle SM}_D}\right) - 
3{\bf Y^{\scriptscriptstyle SM}_D}{\bf Y^{\scriptscriptstyle SM}_U}\left({\bf Y_U}-{\bf Y^{\scriptscriptstyle SM}_U}\right)
+\frac{3}{2} {\bf Y_D}\left({\bf Y^{}_D}{\bf Y_D} -
{\bf Y^{\scriptscriptstyle SM}_U}{\bf Y^{\scriptscriptstyle SM}_U} \right)
\right.
\nonumber\\
&-& \left. 
{\bf Y_D}\left(\frac{1}{4} g_1^2 +\frac{9}{4}g_2^2+8g_3^2
-{\bf Tr(Y)}\right) 
\right\}\, ,
\label{RGE2}\\
\nonumber\\
\frac{d {\bf Y_E}}{d t}&=&\frac{1}{16 \pi^2}\left\{
3\, \xi_H^2 \left( {\bf Y_E}- {\bf Y^{\scriptscriptstyle SM}_E}\right) 
+\frac{3}{2} {\bf Y_E}{\bf Y^{}_E}{\bf Y_E}
-{\bf Y_E}\left(\frac{9}{4}\left( g_1^2 +g_2^2\right)
-{\bf Tr(Y)} \right)
\right\}\, ,
\label{RGE3}
\eea
where ${\bf Tr}$ stands for the trace, and the matrix ${\bf Y }$ is defined as
\bea
{\bf Y }&\equiv &N_c {\bf Y^{}_U}{\bf Y_U}+
N_c {\bf Y^{}_D}{\bf Y_D}+{\bf Y^{}_E}{\bf Y_E}\, .
\eea
In the above equations, we used the  notation ${\bf Y_{U,D,E}}$ for the Yukawa couplings, 
 that stands for 
diagonal $3\times3$ real matrices in flavor space (where ${\bf U,D,E}$ 
stand for up-quarks, down-quarks and charged leptons, respectively). Then, products of 
${\bf Y_{U,D,E}}$ are understood in the matrix space. 
The remaining symbols  in eqs.(\ref{RGE1})-(\ref{RGE3})  are defined 
as
\bea
\xi_H\equiv \frac{g_2 m_H}{2 M_W},~~~~~ {\bf Y^{\scriptscriptstyle SM}_f}\equiv
\frac{g_2}{\sqrt{2} M_W} {\rm diag}[  m_{\rm f_1},m_{\rm f_2},m_{\rm f_3}],~~~
g_1^2\equiv \frac{5}{3}\frac{e^2}{\cos^2{\theta_W}}\, ,
\label{defin}
\eea
where ${\bf Y^{\scriptscriptstyle SM}_f}$ is a diagonal matrix in flavor space,
$m_{\rm f_{ i}}$ being the fermion pole-masses,
with ${\bf f=U,D,E}$,  and $N_c=3$ the number of colors. 
The RG equations for the gauge couplings are the same as in the SM\cite{arason}
\bea
\frac{d g_i}{dt}= -b_i \frac{g_i^3}{16 \pi^2}\, ,
\label{RGEg}
\eea
with
\bea
b_1&=&-\frac{4}{3} n_{g} -\frac{1}{10}\, ,
\nonumber\\
b_2&=&\frac{22}{3}-\frac{4}{3} n_{g} -\frac{1}{6}\, ,
\nonumber\\
b_3&=&11-\frac{4}{3}n_g\, ,
\nonumber
\eea
and $n_g=3$ is the number of fermion generations.
The parameters
${\bf Y^{\scriptscriptstyle SM}_f}$ in eq.(\ref{defin}) 
are the explicit chiral-symmetry breaking parameters, that depend
 on the running scale $t=\log{\mu}$ through 
 the weak-gauge coupling $g_2(t)$. They coincide with 
  the tree-level SM 
Yukawa couplings. Vertex and self-energies diagram contributions
entering in eqs.(\ref{RGE1})-(\ref{RGE3}) have been cross-checked 
with the SM results for the unitary gauge in \cite{BP}.

We stress that $ {\bf Y^{\scriptscriptstyle SM}_f}$ in eqs.(\ref{RGE1})-(\ref{RGE3}) is kept independent from the Yukawa couplings.
For 
$ {\bf Y^{\scriptscriptstyle SM}_f} \to {\bf Y_f}$, 
 the Higgs mechanism for the fermion mass generation,
and  the corresponding SM RG equations \cite{arason} are  recovered.

Another interesting limit is when all the Yukawa couplings
are set to zero at some scale. Then, the diagrams
with two $W$ and two  $Z$ exchange in the loop [cf. Fig.1(a)] give the leading contribution to 
the $\beta$ function, since they are both divergent in the unitary gauge.
Indeed, in the limit ${\bf Y_f}\to 0$, 
the $\beta$ functions are nonvanishing, and the leading 
contribution is proportional to 
corresponding fermion mass times the Higgs mass squared
\bea
\beta_{\bf Y_U}({\bf Y_f}\to 0) &=& 
-\frac{3 {\bf Y^{\scriptscriptstyle SM}_U}  }{16 \pi^2} \left(\xi_H^2\,-\, ({\bf Y^{\scriptscriptstyle SM}_D})^2
\right)\, ,
\label{betalim1}
\\
\beta_{\bf Y_D}({\bf Y_f}\to 0) &=& 
-\frac{3 {\bf Y^{\scriptscriptstyle SM}_D}  }{16 \pi^2} \left(\xi_H^2\,-\, 
({\bf Y^{\scriptscriptstyle SM}_U})^2\right)\, ,
\label{betalim2}
\\
\beta_{\bf Y_E}({\bf Y_f}\to 0) &=& 
-\frac{3 {\bf Y^{\scriptscriptstyle SM}_E}  }{16 \pi^2} \,\,\xi_H^2\, .
\label{betalim3}
\eea
\\
The structure of the r.h.s. of eqs.(\ref{betalim1})-(\ref{betalim3}) 
is crucial  to radiatively induce effective Yukawa couplings 
at low energy.
Indeed, due to the {\it explicit}
breaking of  chiral symmetry by fermion masses, the right-hand side of eqs.(\ref{RGE1})-(\ref{RGE3})  does not
vanish in the ${\bf Y_f}\to 0$ limit. Hence,
Yukawa couplings  receive  radiative logarithmic contributions at low energy, even if they are vanishing at
some high-energy scale.

The appearance of  terms proportional to $\,\xi_H^2\sim m_H^2/M_W^2$  in the right-hand side of eqs.(\ref{RGE1})-(\ref{RGE3}) 
can be interpreted as a manifestation of the nonrenormalizability 
of the theory. These terms come from the divergent part of the vertex
diagrams with two $W$ or two $Z$ running in the loop  
[cf. Fig.1(a)], and from the $W$ and $Z$ loop 
contribution to the Higgs self-energy diagram 
[cf. Fig.1(g)]. Their contribution 
vanishes when 
$
{\bf Y^{\scriptscriptstyle SM}_f} \to {\bf Y_f}$ in 
eqs.(\ref{RGE1})-(\ref{RGE3}).
When considering the renormalization of the 
Yukawa operator with an {\it off-shell} Higgs field, an analogous 
divergent term proportional to the Higgs-momentum square $q^2$ 
will appear from the same diagrams. 
This term can be reabsorbed in the renormalization of
the dimension-6 operator $\sim (\partial_{\mu} \bar{\psi_f}) 
(\partial^{\mu}\psi_f) H$, that explicitly breaks the renormalizability
of the theory.
In general, due to the nonrenormalizability of the effective theory, 
at higher orders in perturbation theory 
we expect new contributions to the
beta-function proportional to higher powers of $q^2/M^2$,  
where M is an effective scale (inversely proportional 
to the fermion masses), that is about 3 TeV for the top-quark case,
and much higher for the lighter fermions [cf. eqs.(\ref{RGE1})-(\ref{RGE3}), when $ m_H^2\sim q^2 $].
Therefore, if we assume that the characteristic energy of the process
(that, in Higgs decays, is $q^2 = m_H^2$) is well below this scale,
higher-order effects coming from the truncation of the 
perturbative series can be safely neglected. 
Clearly, this 
statement is automatically fulfilled for the light Higgs boson scenario, 
where $m_H \le 150$ GeV.

In order to connect the Yukawa couplings ${\rm Y}_f(m_H)$ at the low energy scale $m_H$ with their values ${\rm Y}_f(\Lambda)$
at some high-energy scale $\Lambda$, 
we will numerically integrate the RG equations in 
eqs.(\ref{RGE1})-(\ref{RGE3}) from 
$\mu=\Lambda$ down to $\mu=m_H$. 
The  actual solution will depend on the choice of
 boundary conditions for the Yukawa couplings ${\rm Y}_f(\Lambda)$,
while for the gauge
couplings we assume as input their experimental central values at the  scale $M_Z$.

We define 
$\Lambda$ as the renormalization scale where all the 
Yukawa couplings  vanish
\footnote{Notice that in the SM this condition would have trivial 
implications.
Indeed, due to the Higgs mechanism,
if the Yukawa couplings (and hence fermion mass parameters) are set to zero at some renormalization 
scale, they would be zero also at any other scale, and fermion mass terms would
not be generated. Moreover, the fact that all the Yukawa are required to vanish 
at the same scale follows from some sort of flavor 
universality in the mechanism of fermion mass generation}. 
This choice is motivated by the present  guess that the 
Higgs mechanism is decoupled from the chiral-symmetry-breaking 
mechanism at some high-energy scale.
In principle, the condition of vanishing Yukawa couplings at the scale $\Lambda$ could seem  an  oversimplification, when considered in the framework 
of a realistic NP model that could explain the 
 fermion-mass generation above the scale $\Lambda$.
Indeed, since Yukawa couplings should vanish 
in the limit $m_f\to 0$, one expects, on dimensional grounds, 
a suppression ${\rm Y}_f(\Lambda)\sim {\cal O}({m_f}/\Lambda)$, in case 
the fermion mass parameters 
 are the only chiral-symmetry-breaking parameters of the
theory\footnote{Different scenarios could also be envisaged.}.
Then it is straightforward  to check that, 
if $\Lambda$ is a few orders of magnitude 
above the EW scale, the solution of the RG equations with 
${\rm Y}_f(\Lambda)\sim {\cal O}({m_f}/\Lambda)$ is well approximated 
by   vanishing boundary conditions for the Yukawa couplings.

\begin{figure}[tpb]
\begin{center}
\dofigs{3.1in}{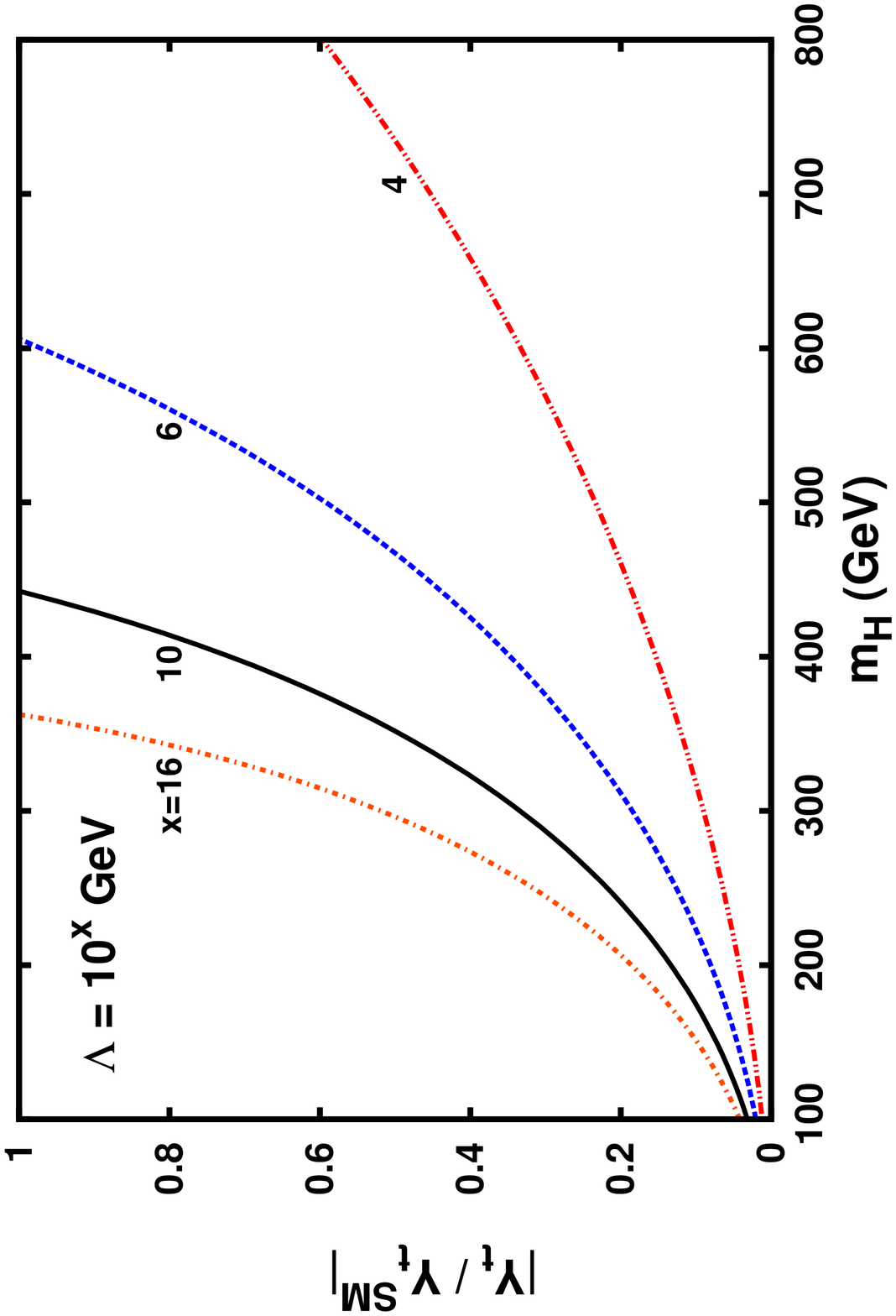}{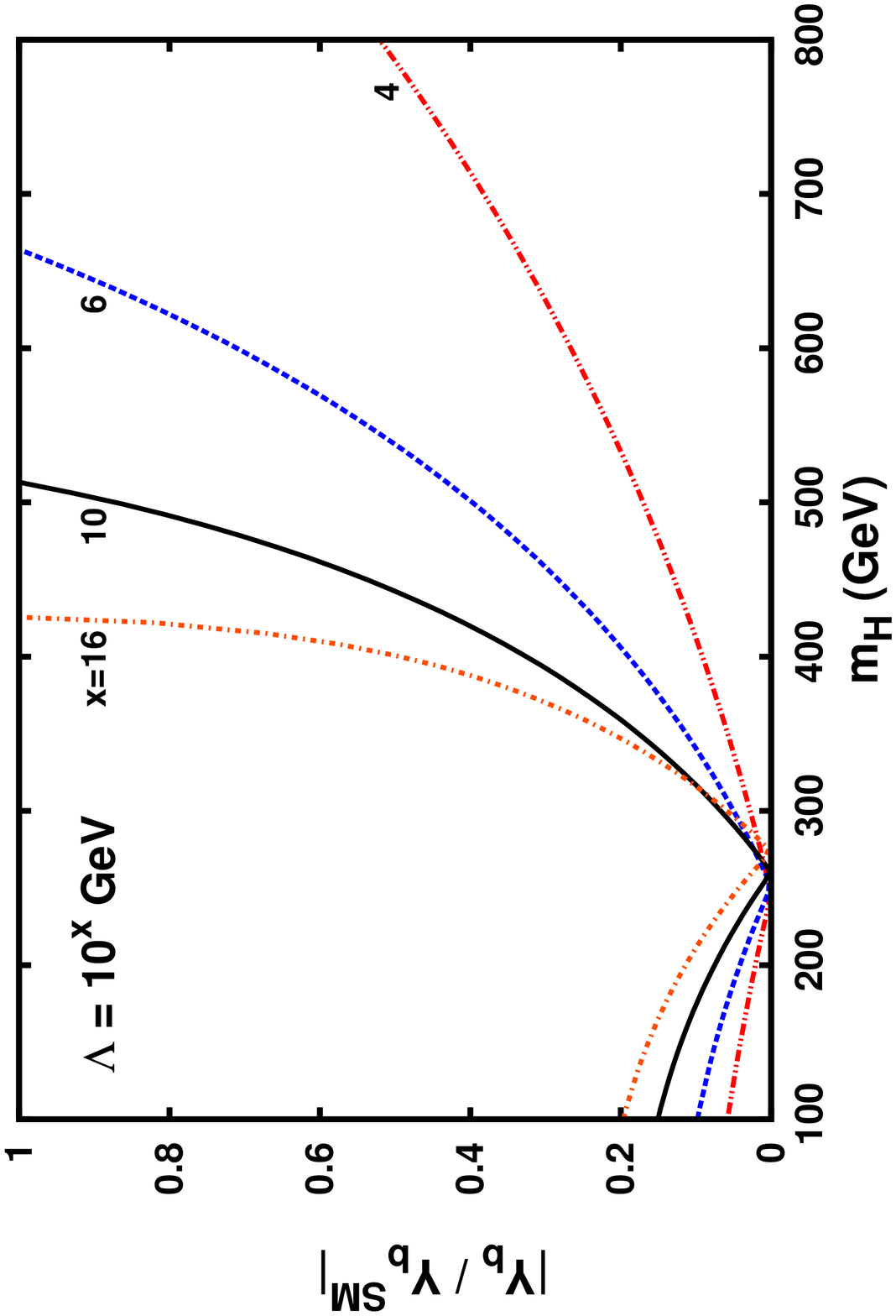}
\end{center}
\caption{\small Absolute values of the effective Yukawa couplings
for top quarks (left) and bottom  quarks (right), 
$| {\rm Y}_f(m_H)/{\rm Y}^{\rm {\scriptscriptstyle SM}}_f|$,
evaluated at the $m_H$ scale, and normalized to their 
respective tree-level SM values, versus $m_H$, and for different 
 $\Lambda$'s.
We assume 
$ {\rm Y}^{\rm {\scriptscriptstyle SM}}_t=0.997$ and 
$ {\rm Y}^{\rm {\scriptscriptstyle SM}}_b=0.0284$, corresponding,
respectively, to the pole masses $m_t$=171.3 GeV and $m_b$=4.88 GeV, and $g_2=g_2(M_Z)$ in eq.(\ref{defin}).}
\label{Yukawas}
\end{figure}

Because of the term proportional to $\,\xi_H^2\sim m_H^2/M_W^2$ in eq.(\ref{betalim1}), for large $m_H$, the 
renormalized top Yukawa coupling ${\rm Y}_{t}(m_H)$, 
evaluated at the scale $\mu=m_H$,  
could become very large due to  RG equation effects, and could spoil the perturbative approach in the Yukawa
sector. 
The requirement that the top Yukawa coupling 
does not exceed  unity at the EW scale can be translated   into
 an upper bound on the Higgs boson mass as a function 
of $\Lambda$.
In particular, if we  retain only the leading contribution in the
$\beta$ function of the top Yukawa coupling, the  
RG equation in eq.(\ref{RGE1}) becomes
\bea
\frac{d {\rm Y}_{t}}{d t}&\simeq &\frac{3}{16 \pi^2}
\,  \xi^2_H \left( {\rm Y}_{t}- {\rm Y}^{\rm {\scriptscriptstyle SM}}_{t}
\right) \, .
\label{RGEred}
\eea
Equation (\ref{RGEred}) can be easily solved 
in the approximation of assuming  $g_2(t)$ as a 
constant. Setting the $g_2(t)$ value at the $M_Z$ scale,
by requiring  ${\rm Y}_{t}(M_Z)< 1$, one then 
 obtains the following upper bound on the Higgs mass
\bea
m_H < 2\pi \sqrt{\log{\left(\frac{1+{\rm Y}^{\rm {\scriptscriptstyle SM}}_{t}}{{\rm Y}^{\rm {\scriptscriptstyle SM}}_{t}}\right)}
\frac{2\sqrt{2}}{\rm 3\, G_F \log(\frac{\Lambda}{M_Z})}}\, .
\label{bounds}
\eea
Equation (\ref{bounds}) sets up the $m_H$ range, versus $\Lambda$, where perturbation theory in the Yukawa sector 
can be still reliable in the present scenario.
The numerical values of the upper bounds obtained from  eq.(\ref{bounds}) are  
$m_H^{\rm max} \simeq 687,\, 488,\, 346,\, 261$ GeV, 
for $\Lambda=10^{\{4,6,10,16\}}$,  respectively. These bounds are rather stronger than the ones 
derived from the exact solution of top-Yukawa eq.(\ref{RGE1})
evaluated at the $m_H$ scale. In particular, the exact $m_H$ bounds 
for ${\rm Y}_{t}(m_H)< 1$  are $m_H^{\rm max}\simeq \{611,\, 443,\, 366 \}$ GeV, 
for $\Lambda=10^{\{6,10,16\}}$ GeV, respectively,
while for $\Lambda=10^4$ GeV, there is 
basically no upper limits (below 1 TeV) on the Higgs mass.
The latter bounds can also  be worked out from Fig.\ref{Yukawas},
where we plot, versus $m_H$ and 
for different   $\Lambda$'s in the range $10^{4-16}$ GeV,  
the effective Yukawa couplings of bottom and top quarks,
evaluated at the $\mu=m_H$ scale,  and normalized 
to their tree-level SM values.

Figure \ref{Yukawas}  also shows  that for  
$m_H\sim (100-250)$ GeV, 
the absolute value of the $b$ quark effective Yukawa coupling 
decreases when $m_H$ increases, 
and  vanishes in the range $m_H\simeq (250-275)$ GeV, 
the exact $m_H$ value depending on  $\Lambda$.
The effective top-Yukawa coupling is 
always negative for $m_H> 100$ GeV, 
while the bottom-Yukawa coupling is positive (negative) for values of the Higgs mass below (above) the minimum of its 
absolute value.

\begin{table}[htbp]
\begin{center}
                                                                                           \begin{tabular}{|c|c|c|c|c|c|c|}                                   \hline { $m_H({\rm GeV})$} &                                      { $\Lambda({\rm GeV})$}                                           & { $|{\rm Y}_t(m_H)|$} & {                                       $|{\rm Y}_b(m_H)|$}  & { $|{\rm Y}_c(m_H)|$} &                    { $|{\rm Y}_{\tau}(m_H)|$} &                                      { $|{\rm Y}_{\mu}(m_H)|$}                                       \\ \hline \hline \multirow{4}{*}{100} & $10^4$ &
   1.2$\times 10^{-2}$                                   &
   1.6$\times 10^{-3}$                                   &
   1.2$\times 10^{-4}$                                   &
   1.4$\times 10^{-4}$                                   &
   8.5$\times 10^{-6}$                                  \\
     & $10^6$ &
   2.0$\times 10^{-2}$                                   &
   2.8$\times 10^{-3}$                                   &
   2.0$\times 10^{-4}$                                   &
   2.7$\times 10^{-4}$                                   &
   1.6$\times 10^{-5}$                                  \\
  & $10^{10}$ &
   3.2$\times 10^{-2}$                                   &
   4.3$\times 10^{-3}$                                   &
   3.0$\times 10^{-4}$                                   &
   4.7$\times 10^{-4}$                                   &
   2.8$\times 10^{-5}$                                  \\
  & $10^{16}$ &
   4.2$\times 10^{-2}$                                   &
   5.6$\times 10^{-3}$                                   &
   4.0$\times 10^{-4}$                                   &
   7.0$\times 10^{-4}$                                   &
   4.2$\times 10^{-5}$                                  \\
                                                                                                                                                                                                                                                                                                                                                                                                                                                                                                                                                                          \hline \hline \multirow{4}{*}{110}  & $10^4$ &
   1.4$\times 10^{-2}$                                   &
   1.6$\times 10^{-3}$                                   &
   1.4$\times 10^{-4}$                                   &
   1.7$\times 10^{-4}$                                   &
   1.0$\times 10^{-5}$                                  \\
     & $10^6$ &
   2.5$\times 10^{-2}$                                   &
   2.7$\times 10^{-3}$                                   &
   2.4$\times 10^{-4}$                                   &
   3.2$\times 10^{-4}$                                   &
   1.9$\times 10^{-5}$                                  \\
  & $10^{10}$ &
   3.8$\times 10^{-2}$                                   &
   4.1$\times 10^{-3}$                                   &
   3.7$\times 10^{-4}$                                   &
   5.7$\times 10^{-4}$                                   &
   3.4$\times 10^{-5}$                                  \\
  & $10^{16}$ &
   5.1$\times 10^{-2}$                                   &
   5.4$\times 10^{-3}$                                   &
   4.9$\times 10^{-4}$                                   &
   8.5$\times 10^{-4}$                                   &
   5.1$\times 10^{-5}$                                  \\
                                                                                                                                                                                                                                                                                                                                                                                                                                                                                                                                                                          \hline \hline \multirow{4}{*}{120}  & $10^4$ &
   1.7$\times 10^{-2}$                                   &
   1.5$\times 10^{-3}$                                   &
   1.6$\times 10^{-4}$                                   &
   2.0$\times 10^{-4}$                                   &
   1.2$\times 10^{-5}$                                  \\
     & $10^6$ &
   2.9$\times 10^{-2}$                                   &
   2.5$\times 10^{-3}$                                   &
   2.8$\times 10^{-4}$                                   &
   3.8$\times 10^{-4}$                                   &
   2.3$\times 10^{-5}$                                  \\
  & $10^{10}$ &
   4.6$\times 10^{-2}$                                   &
   4.0$\times 10^{-3}$                                   &
   4.4$\times 10^{-4}$                                   &
   6.8$\times 10^{-4}$                                   &
   4.0$\times 10^{-5}$                                  \\
  & $10^{16}$ &
   6.1$\times 10^{-2}$                                   &
   5.3$\times 10^{-3}$                                   &
   5.9$\times 10^{-4}$                                   &
   1.0$\times 10^{-3}$                                   &
   6.1$\times 10^{-5}$                                  \\
                                                                                                                                                                                                                                                                                                                                                                                                                                                                                                                                                                          \hline \hline \multirow{4}{*}{130}  & $10^4$ &
   1.9$\times 10^{-2}$                                   &
   1.4$\times 10^{-3}$                                   &
   1.9$\times 10^{-4}$                                   &
   2.3$\times 10^{-4}$                                   &
   1.4$\times 10^{-5}$                                  \\
     & $10^6$ &
   3.4$\times 10^{-2}$                                   &
   2.4$\times 10^{-3}$                                   &
   3.3$\times 10^{-4}$                                   &
   4.4$\times 10^{-4}$                                   &
   2.6$\times 10^{-5}$                                  \\
  & $10^{10}$ &
   5.4$\times 10^{-2}$                                   &
   3.8$\times 10^{-3}$                                   &
   5.2$\times 10^{-4}$                                   &
   8.0$\times 10^{-4}$                                   &
   4.8$\times 10^{-5}$                                  \\
  & $10^{16}$ &
   7.2$\times 10^{-2}$                                   &
   5.1$\times 10^{-3}$                                   &
   6.9$\times 10^{-4}$                                   &
   1.2$\times 10^{-3}$                                   &
   7.2$\times 10^{-5}$                                  \\
                                                                                                                                                                                                                                                                                                                                                                                                                                                                                                                                                                          \hline \hline \multirow{4}{*}{140}  & $10^4$ &
   2.2$\times 10^{-2}$                                   &
   1.3$\times 10^{-3}$                                   &
   2.1$\times 10^{-4}$                                   &
   2.6$\times 10^{-4}$                                   &
   1.6$\times 10^{-5}$                                  \\
     & $10^6$ &
   3.9$\times 10^{-2}$                                   &
   2.3$\times 10^{-3}$                                   &
   3.8$\times 10^{-4}$                                   &
   5.1$\times 10^{-4}$                                   &
   3.0$\times 10^{-5}$                                  \\
  & $10^{10}$ &
   6.3$\times 10^{-2}$                                   &
   3.6$\times 10^{-3}$                                   &
   6.0$\times 10^{-4}$                                   &
   9.3$\times 10^{-4}$                                   &
   5.5$\times 10^{-5}$                                  \\
  & $10^{16}$ &
   8.5$\times 10^{-2}$                                   &
   4.9$\times 10^{-3}$                                   &
   8.1$\times 10^{-4}$                                   &
   1.4$\times 10^{-3}$                                   &
   8.4$\times 10^{-5}$                                  \\
                                                                                                                                                                                                                                                                                                                                                                                                                                                                                                                                                                          \hline \hline \multirow{4}{*}{150}  & $10^4$ &
   2.5$\times 10^{-2}$                                   &
   1.2$\times 10^{-3}$                                   &
   2.4$\times 10^{-4}$                                   &
   3.0$\times 10^{-4}$                                   &
   1.8$\times 10^{-5}$                                  \\
     & $10^6$ &
   4.5$\times 10^{-2}$                                   &
   2.1$\times 10^{-3}$                                   &
   4.3$\times 10^{-4}$                                   &
   5.9$\times 10^{-4}$                                   &
   3.5$\times 10^{-5}$                                  \\
  & $10^{10}$ &
   7.3$\times 10^{-2}$                                   &
   3.4$\times 10^{-3}$                                   &
   7.0$\times 10^{-4}$                                   &
   1.1$\times 10^{-3}$                                   &
   6.4$\times 10^{-5}$                                  \\
  & $10^{16}$ &
   9.8$\times 10^{-2}$                                   &
   4.6$\times 10^{-3}$                                   &
   9.4$\times 10^{-4}$                                   &
   1.6$\times 10^{-3}$                                   &
   9.8$\times 10^{-5}$                                  \\
                                                                                                                                                                                                                                                                                                                                                                                                                                                                                                                                                                          \hline \hline \multirow{4}{*}{160}  & $10^4$ &
   2.8$\times 10^{-2}$                                   &
   1.1$\times 10^{-3}$                                   &
   2.7$\times 10^{-4}$                                   &
   3.3$\times 10^{-4}$                                   &
   2.0$\times 10^{-5}$                                  \\
     & $10^6$ &
   5.1$\times 10^{-2}$                                   &
   1.9$\times 10^{-3}$                                   &
   4.9$\times 10^{-4}$                                   &
   6.7$\times 10^{-4}$                                   &
   4.0$\times 10^{-5}$                                  \\
  & $10^{10}$ &
   8.3$\times 10^{-2}$                                   &
   3.2$\times 10^{-3}$                                   &
   8.0$\times 10^{-4}$                                   &
   1.2$\times 10^{-3}$                                   &
   7.3$\times 10^{-5}$                                  \\
  & $10^{16}$ &
   1.1$\times 10^{-1}$                                   &
   4.4$\times 10^{-3}$                                   &
   1.1$\times 10^{-3}$                                   &
   1.9$\times 10^{-3}$                                   &
   1.1$\times 10^{-4}$             \\ \hline \end{tabular}

\end{center} 
\caption[]{Absolute values of the
effective Yukawa couplings for $t,b,c$ quarks and $\tau, \mu$ leptons,
evaluated at the 
scale $m_H$, for different values of the Higgs mass and  the energy scale $\Lambda$.
The corresponding values of the SM Yukawa couplings
${\rm Y}^{\rm {\scriptscriptstyle SM}}_f$ [as defined in eq.(\ref{defin}), with 
 $g_2=g_2(M_Z)$] 
are given by ${\rm Y}^{\rm {\scriptscriptstyle SM}}_t=0.997$, 
${\rm Y}^{\rm {\scriptscriptstyle SM}}_b=0.0284$, 
${\rm Y}^{\rm {\scriptscriptstyle SM}}_c=9.54\cdot 10^{-3}$, 
${\rm Y}^{\rm {\scriptscriptstyle SM}}_{\tau}=0.0103$, 
${\rm Y}^{\rm {\scriptscriptstyle SM}}_{\mu}=6.15\cdot 10^{-4}$.
}
\label{Yeff} 
\end{table}
In Table \ref{Yeff},
we present the values of the effective Yukawa couplings
for  $t,b,c$ quarks and $\tau, \mu$ leptons, as obtained by numerically solving eqs.(\ref{RGE1})-(\ref{RGE3}), 
versus  $m_H$, and for $\Lambda=10^{4,6,10,16}$ GeV. For reference, the corresponding Yukawa-coupling SM values [defined as in  
  eq.(\ref{defin}), with $g_2=g_2(M_Z)$], 
are given by ${\rm Y}^{\rm {\scriptscriptstyle SM}}_t=0.997$, 
${\rm Y}^{\rm {\scriptscriptstyle SM}}_b=0.0284$, 
${\rm Y}^{\rm {\scriptscriptstyle SM}}_c=9.54\cdot 10^{-3}$, 
${\rm Y}^{\rm {\scriptscriptstyle SM}}_{\tau}=0.0103$, 
${\rm Y}^{\rm {\scriptscriptstyle SM}}_{\mu}=6.15\cdot 10^{-4}$.
One can see that effective bottom Yukawa coupling is of the order   $10\%$ of ${\rm Y}^{\rm {\scriptscriptstyle SM}}_b$, in the 
$m_H$ and $\Lambda$ ranges considered.

Table \ref{Yeff}  also shows that the top-quark Yukawa coupling is 
at most  ${\cal O}(10^{-2})-{\cal O}(10^{-1})$, for  
$m_H\simeq(100-160)$ GeV. Then, in 
 Higgs boson production at hadron colliders, 
the  gluon fusion 
turns out to be quite depleted with respect to the VBF mechanism. 
For larger $m_H$ (cf. Fig.\ref{Yukawas}), the top Yukawa coupling
increases, and the cross section for 
 gluon fusion can still be competitive, and even larger 
than the VBF cross section.

In the following sections, after a few comments on the theoretical consistency of the present approach, we will discuss the phenomenological implications of the present results.

\section{Possible open issues in the present theoretical model}
Before proceeding to a more phenomenological study,  
we will comment on  possible issues that could endanger  
the consistency of  the present theoretical approach to the ChSB. 
In particular, we will address 
the issue of tree-level unitarity, the presence of a possible fine-tuning in the eventual UV completion of the  theory, and the consistency of the model with EWPT.

 One could wonder whether, by  pushing apart the scales of 
the EW symmetry breaking and the fermion mass generation,  problems with 
(perturbative) tree-level unitarity would arise \cite{appelquist}. 
Indeed, 
when fermion masses are put in by hand, the partial-wave 
unitarity in  the tree-level fermion-antifermion
scatterings into massive gauge bosons will be spoiled at some c.m.   energy scale $E_{\rm 0}$, which is  proportional
to the inverse of the fermion masses $m_f$, 
$
E_{\rm 0}\simeq \frac{4\pi \sqrt{2}}{\sqrt{3 N_C} G_F m_f}\, ,
$
where $N_C=1(3)$ for leptons (quarks), and $G_F$ is the Fermi constant \cite{appelquist}.
Then, the corresponding unitarity bounds range from 3 TeV for the top-quark
up to $1.7\times 10^6$ TeV for the electron.
In a more recent analysis \cite{dicus},  multiple massive gauge boson productions
in fermion-antifermion scatterings has been considered, which provide even 
more stringent unitarity bounds for light fermions.
\\
Let us now suppose that
in the SM the scale of fermion mass generation is pushed  
above the scale of unitarity bounds.
As  recalled above, if we switch off the Higgs
coupling to fermions, the SM becomes nonrenormalizable and 
perturbation theory ceases to be a good approximation
for real processes involving energies above 
the scale of unitarity bounds.
Unless we want to restore {\it perturbative} unitarity at all energies  
(which would call for  new degrees of freedom with masses of the order of
the scale of unitarity bounds), the interpretation of these bounds in terms of the scale of fermion mass generation will be questionable.  
\\
In our approach, 
the theory keeps  consistent even at scales above unitarity  bounds, although the validity of 
perturbation theory will of course be subject to restrictions.
We will assume that unitarity will be recovered 
not by new fundamental massive particles, but, for instance,  by new nonperturbative phenomena 
 arising  in scattering processes at energies
above the scale of unitarity bounds. Regarding  processes 
at energies below the scale of unitarity bounds (that we are concerned with in the present analysis),
perturbation theory can still be  trusted, provided the effects of
potential higher-dimensional operators, connected 
 to the nonrenormalizability of the theory,  
are  under control, and do not spoil the perturbative expansion
(as discussed in Section 2).

Next, 
in our approach we implicitly assume that a UV completion of the theory, that is 
free from fine-tuning problems, can be realized. In general, ChSB implies EWSB, 
while EWSB does not necessarily imply ChSB. Therefore, in case 
the scale of fermion-mass generation is taken arbitrarily large 
with respect to the $W$ and $Z$ mass scale, concerns about a
potential fine-tuning  in  the EWSB and ChSB contributions 
to the weak gauge-boson masses are legitimate.
However, these expectations are natural only in case 
the mechanism of fermion mass generation is  perturbative, or, 
analogously, the hidden sector responsible for ChSB is weakly coupled.
There are examples in the literature, in the framework 
of nonperturbative ChSB mechanisms, where the aforementioned 
fine-tuning problem (or hierarchy problem) does not arise 
\cite{Miransky,Anselmi,Antola}.
In the model in \cite{Miransky}, 
a dynamical ChSB mechanism with large anomalous dimensions for the top-quark condensate is proposed, where the scale of ChSB can be as high as the Planck scale,
while no fine-tuning  in the $W$ and $Z$ masses at the EW 
scale are expected to arise. 
\\
More exotic interactions involving, for instance, Lorentz-violating  four-fermion operators suppressed by a scale above the GUT scale, have been conjectured in \cite{Anselmi} to dynamically trigger ChSB at low energy.
\\
An approach similar to our scenario has also been recently explored 
in the context of Technicolor models \cite{ETC,Antola}. 
In \cite{Antola},
a fundamental light Higgs is introduced to induce EWSB, while fermion masses are generated by extended-Technicolor  interactions well above the EWSB scale.

Finally, in our effective theory, quadratic divergencies in the
$W$ and $Z$ self-energies could in principle appear due to radiative corrections.
Although the latter will be in general proportional to fermion masses, 
they could spoil the EWPT 
even for moderate values of $\Lambda$, in case the fermion mass involved 
is the top-quark mass.
However, these corrections are not 
expected to arise at one loop, since 
Yukawa interactions are not entering the $W$ and $Z$ self-energies at this order.
Indeed, at one loop the 
oblique corrections  (S,T,U variables
\cite{Peskin}) are not affected when switching from the SM to our model. 
At two loop, an estimate of oblique corrections is less straightforward.
By a closer inspection of the superficial degree of divergency
in the corresponding two-loop diagrams, we expect that
terms proportional to the square of the cutoff $\Lambda$ be absent, hence 
taming a potential tension in the EWPT for large $\Lambda$.
This expectation is supported  by the well-known SM results
on the two-loop EW contributions to the $\rho$-parameter.
In this case, possible terms proportional 
to $g^4 m_H^2 m_t^2$ do not appear
in the asymptotic limit $m_H \gg m_t$, and the leading terms 
are only of order $g^4 m_t^2 \log^2(m_H/M_W)$ (see for instance 
\cite{Fleischer}). On this basis,  
we could conclude that 
terms proportional to $m_t^2 \Lambda^2$ should not arise at 
two-loop level, either.
Indeed, the Higgs mass appearing in the aforementioned
leading terms ${\cal  O}(g^4 m_t^2 \log^2(m_H/M_W))$ can be interpreted as the effective 
scale of ChSB for $m_H \gg M_W$. 
Clearly, a full understanding of this 
issue in our effective theory needs careful investigations 
at two-loop level.
\section{Higgs boson decay width and branching ratios}

We present now  numerical results for the Higgs branching ratios and width, for 
  $m_H\lsim 150$ GeV, as obtained on the basis of the effective Yukawa couplings shown in Table \ref{Yeff}.
We concentrate on  the Higgs mass range where the Higgs branching ratios show sensitivity to the scale $\Lambda$. This   occurs for  $m_H$ not too close to the threshold for decaying into two real $W$ bosons, where the radiative Yukawa contributions to the Higgs total width become negligible. 
We will see that, when $m_H$ approaches 150 GeV, the  phenomenological features of the effective Yukawa scenario 
can not be 
 distinguished from the  {\it pure} fermiophobic-Higgs scenario, where all the decays and production channels are computed by assuming that the Higgs boson couples to vector bosons as in the SM, while it  has vanishing  couplings to all fermions. In all the following tables and figures, the results corresponding to the {\it pure} fermiophobic-Higgs scenario will be labeled  ``FP".

As in the SM, the main Higgs decay channels can be classified into  two categories: i) 
the decays generated at  tree-level, $H\to f\bar{f}$
\cite{Hfftree,HffNLO,HdecayggA}, and $H\to VV$ \cite{HVVtree}, 
with  $V=W,Z$, and ii)
the loop-induced ones, $H\to \gamma \gamma$ \cite{HdecayggA,HdecayggB},
$H\to Z\gamma$ \cite{HZg}, and $H\to g g$ \cite{Hgg,Hglgl}.
Since  our analysis will be concerned with $m_H$ values 
below the $WW$ and $ZZ$ kinematical thresholds, 
the relevant tree-level Higgs decays will be the ones mediated by either on-shell or off-shell EW gauge bosons, namely $H\to WW^* \to W f\bar{f}^\prime,
\,\, H\to ZZ^*\to Z f\bar{f}$, and 
$H\to W^*W^* \to  f\bar{f}^\prime f^{\prime\prime}\bar{f}^{\prime
\prime\prime}, \,\,
H\to Z^*Z^*\to  f\bar{f} f^{\prime}\bar{f}^{\prime}$,
where  fermions $f$ are 
summed over all allowed species, and $W^*$ and $Z^*$ are understood to be 
off shell\footnote{ 
From now on, we will label all these final states  as $WW$ and $ZZ$, and omit the $^*$ symbol for
 {\it off shell} $W$ and $Z$.}.

For Higgs decays into a fermion-pair $H\to f \bar{f}$, 
we will use the tree-level SM widths \cite{Hfftree,HdecayggA},  with the SM 
Yukawa coupling replaced by the 
effective one evaluated at the $m_H$ scale [cf. eq.(\ref{Gammaff}) 
in the Appendix], thus 
 resumming all the leading logarithmic contributions 
$g_i^{2n}(\log{(\Lambda/m_H)})^n$ in the radiative corrections.

The one-loop Higgs decays will all be affected by the suppression 
 of  the top Yukawa coupling.
The amplitude for the 
Higgs boson into two gluons ($H\to gg$) decreases conspicuously
(cf. Fig.\ref{Yukawas}), with dramatic consequences 
on the Higgs production mechanisms at hadron colliders. The leading 
loop-induced Higgs decays are  then $H\to \gamma \gamma$
and $H\to Z \gamma$, that  receive  
contribution not only from  the top-quark, but also from 
 $W$ loops 
\cite{HdecayggA,HdecayggB,HZg}.

In the following, we will always take into account
the moderate effect of the radiatively generated
top-quark Yukawa coupling in  all the one-loop 
decay channels $H\to \gamma \gamma$,
 $H\to Z \gamma$, and  $H\to g g $,
by rescaling the SM top-quark loop amplitude by the ratio
${\rm Y}_t(m_H)/{\rm Y}_t^{\rm{\scriptscriptstyle SM}}$. 

Because of the smallness of the effective top-quark Yukawa coupling, 
neglecting two-loop effects 
in the $H\to gg$ decay amplitude will be  safe in the present analysis.
The analytical form for all  Higgs decay widths,  used in the following, are reported in the Appendix.

For the SM branching ratios (BR's), we adopted the widths 
of $H\to b\bar{b}$ and $H\to c\bar{c}$ at the (resummed) QCD next to leading order 
\cite{HffNLO},  while for  Higgs decays into other fermions 
we kept only the Born approximation with the
pole fermion masses \cite{Hfftree,HdecayggA}. Regarding the SM widths of 
one-loop-induced decays, we neglected higher-order corrections, 
while for the $H\to WW$ and $H\to ZZ$ decays we used the tree-level 
expression as in eq.(\ref{GammaVV}), which includes
the 2-, 3-, and 4-bodies decays \cite{HVV4f}.
\begin{figure}[tpb]
\begin{center}
\dofigs{3.1in}{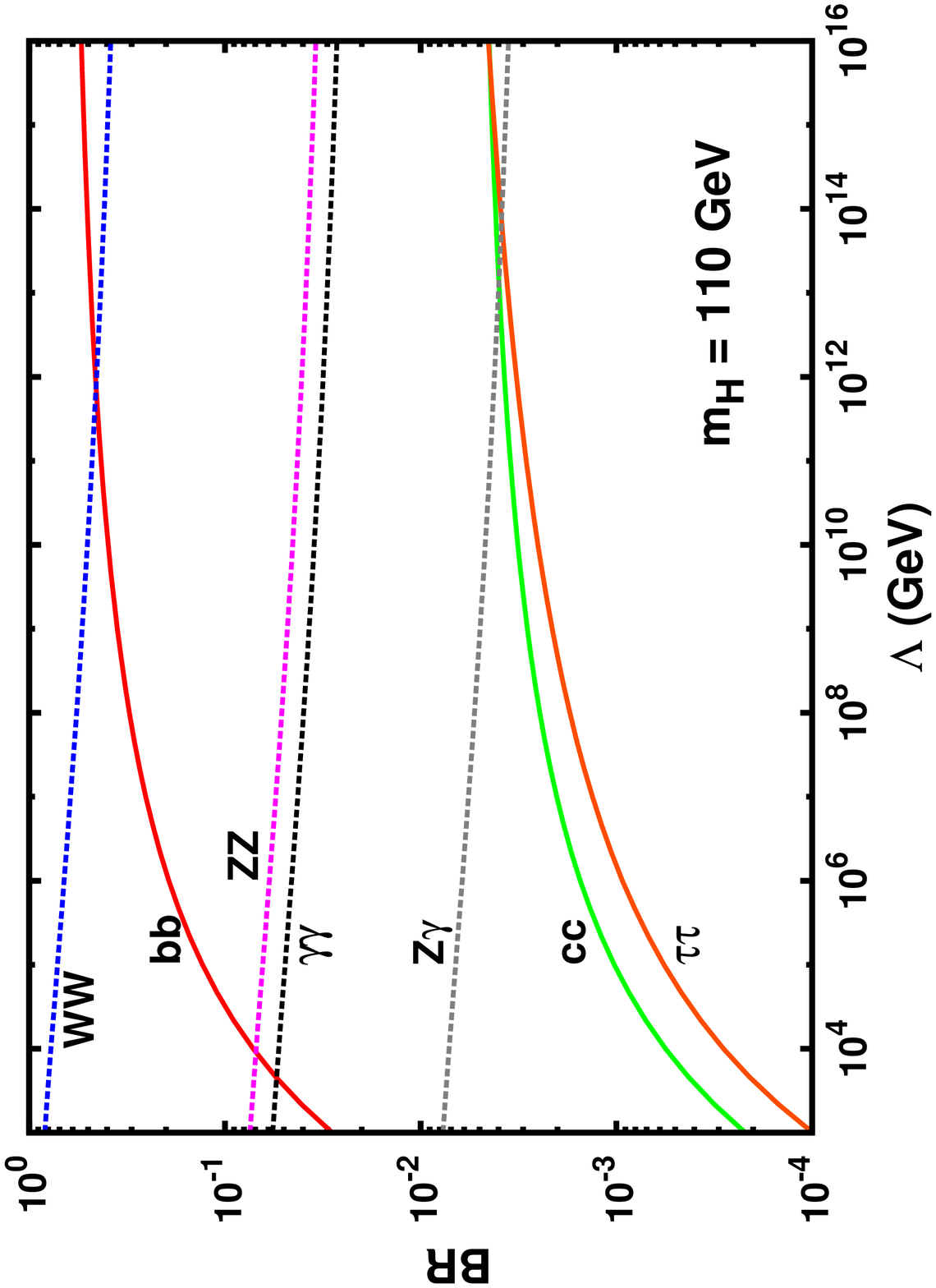}{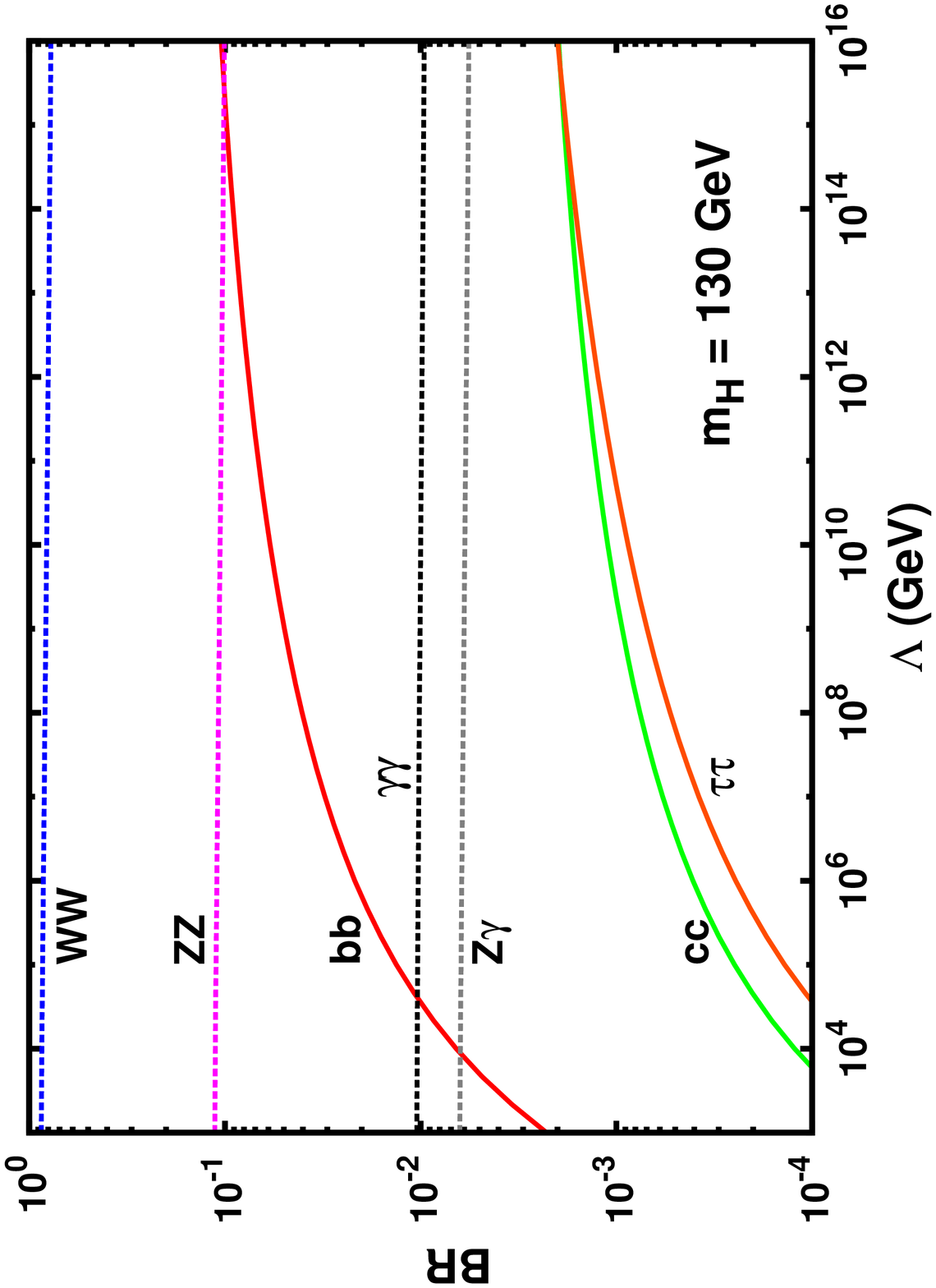}
\end{center}
\caption{\small Branching ratios for the Higgs-boson  decay into dominant channels, as a function of $\Lambda$, for $m_H=110$ GeV
(left), and $m_H=130$ GeV (right). }
\label{BR_scale}
\end{figure}

\begin{table}[htbp]
\begin{center}
\begin{tabular}{|c|c|c|c|c|c|c|c|c|}                
 \hline {\large
 $m_H$} & {$\Lambda$} &  $\gamma\gamma$ & $WW$ &  $ZZ$ & $Z\gamma$ & $b\bar{b}$ & $c\bar{c}$ & $\tau\bar{\tau}$ \\ \hline
 $({\rm GeV})$ & $({\rm GeV})$ & BR(\%) & BR(\%)  & BR(\%) & BR(\%) & BR(\%) & BR(\%) & BR(\%) \\ \hline \hline                       \multirow{4}{*}{100}                                              & $10^4$ &
  12                                   &
  52                                   &
         5.1                                   &
        0.26                                   &
  30                                   &
        0.15                                   &
       0.076                                  \\
     & $10^6$ &
         8.0                                   &
  33                                   &
         3.3                                   &
        0.17                                   &
  55                                   &
        0.28                                   &
        0.17                                  \\
  & $10^{10}$ &
         4.6                                   &
  19                                   &
         1.9                                   &
       0.094                                   &
  74                                   &
        0.38                                   &
        0.30                                  \\
  & $10^{16}$ &
         3.0                                   &
  12                                   &
         1.2                                   &
       0.062                                   &
  82                                   &
        0.43                                   &
        0.44          \\ \hline \multirow{2}{*}{  100     }
         & FP &
  18                                   &
  74                                   &
         7.4                                   &
        0.37                                   &
         0          &
         0          &
         0         \\
         & SM &
        0.15                                   &
         1.1                                   &
        0.11                                   &
       0.005                                   &
  82                                   &
         3.8                                   &
         8.3                                  \\
                                                                                                                                                                                                                                                                                                                                                                                                                                                                                                                                                                          \hline \hline \multirow{4}{*}{110}  & $10^4$ &
         5.3                                   &
  78                                   &
         7.0                                   &
        0.72                                   &
         9.1                                   &
       0.071                                   &
       0.036                                  \\
     & $10^6$ &
         4.6                                   &
  66                                   &
         5.9                                   &
        0.61                                   &
  22                                   &
        0.18                                   &
        0.11                                  \\
  & $10^{10}$ &
         3.5                                   &
  50                                   &
         4.5                                   &
        0.46                                   &
  41                                   &
        0.33                                   &
        0.26                                  \\
  & $10^{16}$ &
         2.7                                   &
  38                                   &
         3.4                                   &
        0.36                                   &
  54                                   &
        0.45                                   &
        0.45          \\ \hline \multirow{2}{*}{  110     }
         & FP &
         5.8                                   &
  86                                   &
         7.7                                   &
        0.79                                   &
         0          &
         0          &
         0         \\
         & SM &
        0.18                                   &
         4.6                                   &
        0.41                                   &
       0.037                                   &
  78                                   &
         3.6                                   &
         7.9                                  \\
                                                                                                                                                                                                                                                                                                                                                                                                                                                                                                                                                                          \hline \hline \multirow{4}{*}{120}  & $10^4$ &
         2.2                                   &
  85                                   &
         9.4                                   &
        0.75                                   &
         2.6                                   &
       0.032                                   &
       0.016                                  \\
     & $10^6$ &
         2.1                                   &
  81                                   &
         8.9                                   &
        0.72                                   &
         7.5                                   &
       0.092                                   &
       0.056                                  \\
  & $10^{10}$ &
         1.9                                   &
  72                                   &
         8.0                                   &
        0.64                                   &
  17                                   &
        0.21                                   &
        0.16                                  \\
  & $10^{16}$ &
         1.7                                   &
  64                                   &
         7.1                                   &
        0.57                                   &
  26                                   &
        0.32                                   &
        0.33          \\ \hline \multirow{2}{*}{  120     }
         & FP &
         2.3                                   &
  87                                   &
         9.7                                   &
        0.77                                   &
         0          &
         0          &
         0         \\
         & SM &
        0.21                                   &
  13                                   &
         1.5                                   &
        0.11                                   &
  69                                   &
         3.2                                   &
         7.0                                  \\
                                                                                                                                                                                                                                                                                                                                                                                                                                                                                                                                                                          \hline \hline \multirow{4}{*}{130}  & $10^4$ &
         1.0                                   &
  86                                   &
  11                                   &
        0.63                                   &
        0.84                                   &
       0.016                                   &
       0.008                                  \\
     & $10^6$ &
         1.0                                   &
  85                                   &
  11                                   &
        0.62                                   &
         2.6                                   &
       0.048                                   &
       0.029                                  \\
  & $10^{10}$ &
        1.0                                   &
  81                                   &
  11                                   &
        0.59                                   &
         6.1                                   &
        0.12                                   &
       0.092                                  \\
  & $10^{16}$ &
        0.96                                   &
  77                                   &
  10                                   &
        0.57                                   &
  10                                   &
        0.20                                   &
        0.20          \\ \hline \multirow{2}{*}{  130     }
         & FP &
         1.0                                   &
  87                                   &
  11                                   &
        0.63                                   &
         0          &
         0          &
         0         \\
         & SM &
        0.21                                   &
  29                                   &
         3.8                                   &
        0.19                                   &
  54                                   &
         2.5                                   &
         5.4                                  \\
                                                                                                                                                                                                                                                                                                                                                                                                                                                                                                                                                                          \hline \hline \multirow{4}{*}{140}  & $10^4$ &
        0.53                                   &
  87                                   &
  12                                   &
        0.48                                   &
        0.29                                   &
       0.008                                   &
       0.004                                  \\
     & $10^6$ &
        0.53                                   &
  86                                   &
  12                                   &
        0.48                                   &
        0.90                                   &
       0.026                                   &
       0.016                                  \\
  & $10^{10}$ &
        0.53                                   &
  85                                   &
  12                                   &
        0.47                                   &
         2.3                                   &
       0.064                                   &
       0.051                                  \\
  & $10^{16}$ &
        0.52                                   &
  83                                   &
  11                                   &
        0.46                                   &
         4.1                                   &
        0.11                                   &
        0.12          \\ \hline \multirow{2}{*}{  140     }
         & FP &
        0.53                                   &
  87                                   &
  12                                   &
        0.48                                   &
         0          &
         0          &
         0         \\
         & SM &
        0.19                                   &
  48                                   &
         6.6                                   &
        0.24                                   &
  36                                   &
         1.6                                   &
         3.6             \\ \hline \end{tabular}
\end{center} 
\caption[]{Branching ratios (in percentage) for  dominant Higgs boson decay channels, at different values of 
the Higgs mass and  $\Lambda$. The SM and FP rows present
the  SM  and  fermiophobic-Higgs scenario results, respectively.}
\label{TabBR} 
\end{table}

In Fig.\ref{BR_scale}, 
we show the results for  Higgs BR's into different final states,
for two representative values  $m_H=110$ GeV and $m_H=130$ GeV, versus the 
scale $\Lambda$, in the range $(10^{3}-10^{16})$ GeV. The numerical values
for the same BR's  are also reported in table \ref{TabBR}, for 
$\Lambda=10^{4,6,10,16}$ GeV. 
From   Fig.\ref{BR_scale}  (left), we can see that 
for $m_H=110$ GeV 
the ${\rm BR}(b\bar{b})$ can still be conspicuous, and even dominant over the other channels.
Its value is about 9.1\% for  $\Lambda=10^4$ GeV but can increase up to
40-50\%, for $\Lambda>10^{10}$ GeV. This is a clear effect of the 
resummation of the leading-log terms, that in BR($b\bar{b}$)
are particularly large. 
As a consequence, all the  BR's into two gauge bosons are also quite sensitive to the scale $\Lambda$, due to the  $\Lambda$ dependence of the Higgs total width.

The $H\to WW$ channel has a dominant role, too.
For $m_H=110$ GeV, ${\rm BR}(WW)$ 
 varies from 78\%, at $\Lambda=10^{4}$ GeV, down to 38\%, at $\Lambda=10^{16}$ GeV.
\\ 
 Regarding the $H\to \gamma\gamma$ channel,
its BR is about 5.3\% for $\Lambda =10^{4}$ GeV, and falls  
down to 2.7\%, at $\Lambda=10^{16}$ GeV
(cf. table \ref{TabBR}). 
Although smaller than  BR$(WW)$ and BR$(\bar{b}b)$, the BR($\gamma\gamma$) is 
strongly enhanced with respect to the SM value [${\rm BR}_{\rm SM}(\gamma\gamma)=0.18\%$], 
when fermion decay modes
are radiatively generated.

 BR$(ZZ)$ and BR$(Z\gamma)$  are also quite sensitive to the 
scale $\Lambda$. They are enhanced with respect to  their SM values, and depleted with respect to the fermiophobic-Higgs, for $m_H=110$ GeV (cf. table \ref{TabBR}). 
Note that ${\rm BR}(c\bar{c})$ and
${\rm BR}(\tau\bar{\tau})$ become comparable to 
 BR$(Z\gamma)$ when $\Lambda > 10^{10}$ GeV, for $m_H=110$ GeV.

At $m_H=100$ GeV, the enhancement of all the fermionic decays   is even more dramatic (cf. table \ref{TabBR}). 
On the other hand, 
when $m_H>110$ GeV, the BR hierarchy  for the different 
decay channels gets modified [see Fig.\ref{BR_scale}  (right)]. The fermionic decays get depleted,
while both BR$(WW)$ and BR$(ZZ)$ increase, and lose sensitivity to $\Lambda$, due to the fast grow of the 
$H\to WW$ and $H\to ZZ$ decay widths when approaching the 
real $WW$ threshold.
The rates for one-loop decays  ($H\to \gamma\gamma$, particularly)  also get smaller.

At $m_H\simeq150$ GeV, one can hardly distinguish the effective Yukawa's scenario from the pure fermiophobic-Higgs one, if not at  very large $\Lambda$. One can reach BR$(\bar{b}b)\gsim 0.3\%$ for $\Lambda \gsim 10^{6}$ GeV, and BR$(\bar{b}b)\gsim 1.4\%$ for $\Lambda \gsim 10^{16}$ GeV.
\begin{figure}[tpb]
\begin{center}
\dofourfigs{3.1in}{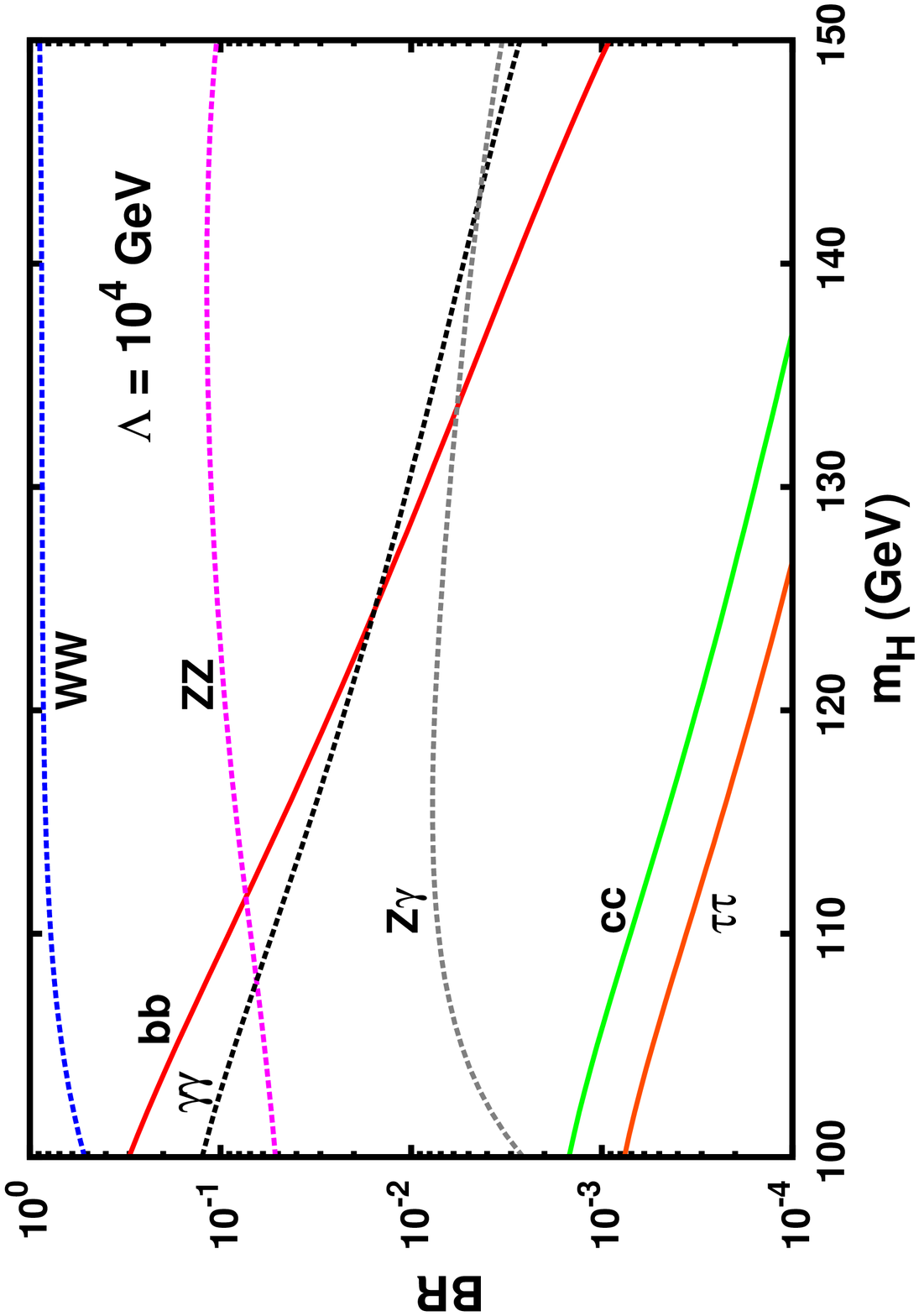}{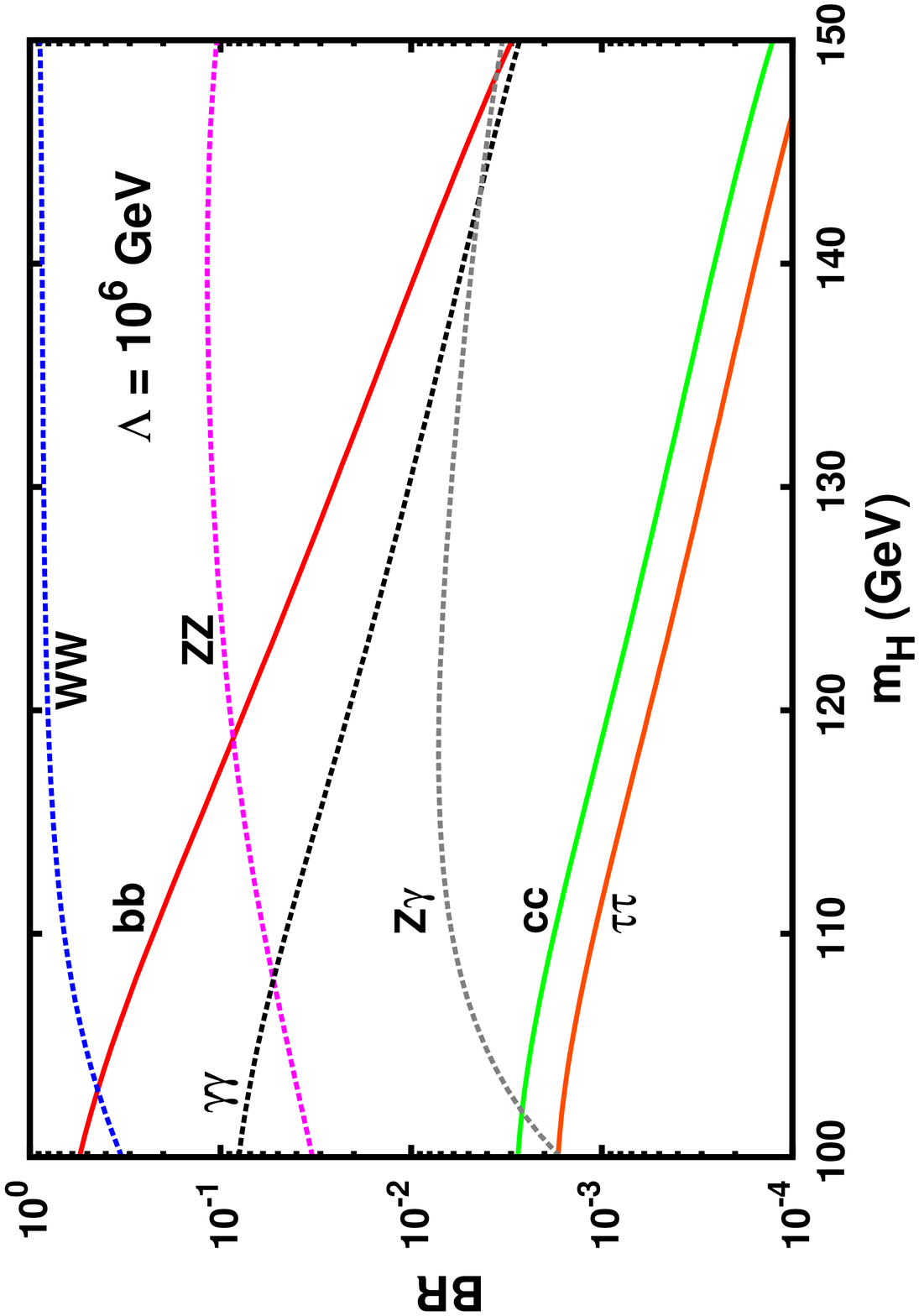}{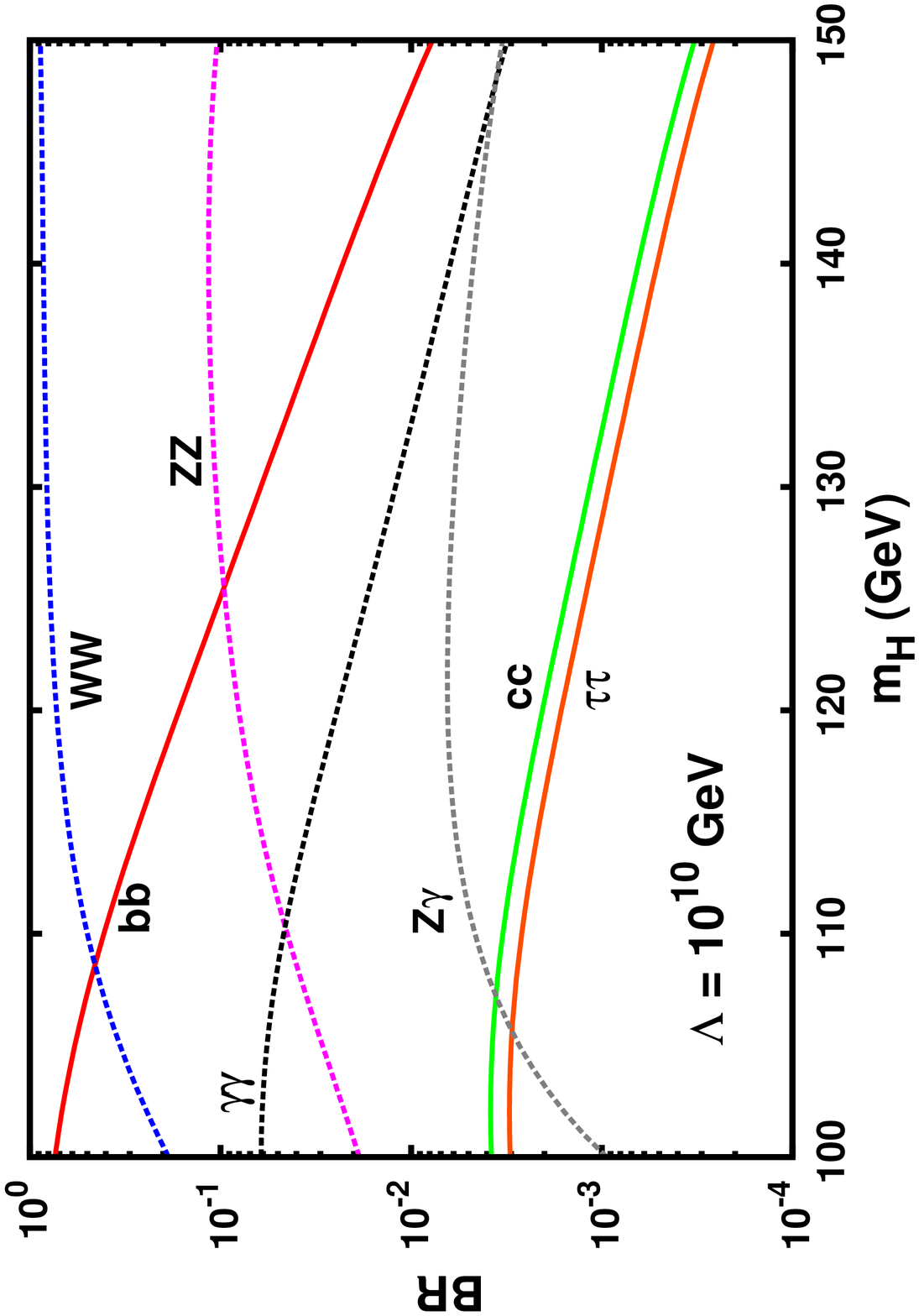}{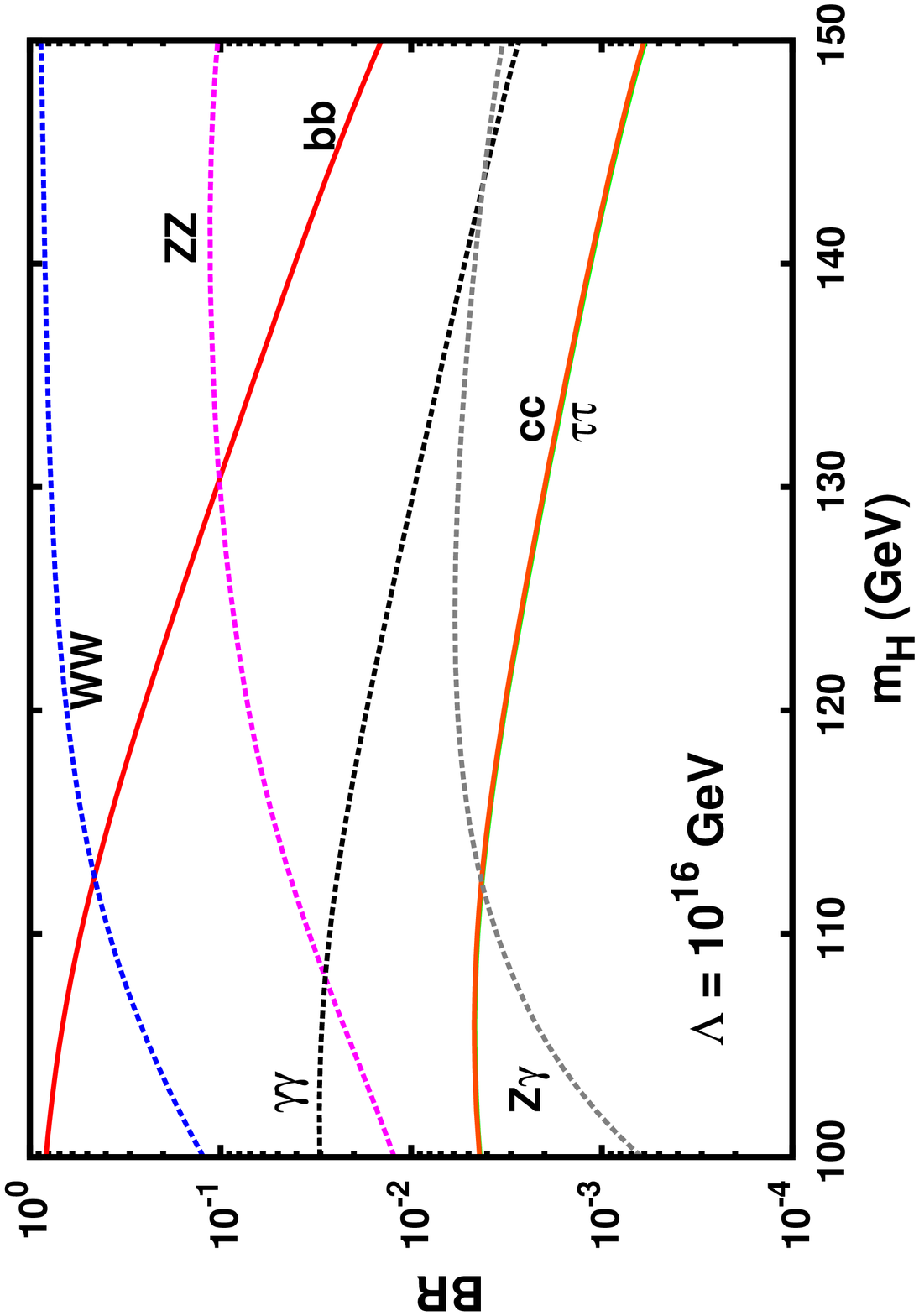}
\end{center}
\caption{\small Higgs branching ratios for different decay channels  versus the Higgs mass, for four representative values 
$\Lambda=10^{4,6,10,16}$ GeV.}
\label{BR_MH}
\end{figure}
This can be checked in Fig.\ref{BR_MH}, where we show the main   Higgs BR's versus the Higgs mass, 
 for   $\Lambda=10^{4,6,10,16}$ GeV. 
 While  the tree-level Higgs BR's in EW gauge bosons
become almost insensitive to the scale $\Lambda$ for Higgs masses larger
than 130 GeV, 
 the Higgs rates for decays into 
$b\bar{b}$, $c\bar{c}$, and $\tau\bar{\tau}$, although depleted 
at $m_H\sim 150$, keep their sensitivity to $\Lambda$ in all the $m_H$ range considered\footnote{BR($\mu{\mu}$), that is  
of order ${\cal O}(10^{-5})$, is also very sensitive to the scale $\Lambda$.}.

\begin{table} \begin{center}

                                                                                                                                                                                                                                                                                                                         \begin{tabular}{|c||c|c|c|c|c|}                                     \hline ${\rm \Lambda({\rm GeV})}$                               & $\Gamma_{\rm H}^{100}({\rm MeV})$ &                             $\Gamma_{\rm H}^{110}({\rm MeV})$  &                              $\Gamma_{\rm H}^{120}( {\rm MeV})$ &                              $\Gamma_{\rm H}^{130}( {\rm MeV})$ &                              $\Gamma_{\rm H}^{140}( {\rm MeV})$                                \\ \hline \hline $10^4$ &     5.3$\times 10^{-2}$          &    1.7$\times 10^{-1}$          &    5.8$\times 10^{-1}$          &    1.7          &    4.6            \\ \hline
                                                                                                                                                                                                                                                                                                                                                                                                                                                                                                           $10^6$ &     8.3$\times 10^{-2}$          &    2.0$\times 10^{-1}$          &    6.1$\times 10^{-1}$          &    1.7          &    4.7                                                                                                                \\ \hline
                                                                                                                                                                                                                                                                                                                                                                                                                                                                                                        $10^{10}$ &     1.5$\times 10^{-1}$          &    2.7$\times 10^{-1}$          &    6.8$\times 10^{-1}$          &    1.8          &    4.7                                                                                                                \\ \hline
                                                                                                                                                                                                                                                                                                                                                                                                                                                                                                        $10^{16}$ &     2.2$\times 10^{-1}$          &    3.5$\times 10^{-1}$          &    7.6$\times 10^{-1}$          &    1.9          &    4.8                                                                                                               \\ \hline 
                                                                                                                                                                                                                                                                                                                                                                                                                                                                                                               FP &     3.7$\times 10^{-2}$          &    1.6$\times 10^{-1}$          &    5.6$\times 10^{-1}$          &    1.7          &    4.6                                                                                                               \\ \hline 
                                                                                                                                                                                                                                                                                                                                                                                                                                                                                                               SM &     2.6          &    3.0          &    3.7          &    5.1          &    8.3                                                                                                 \\ \hline \end{tabular} 
  
\caption[]{Total width $\Gamma_{H}^{m_H}$ of the Higgs boson 
for different $m_H$ (expressed in GeV) and $\Lambda$. 
The last two rows show the fermiophobic-Higgs
(FP) and SM cases, respectively.}
\label{TabWidth} 
\end{center} 
\end{table}

In table \ref{TabWidth}, we show the results for the Higgs 
total width $\Gamma_H^{m_H}$  for different values of 
the Higgs mass and $\Lambda$.
For comparison, the corresponding FP and SM values are shown in the
last two rows of the same table. 
At $m_H=110$ GeV, a considerable depletion of 
the total width  is seen with respect to the SM one 
($\Gamma_{\rm SM}=3.0$ MeV). The corresponding $\Gamma_H$
ranges from $\Gamma_H=0.17$ MeV
for $\Lambda=10^4$ GeV up to $\Gamma_H=0.35$ MeV 
for  $\Lambda=10^{16}$ GeV. This is mainly due to the 
 $\Gamma(H\to b\bar{b})$ suppression by Yukawa effective couplings.
For larger $m_H$, the total width substantially approaches
the SM value (and matches the FB value), since  the tree-level decays  $H\to WW,ZZ$ become dominant near the  
$WW$ threshold.

\begin{figure}[tpb]
\begin{center}
\dofourfigs{3.1in}{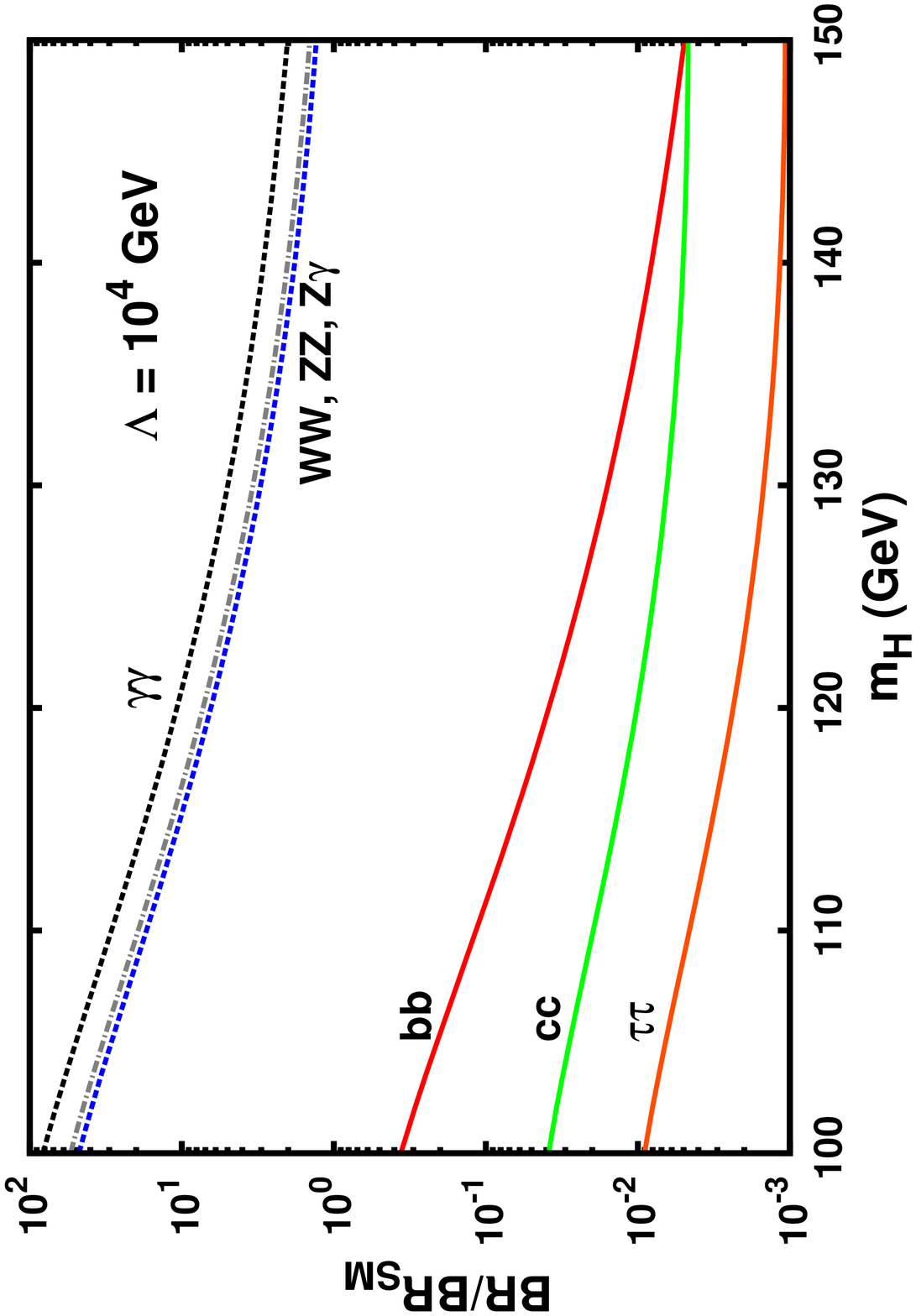}{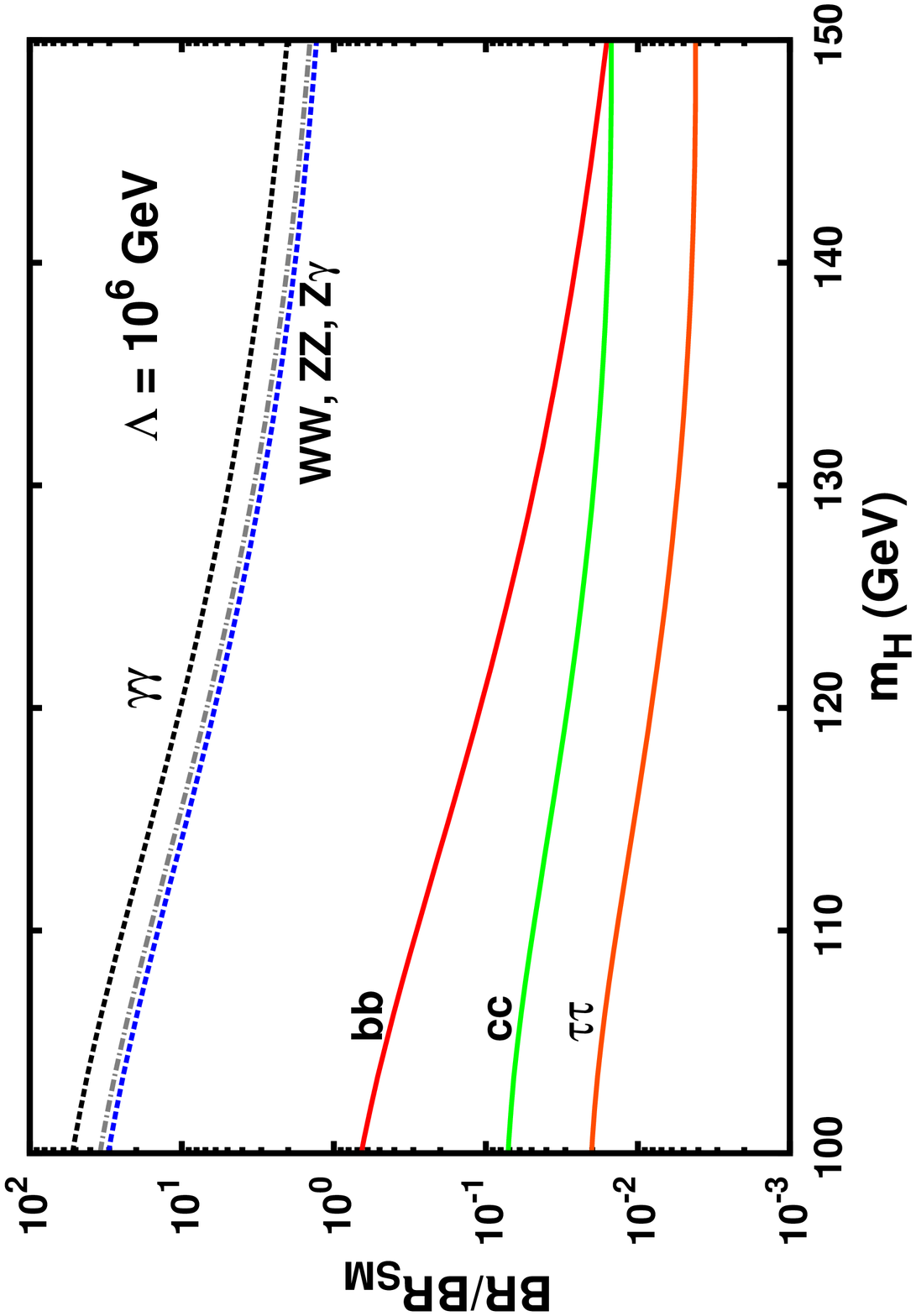}{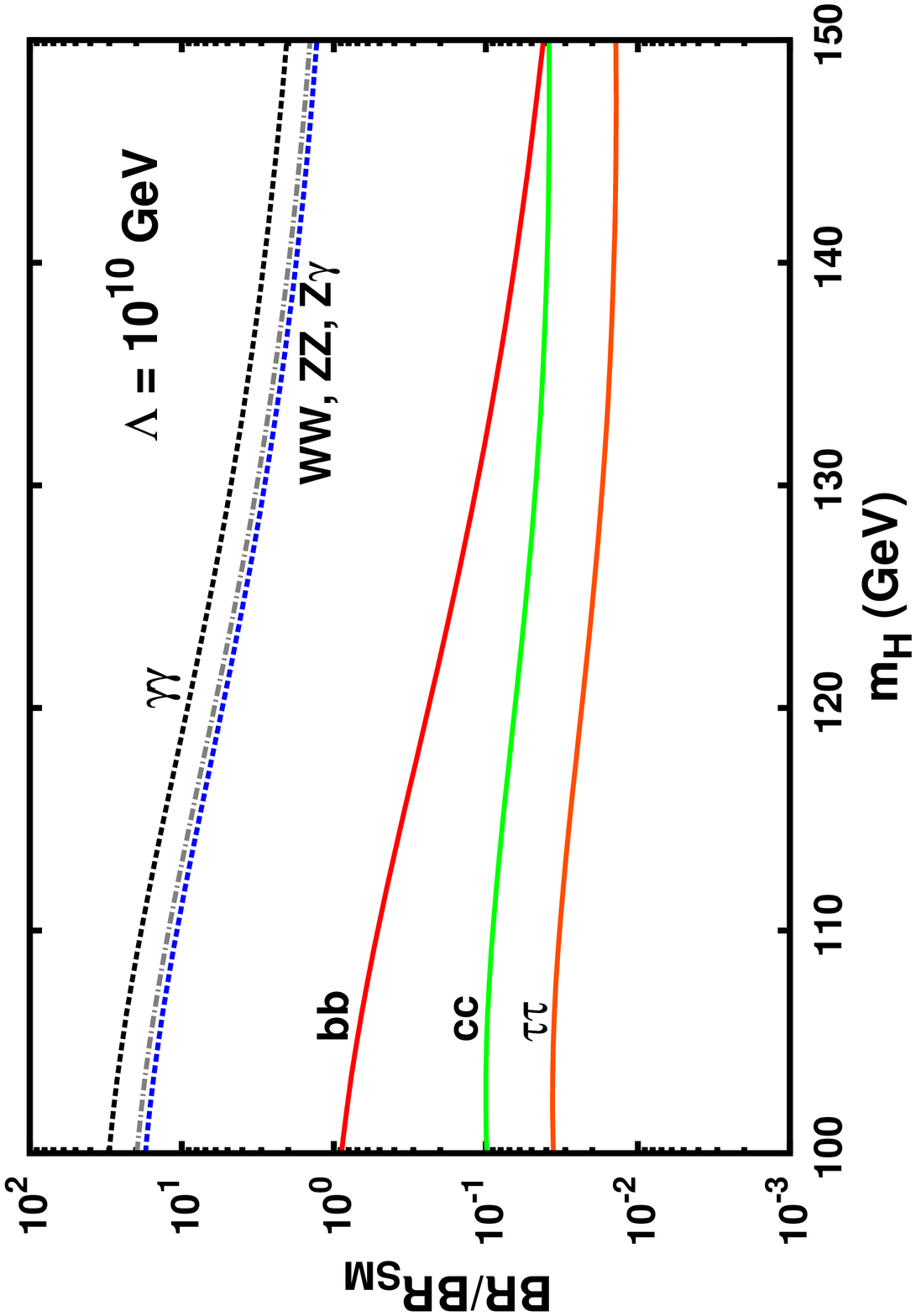}{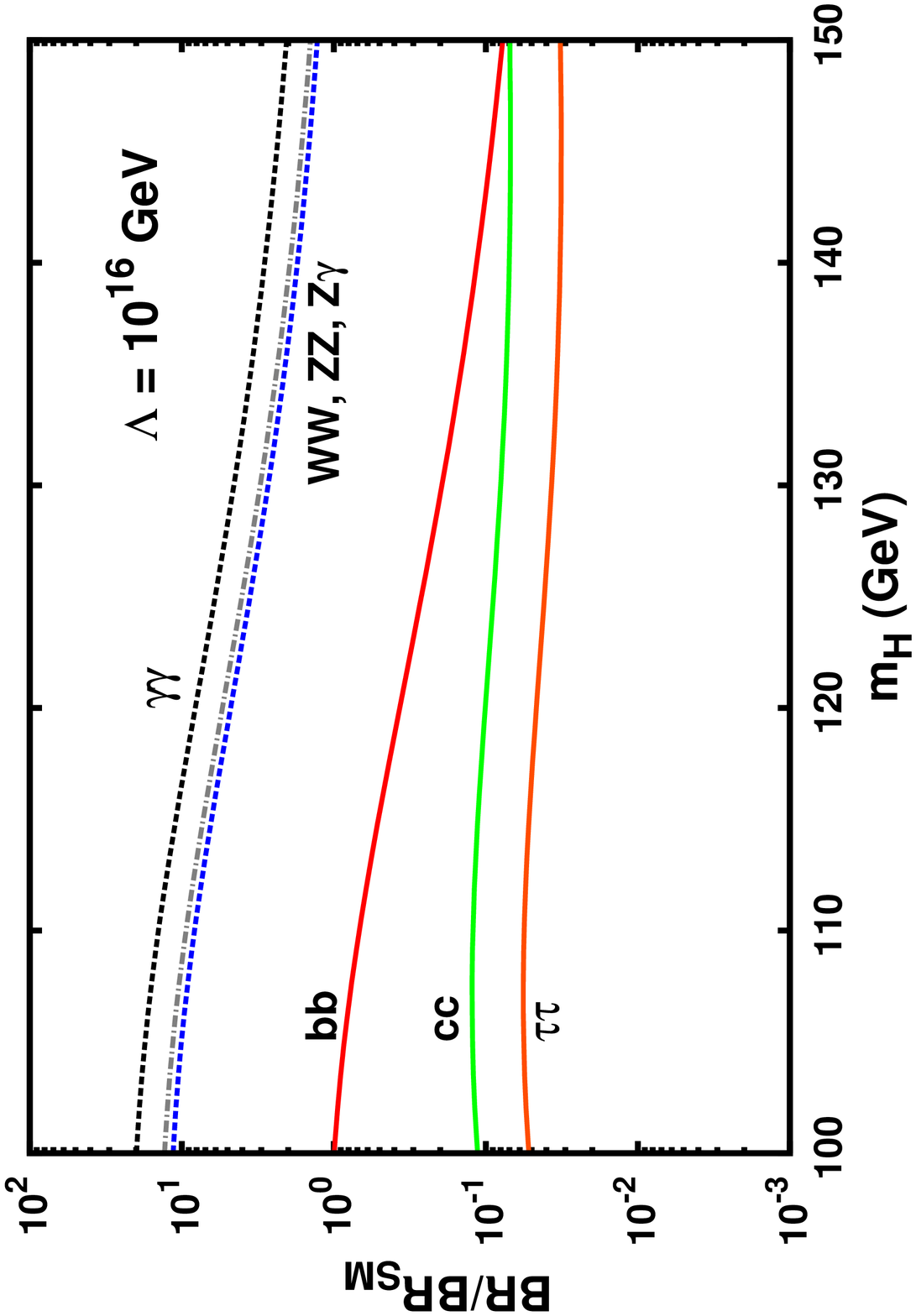}
\end{center}
\caption{\small Higgs branching ratios  normalized to the SM values, 
for the dominant decay channels, versus $m_H$, and for a few representative  values of
$\Lambda$. Note that the $Z\gamma$ curves are the upper ones among the almost degenerate $WW$, $ZZ$, and $Z\gamma$ sets.}
\label{R_MH}
\end{figure}
In Fig.\ref{R_MH},   the Higgs BR's, 
normalized to the corresponding SM values, are shown.
Although all the Higgs decays in EW gauge bosons are enhanced with respect to the SM, 
the most substantial effect is observed in the 
$H\to \gamma\gamma$  channel. For instance,  BR$(\gamma \gamma)$, at $m_H=110$ GeV
and for $\Lambda=10^{4 (16)}$ GeV,
is enhanced up to 29 (14) times the SM value,
while for 
BR$(WW)$, BR$(ZZ)$, and BR$(Z\gamma)$
 the enhancement factor is about 
 19 (9.5). As for  the $\Lambda$ dependence, 
at large  $\Lambda$ 
 the decay widths into fermion pairs tend to be wider, 
and  the enhancement of the decays 
into EW gauge bosons drops.

The larger enhancement for  BR$(\gamma \gamma)$
 with respect to  the BR$(WW)$ and BR$(ZZ)$ cases is due to the destructive interference, in the SM,  between 
the top-quark loop and the $W$ loop  in $H\to \gamma\gamma$, that  is suppressed in the present scenario. The same holds for the smaller enhancement in the  $H\to Z\gamma$ BR, that is however less affected by the top-quark loop.

\section{ Production cross sections for different Higgs boson signatures}
In this section, we will present Higgs production cross sections at hadron colliders, corresponding to different Higgs decay channels, in the effective Yukawa scenario.

Experimental strategies to constrain this scenario can be elaborated on the basis of existing LEP and Tevatron data. 
As already mentioned, the most stringent 
bounds on $m_H$ in the {\it pure} fermiophobic Higgs scenario 
have been obtained at LEP ($m_H > $ 109.7 GeV at 95\% C.L. \cite{LEPFP}) by studying the process $e^+e^-\to HZ$, with 
$H\to \gamma\gamma$, and at the Tevatron ($m_H > $ 106 GeV at 95\% C.L., corresponding to  $3.0 \,{\rm fb}^{-1}$ of analyzed data \cite{D0,CDF}), via both Higgsstrahlung 
$p\bar{p}\to H V\to \gamma \gamma + X$,
and VBF  
$p\bar{p}\to V V+X\to H +X\to \gamma \gamma + X$.
Effective Yukawa couplings  deplete  BR$(H\to\gamma\gamma)$ with respect to the fermiophobic-Higgs scenario,  due to the 
nonvanishing 
  $\Gamma(H\to\bar{f} f)$ contribution to the total width
  (cf. table \ref{TabBR}). Then,
we expect  the analysis in \cite{LEPFP,D0,CDF}, 
 when applied to our model, should end up into weaker 
bounds on the Higgs mass, depending on the scale $\Lambda$. Note, however, that the  fermionic decays  would play an extra role in  excluding  $m_H$ ranges at LEP, with respect to 
a {\it pure} fermiophobic Higgs scenario.
In the following, we will present our results for  $m_H \ge 100$ GeV, that, we expect on the basis of the results in table \ref{TabBR}, should cover all the experimentally allowed region.
\begin{figure}[tpb]
\begin{center}
\dofourfigs{3.1in}{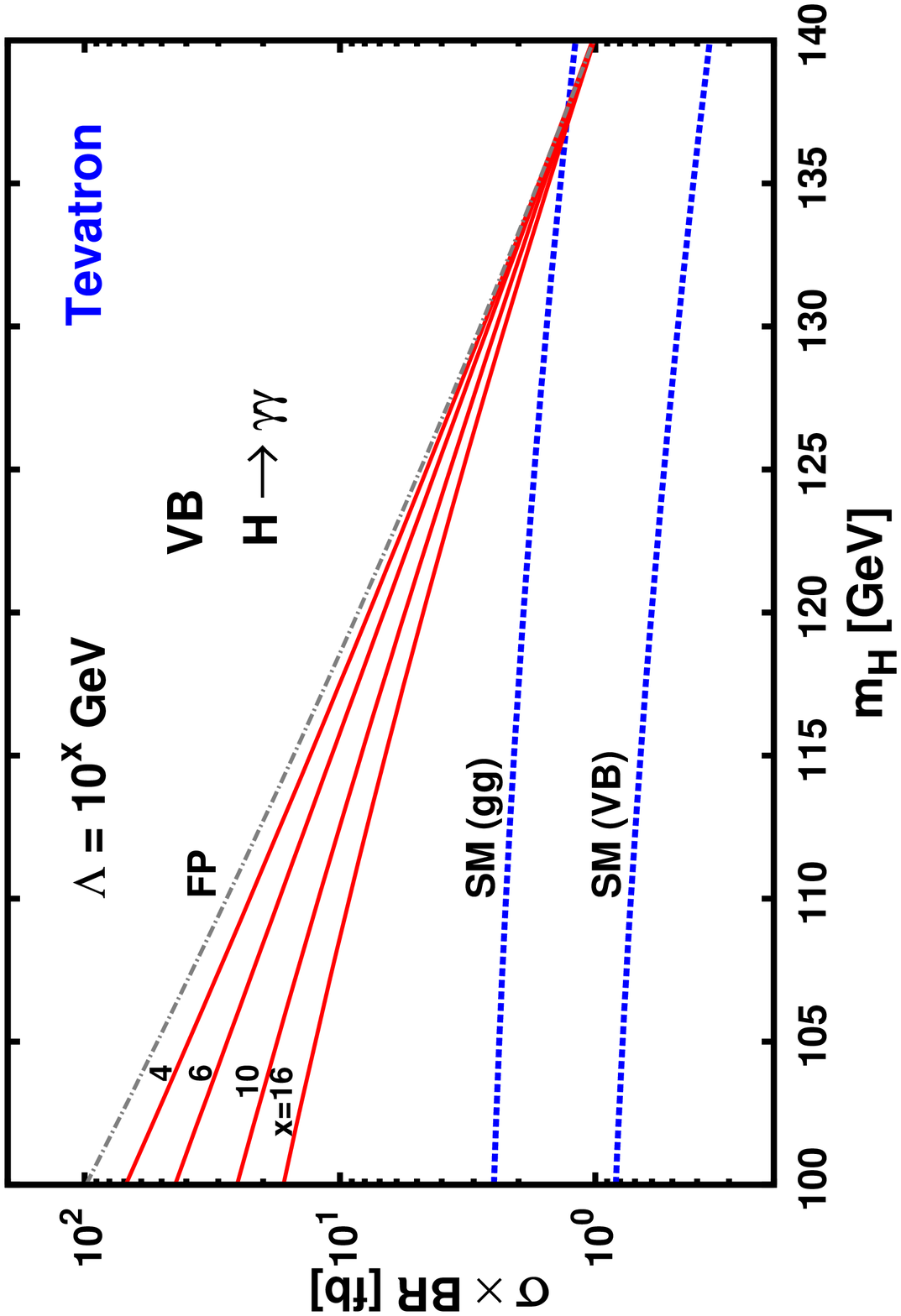}{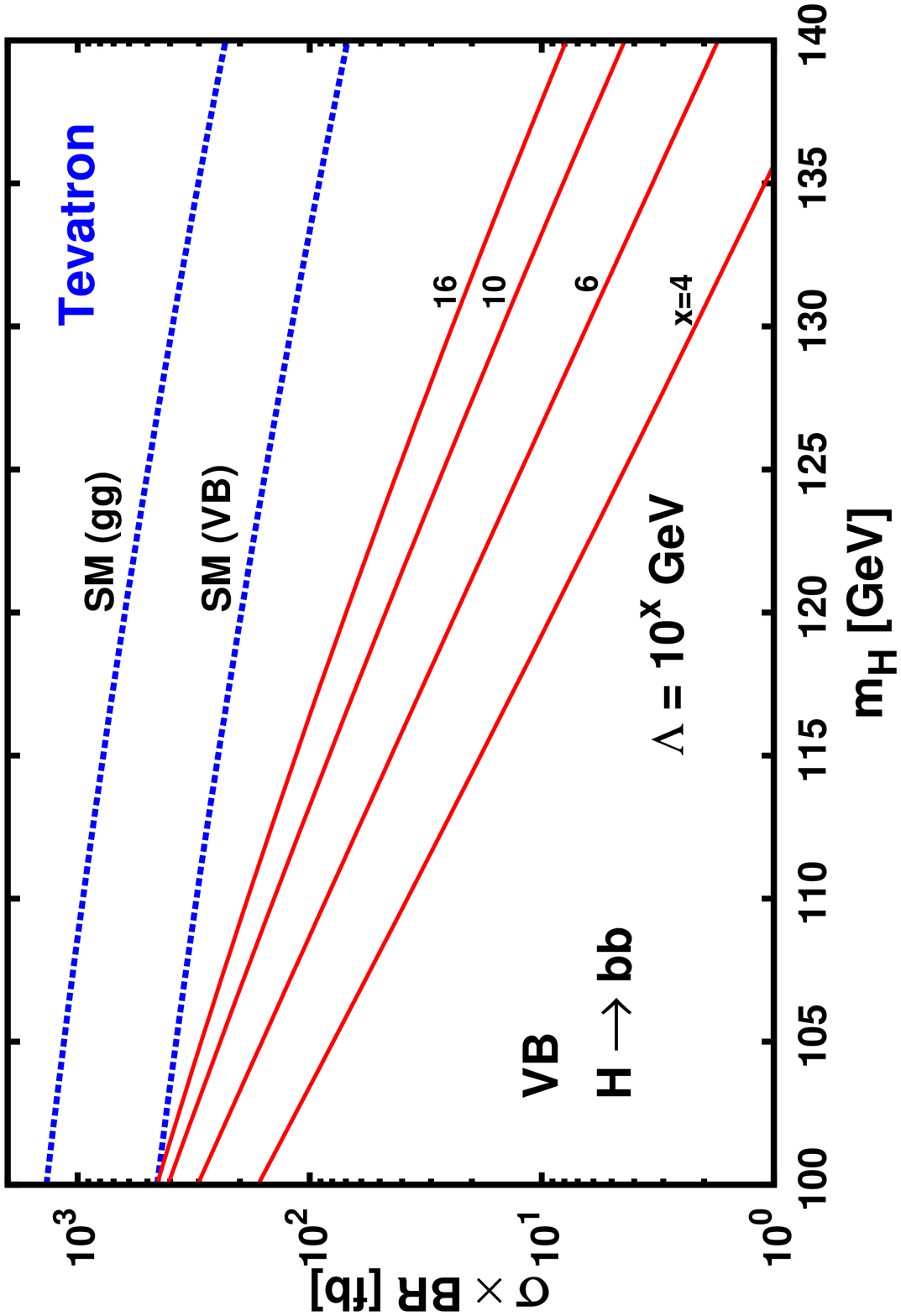}{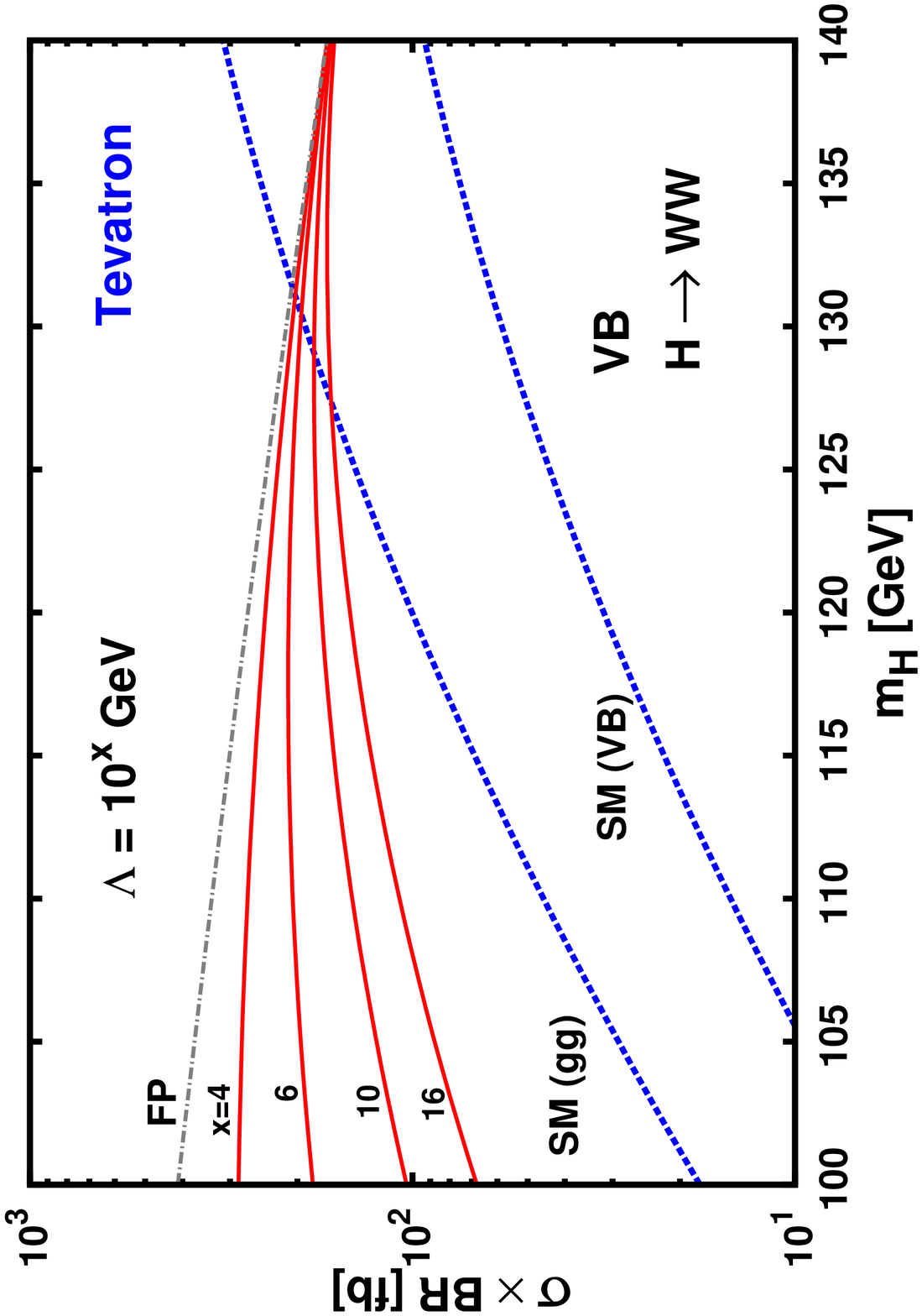}{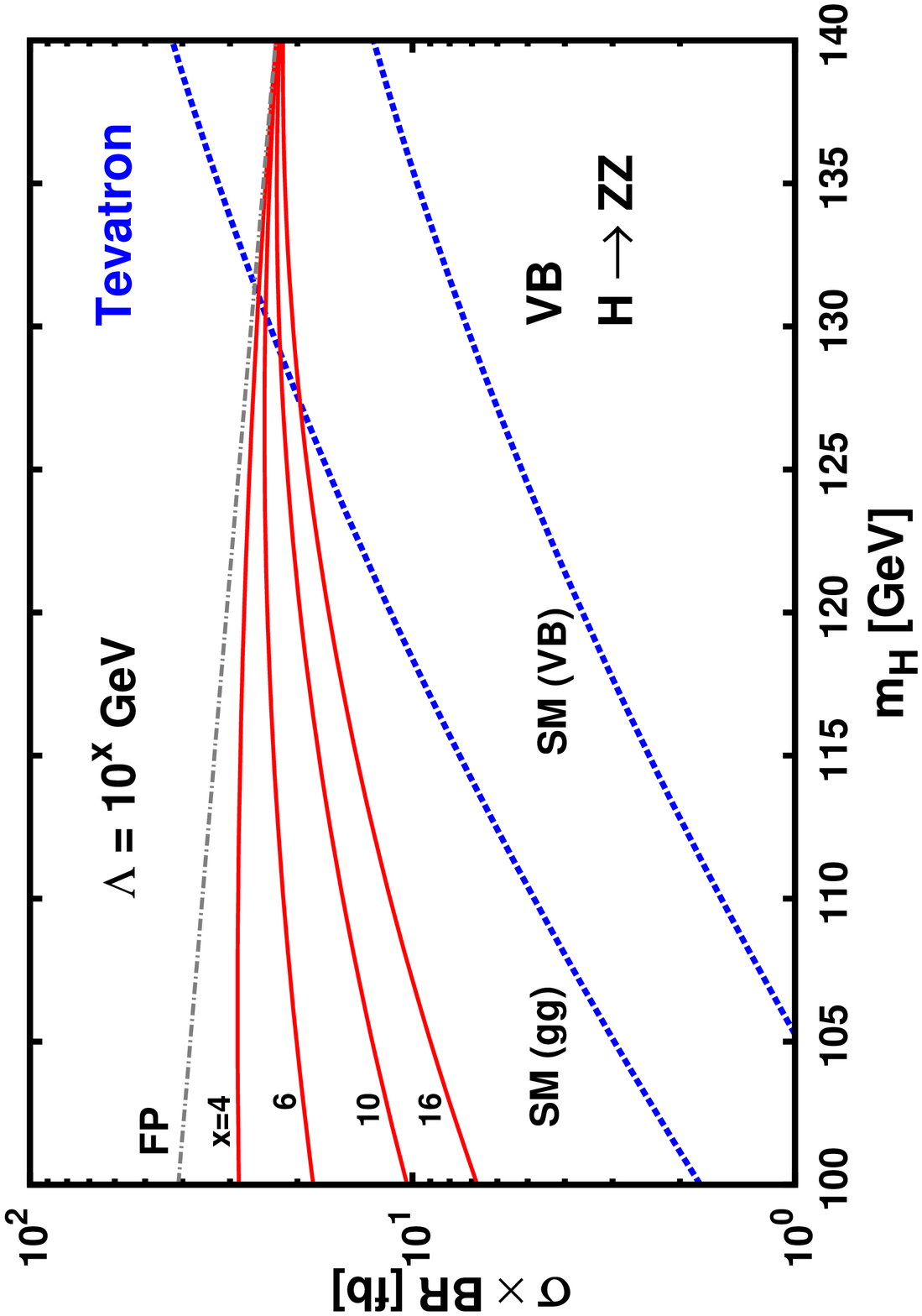}
\end{center}
\caption{\small 
Total cross sections times Higgs branching ratios for $p\bar{p}$
collisions at Tevatron , with c.m
energy $\sqrt{S}=1.96$ TeV, 
for the Higgs decays 
$H\to  \gamma \gamma$, $b\bar b$, $WW$, $ZZ$,  versus the Higgs 
mass. Continuous (red) lines
correspond to the VB predictions  in the effective Yukawas model, for 
 $\Lambda=10^{4,6,10,16}$ GeV, with  
 VB standing for the inclusive cross section
for  VBF + $HW$ +$HZ$.
 The dashed (blue) lines and 
dot-dashed (grey) lines correspond to the SM (mediated by either $gg$ fusion or VB processes) and fermiophobic Higgs  
scenario (FP), respectively.
}
\label{TEV}
\end{figure}

The Tevatron has a considerable potential for constraining models with enhanced BR$(H\to\gamma\gamma)$ \cite{FPTevatron}.
Present searches are  sensitive, in the range 115 GeV$\le m_H \le 130$ GeV, to scenarios where the $H\to\gamma\gamma$ rate is enhanced by at least a factor of about 20 with respect to its SM value \cite{D0}. With $2.7 \,{\rm fb}^{-1}$ of integrated luminosity, 
a 95\% C.L. upper limit on BR$(H\to\gamma\gamma)$
between 14.1\% and 33.9\%, for 100 GeV$\le m_H \le 150$ GeV,
has been derived by D0, for models where the Higgs does not couple to the top quark. This is not yet sufficient to  test the effective Yukawa coupling scenario, that, for $\Lambda\gsim 10^4$ GeV,  predicts
BR$(H\to\gamma\gamma)\lsim 12\%$ in this  $m_H$ range 
 (cf. table \ref{TabBR}).

In Fig.\ref{TEV}, we present
the total cross sections times the Higgs branching ratios for different decay channels, in $p\bar{p}$
collisions at Tevatron, with c.m
energy $\sqrt{S}=1.96$ TeV. In particular, we consider  
 the Higgs decay channels 
$H\to  \gamma \gamma$, $b\bar b$, $WW$, $ZZ$. Solid (red) lines
correspond to the inclusive cross section 
for  VBF + $HW$ +$HZ$ production (labeled  VB in the figure) in the effective Yukawas model, for 
 $\Lambda=10^{4,6,10,16}$ GeV.
 The dashed (blue) lines and 
dot-dashed (grey) lines correspond to the SM (mediated by either $gg$ fusion or VB), and fermiophobic-Higgs  (FP) 
scenario (with vanishing Yukawa couplings), respectively.
We did not include, in the effective Yukawa model predictions, the 
gluon fusion contribution, that is expected to be suppressed by
more that a factor $10^{-2}$ with respect to the SM cross section 
(cf. Y$_t$ values in table \ref{Yeff}). Cross sections are computed at NLO, and presented by their central values, neglecting different theoretical uncertainties, as obtained from \cite{Malto}.
\\
While the $H\to b\bar b$ channel  tends to be considerably more difficult than in the SM, there is a remarkable enhancement, not only in the $H\to  \gamma \gamma$ decay, but also in the 
$H\to WW$, $ZZ$ channels, for masses up to $m_H\sim 125$ GeV. For lower Higgs masses, the sensitivity to the scale $\Lambda$ is quite large.

\begin{figure}[tpb]
\begin{center}
\dofigsV{3.1in}{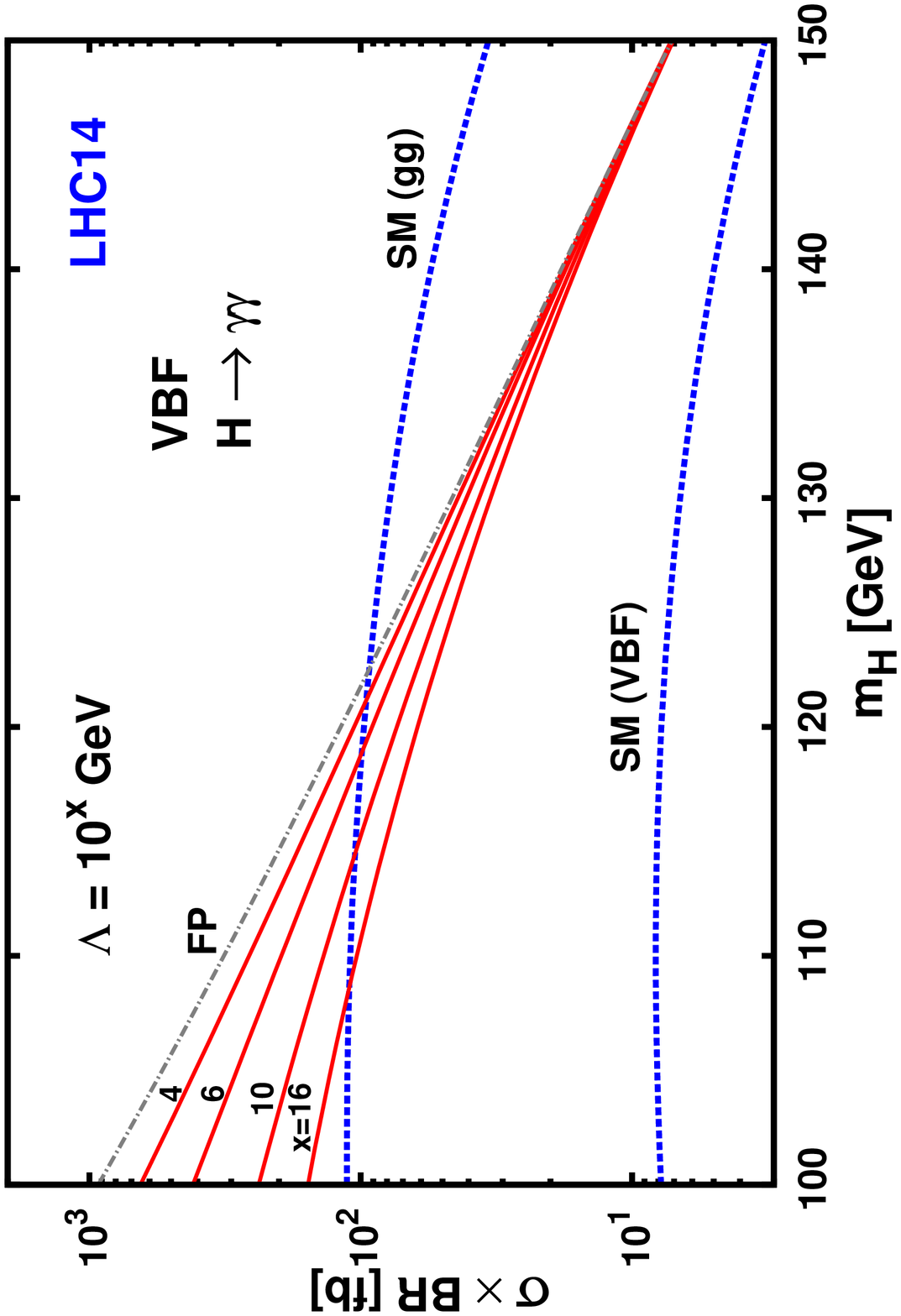}{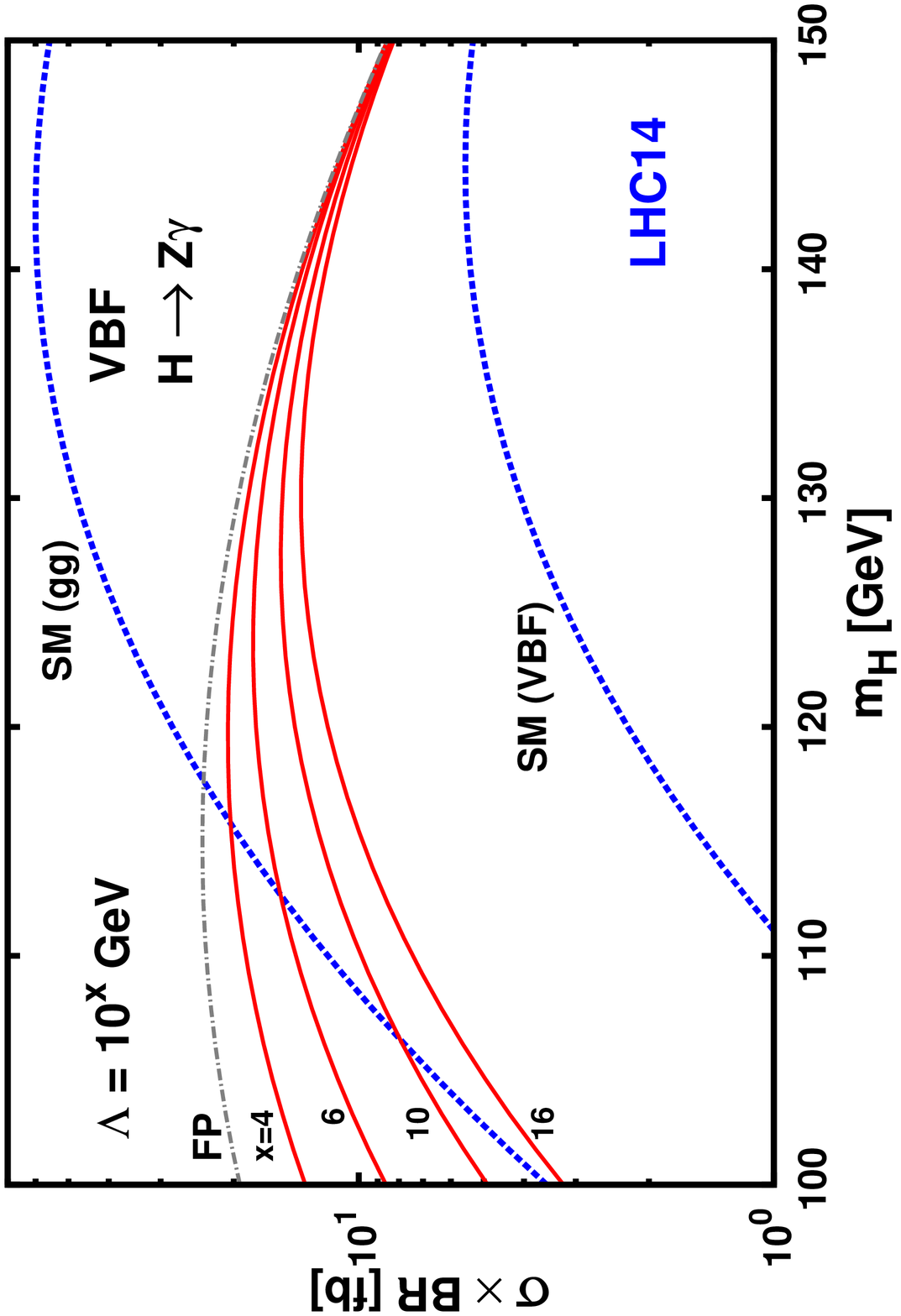}
\end{center}
\caption{\small Total cross sections times Higgs branching ratios, for $pp$ collisions at the LHC
with c.m. energy $\sqrt{S}=14$ TeV, 
for the Higgs decaying 
 into the loop-induced  modes
$H\to \gamma \gamma$ (top) and $H\to Z\gamma$ (bottom), versus the Higgs 
mass. Continuous (red) lines
correspond to the VBF predictions   in the effective Yukawas model, for 
 $\Lambda=10^{4,6,10,16}$ GeV. The dashed (blue) lines and 
dot-dashed (grey) lines correspond to the SM (mediated by either $gg$ fusion or VBF) and fermiophobic Higgs  
scenario (FP), respectively.}
\label{LHC14A}
\end{figure}
\begin{figure}[tpb]
\begin{center}
\dofigsV{3.1in}{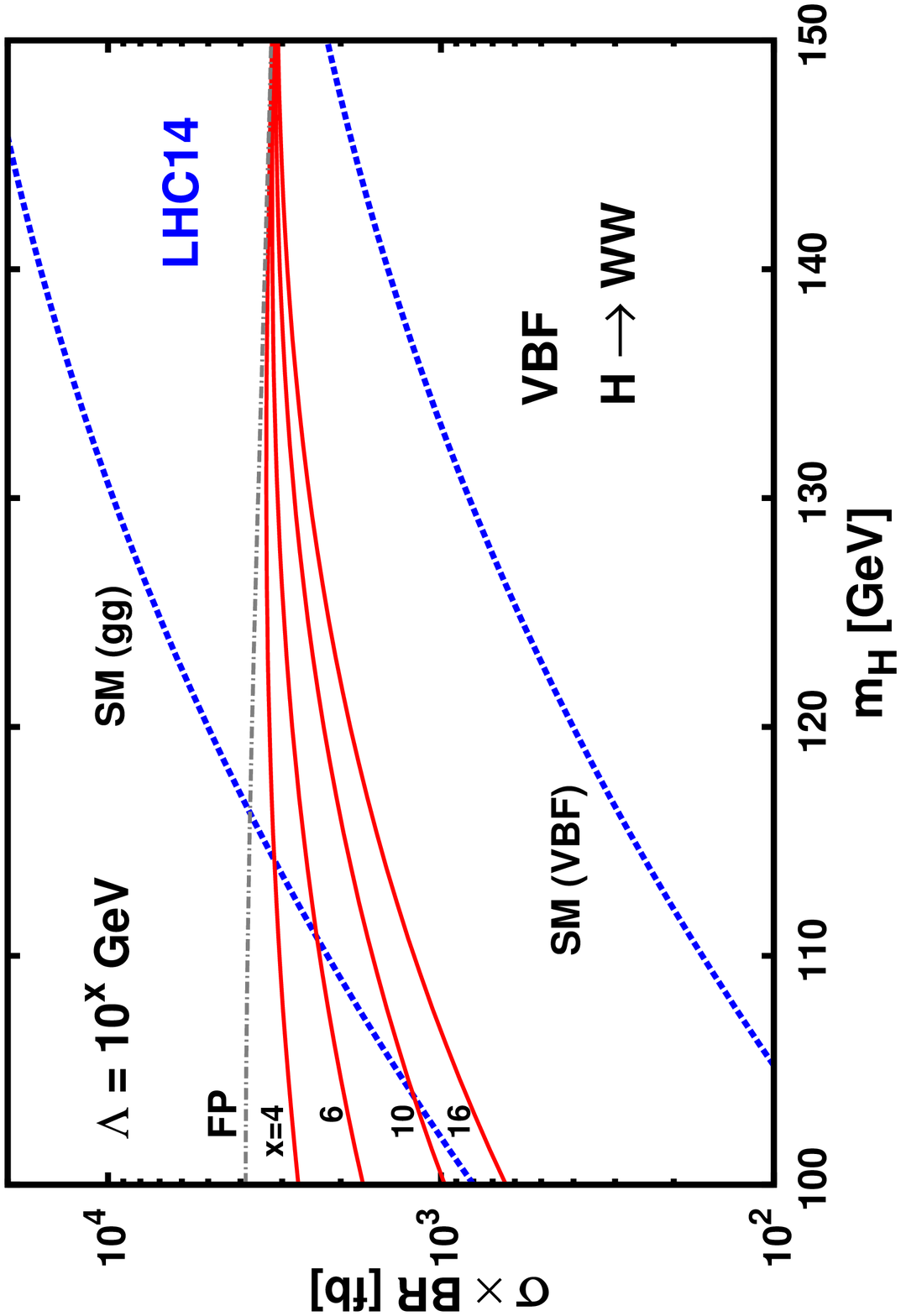}{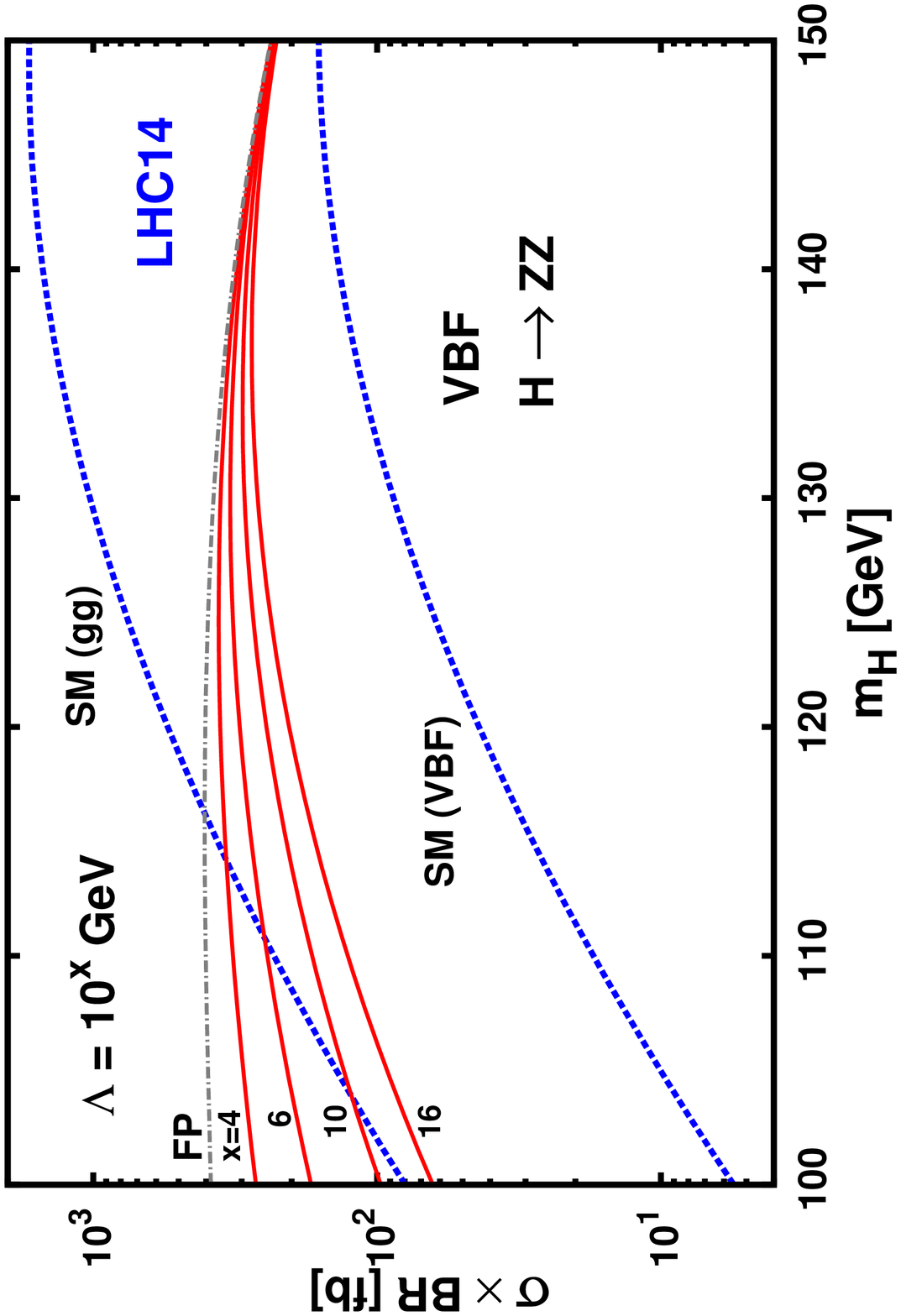}
\end{center}
\caption{\small 
Total cross sections times Higgs branching ratios, for $pp$ collisions at the LHC
with c.m. energy $\sqrt{S}=14$ TeV, for the Higgs decaying 
 into tree-level  modes
$H\to WW$ (top) and $H\to ZZ$ (bottom), versus the Higgs 
mass. Continuous (red) lines
correspond to the VBF predictions   in the effective Yukawas model, for 
 $\Lambda=10^{4,6,10,16}$ GeV. The dashed (blue) lines and 
dot-dashed (grey) lines correspond to the SM (mediated by either $gg$ fusion or VBF) and fermiophobic-Higgs  
scenario (FP), respectively.
 }
\label{LHC14B}
\end{figure}

Expectations for cross sections times branching ratios at the LHC are presented in Figs.\ref{LHC14A} and
\ref{LHC14B}, for the c.m. energy $\sqrt{S}=14$ TeV. 
In particular, Fig.\ref{LHC14A} refers to the one-loop decays
$H\to \gamma \gamma$ (top), and $H\to Z\gamma$ (bottom),
while Fig.\ref{LHC14B} refers to the tree-level decays
$H\to WW$ (top), and $H\to ZZ$ (bottom). In general, conventions in Figs.\ref{LHC14A} and
\ref{LHC14B} are the same as for the Tevatron case in
Fig.\ref{TEV}, but for the LHC we show only the production via VBF, and discuss 
 the associated $HW$ and $HZ$ production later on.
Cross sections are computed as for Fig.\ref{TEV}\footnote{The SM gluon fusion  cross sections   at the (resummed) NNLL+NNLO 
have been obtained by  the online calculator by M. Grazzini, for the parton distribution functions  set MSTW2008 NNLO \cite{Hxsect}. }.
\\
At the LHC the SM gluon fusion production plays a more relevant role than at the Tevatron. It is then interesting to have a look at the depleted 
$gg$ contribution to the total cross section arising in the effective Yukawa scenario. In table \ref{INT}, we present 
the ratio 
${\rm R}_{\sigma}^{m_H}=\tilde\sigma_{gg}/\sigma_{{\rm VBF}}$ of total cross 
sections for Higgs  production via gluon fusion  ($\tilde\sigma_{gg}$)  and
via  
 VBF ($\sigma_{{\rm VBF}}$) \cite{Malto}, in the effective Yukawa model, at the LHC with  $\sqrt{S}=14$ TeV. We present results for   $\Lambda=10^{4,6,10,16}$ GeV, and $m_H=100,120,140$ GeV.
\\
The gluon fusion cross section $\tilde\sigma_{gg}$ has been 
computed by rescaling the SM values  at NNLL+NNLO \cite{Hxsect} 
by the factor  due to the top-quark Yukawa suppression
$({\rm Y}_t(m_H)/{\rm Y}_t^{\rm {\scriptscriptstyle SM}})^2$.
One can see that  the gluon fusion  cross 
section
is at most a few percent of the VBF cross section, in the range of $m_H$ and $\Lambda$ considered. We will neglect here this contribution.

\begin{table} \begin{center}    

                                                                                              \begin{tabular}{|c||c|c|c|}                                         \hline ${\rm \Lambda\,({\rm GeV})}$                               & ${\rm R_{\sigma}^{100}}$ &                                      ${\rm R_{\sigma}^{120}}$  &                                       ${\rm R_{\sigma}^{140}}$                                          \\ \hline \hline $10^4$ &     2.1$\times 10^{-3}$    &    3.5$\times 10^{-3}$    &    5.5$\times 10^{-3}$  \\ \hline
      $10^6$ &     6.0$\times 10^{-3}$    &    1.1$\times 10^{-2}$    &    1.7$\times 10^{-2}$                                                                                            \\ \hline
   $10^{10}$ &     1.5$\times 10^{-2}$    &    2.6$\times 10^{-2}$    &    4.4$\times 10^{-2}$                                                                                            \\ \hline
   $10^{16}$ &     2.5$\times 10^{-2}$    &    4.7$\times 10^{-2}$    &    7.9$\times 10^{-2}$                                                                             \\ \hline \end{tabular}

\caption[]{Ratio 
${\rm R}_{\sigma}^{m_H}=\tilde\sigma_{gg}/\sigma_{{\rm VBF}}$ of total cross 
sections for Higgs  production via gluon fusion  and
via  
 VBF in the effective Yukawa model at the LHC, with  $\sqrt{S}=14$ TeV, for  representative values 
of   $\Lambda$ and $m_H$, expressed in GeV.
 }\label{INT} 
\end{center} \end{table}
As one can see from Figs.\ref{LHC14A} and
\ref{LHC14B}, at the LHC 
the depletion in the total cross section with respect to the SM 
gluon fusion mechanism is, for $m_H\lsim130$ GeV, 
 largely compensated for by 
the enhancements in the BR's. 
The sensitivity to the scale $\Lambda$ is considerable, in this $m_H$ range.  Notice that, while the $H\to  \gamma \gamma$  cross section falls by more than  1 order of magnitude when $m_H$ moves from 100 GeV to 150 GeV, for the 
$H\to  WW$, $ZZ$, $Z\gamma$
 channels the VBF cross section drop  at larger $m_H$ is more than compensated for by the rise in the  corresponding BR's (cf. fig.\ref{BR_MH}).
\\
The combined analysis of the  
$H\to  \gamma \gamma$, $Z\gamma$, $WW$, $ZZ$ channels, at  integrated luminosities of a few 10 fb$^{-1}$, could clearly distinguish an anomalous VBF signal corresponding to effective Yukawa couplings, that should replace the dominant gluon  fusion signature in  the SM cross sections. 

\begin{figure}[tpb]
\begin{center}
\dofourfigs{3.1in}{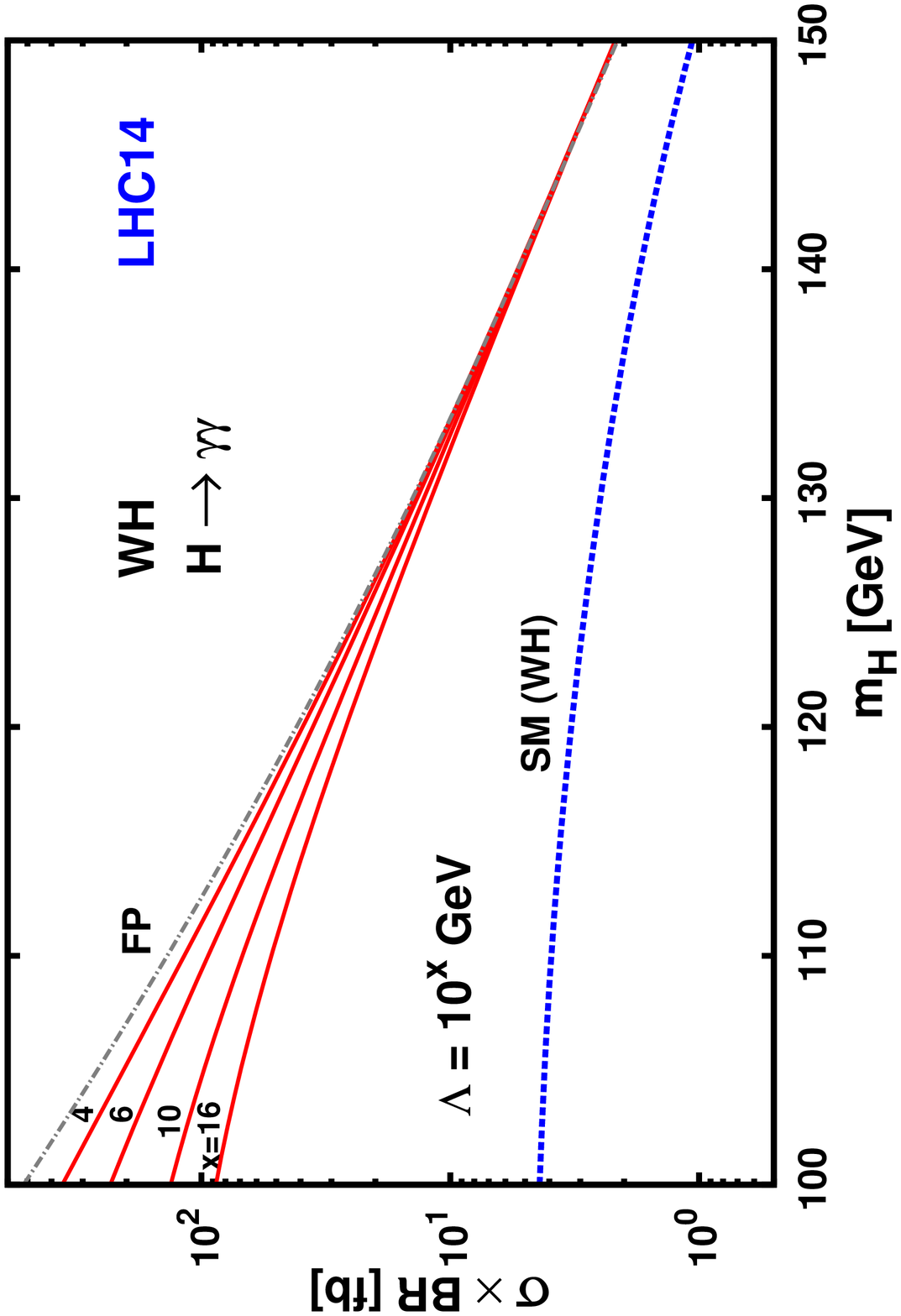}{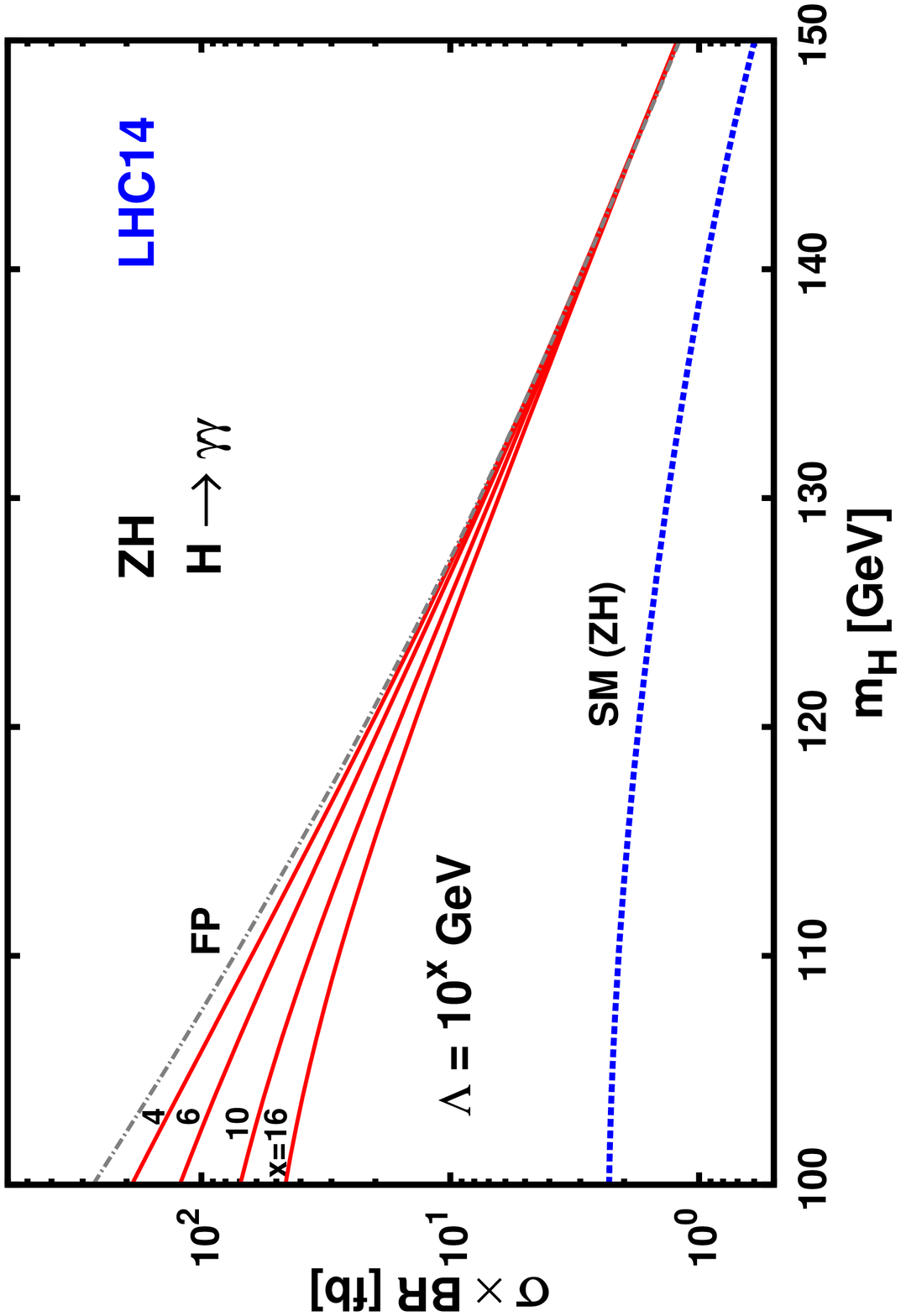}{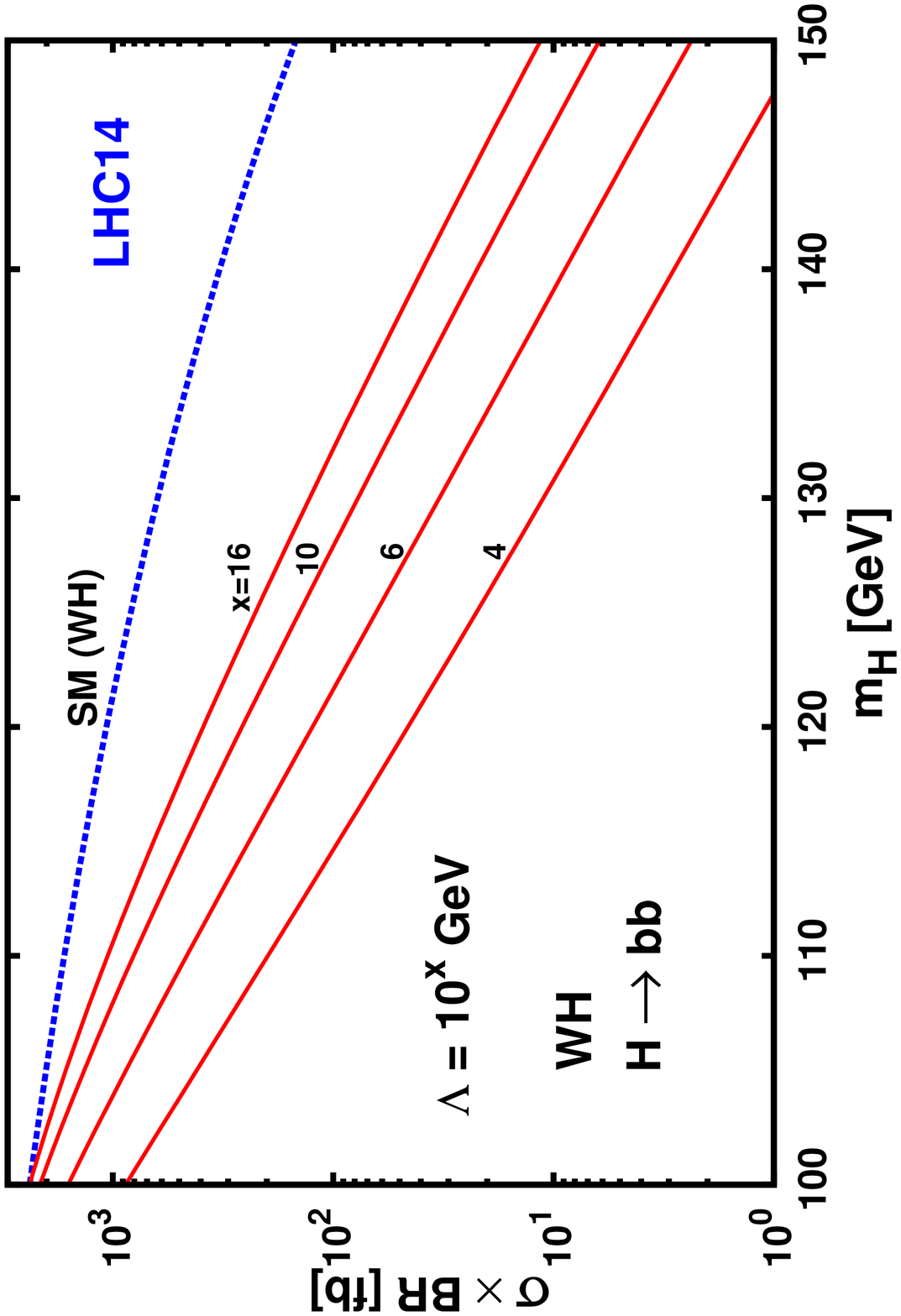}{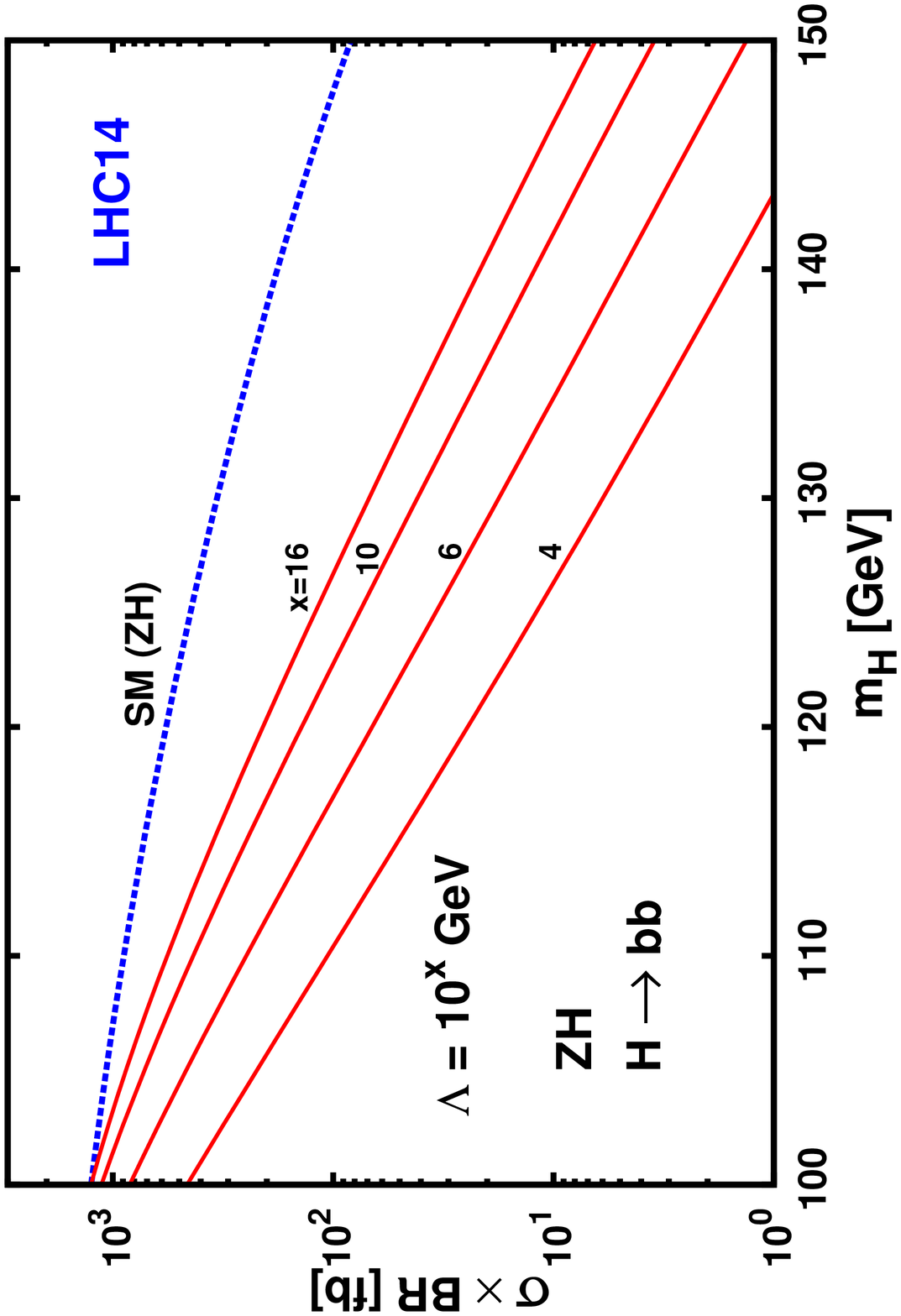}
\end{center}
\caption{\small 
Total cross sections times Higgs branching ratios, for $pp$ collisions at the LHC
with c.m. energy $\sqrt{S}=14$ TeV, via the WH (left) and ZH (right) associated 
Higgs production mechanism, for the Higgs decaying 
 into $\gamma \gamma$ (top) and $b\bar{b}$ (bottom) modes, versus the Higgs 
mass. Continuous (red) lines
correspond to the predictions  in the effective Yukawas model, for 
 $\Lambda=10^{4,6,10,16}$ GeV. The dashed (blue) lines and 
dot-dashed (grey) lines correspond to the SM and fermiophobic Higgs  
scenario (FP), respectively.
}
\label{HVLHC14}
\end{figure}

The enhanced BR($H\to\gamma\gamma$) also makes the $WH/ZH$ associated production with $H\to\gamma\gamma$  at the LHC      a remarkably promising channel, that could be studied for inclusive $W/Z$ decays. Total cross sections times Higgs branching ratios are shown in the upper part of Fig.\ref{HVLHC14},
 for $pp$ collisions at 
 c.m. energy $\sqrt{S}=14$ TeV, for the associated $HW$ (left) and $HZ$ (right) 
Higgs production, with $H\to  \gamma \gamma$.
$WH/ZH$ production rates are according to \cite{Malto}.
The enhancement for the $\gamma\gamma$ signature with respect to the SM scenario, and the easy topology of the $WH/ZH$
events could make the Higgs associated production a crucial handle in the experimental establishment of the effective Yukawa scenario at the LHC.

On the other hand, it could be  harder at the LHC   pinpointing the depleted fermionic Higgs decays like $H\to b\bar b$ and  $\tau\tau$, the first also being challenging in the {\it easier}$\;$ SM scenario, the latter being very much suppressed in the radiative Yukawa framework (cf. Fig. \ref{R_MH}). However, in view of the new techniques for  $b$-jet tagging
in Higgs associated production that are being developed at the LHC \cite{Butterworth:2008iy}, we also plot in the lower part of Fig.\ref{HVLHC14} the corresponding $H\to b\bar b$ cross sections for the 
$HW$ (left) and $HZ$ (right)  associated production.
Further advances in the   $b$-jets tagging technique seem to be required to gain some sensitivity to the $H\to b\bar b$ rates  predicted in our model.

\begin{figure}[tpb]
\begin{center}
\dosixfigs{3.1in}{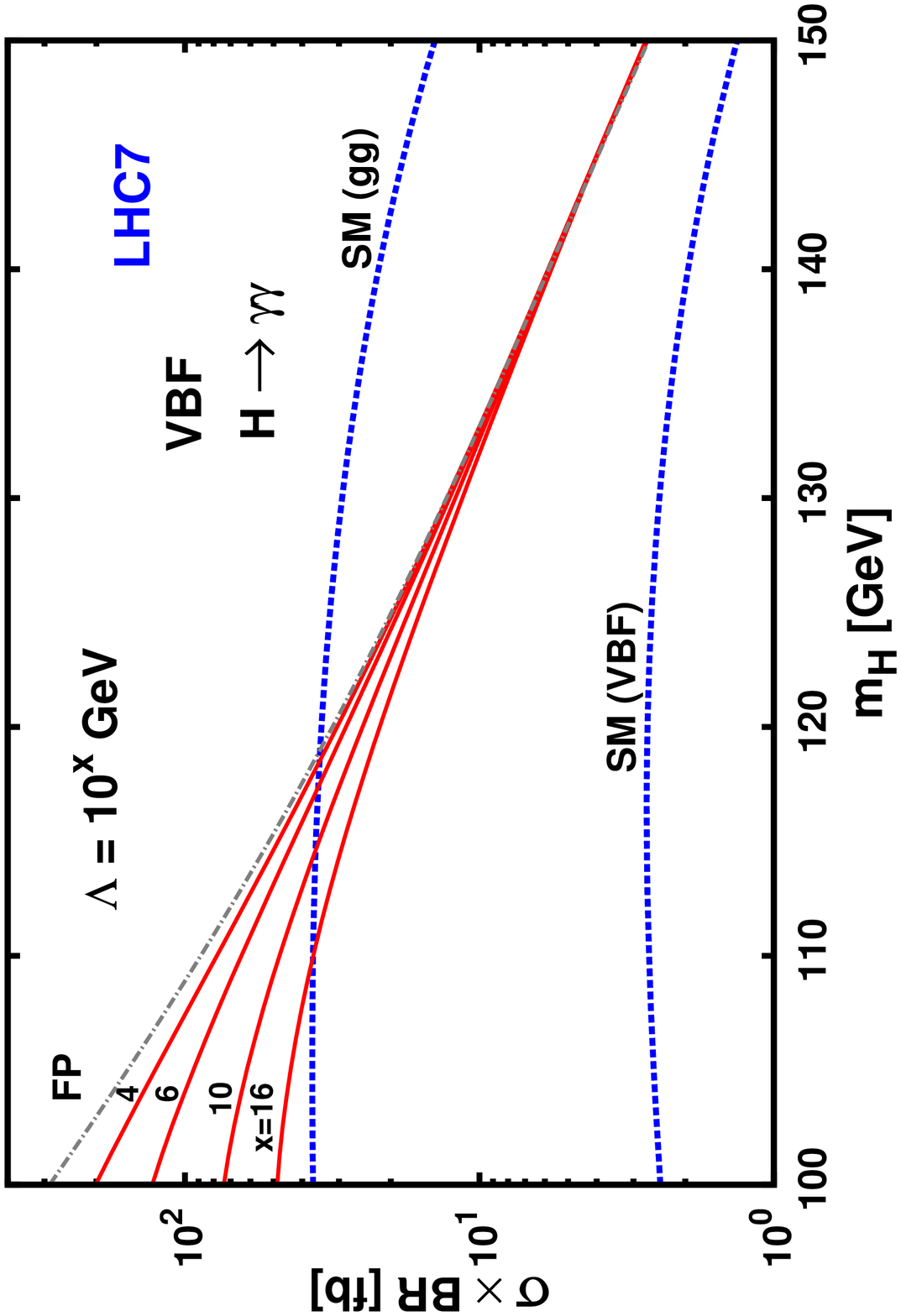}{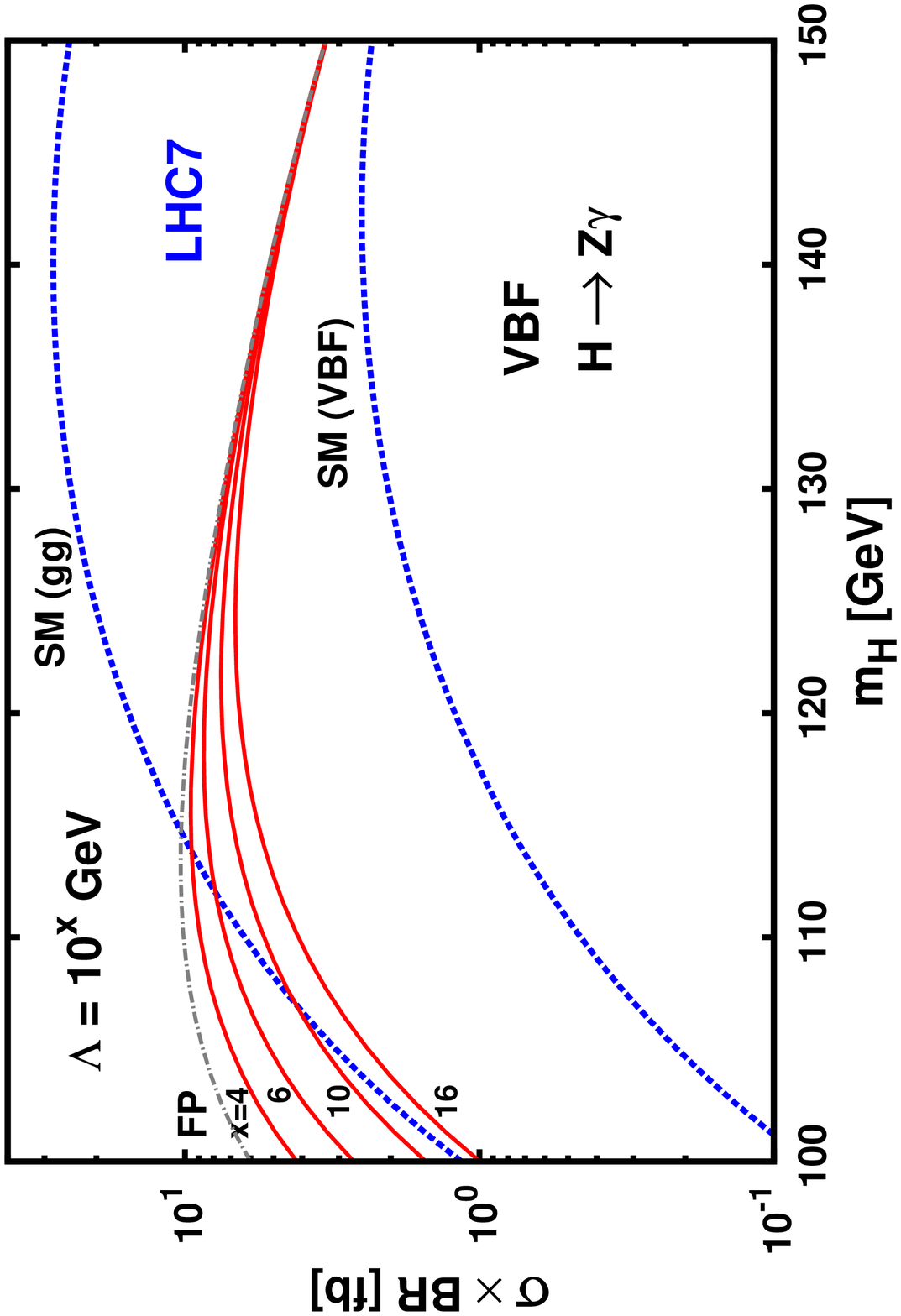}{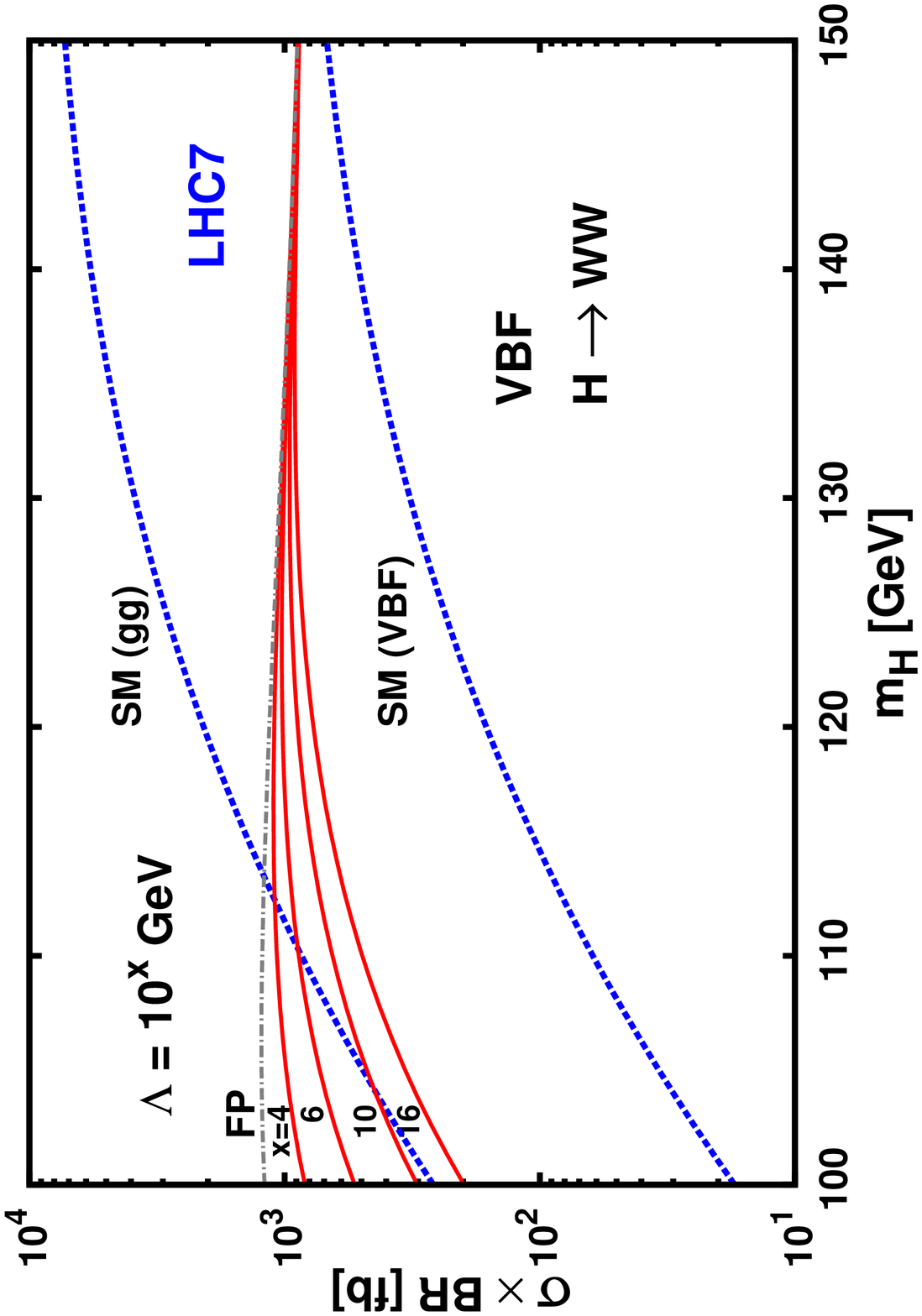}{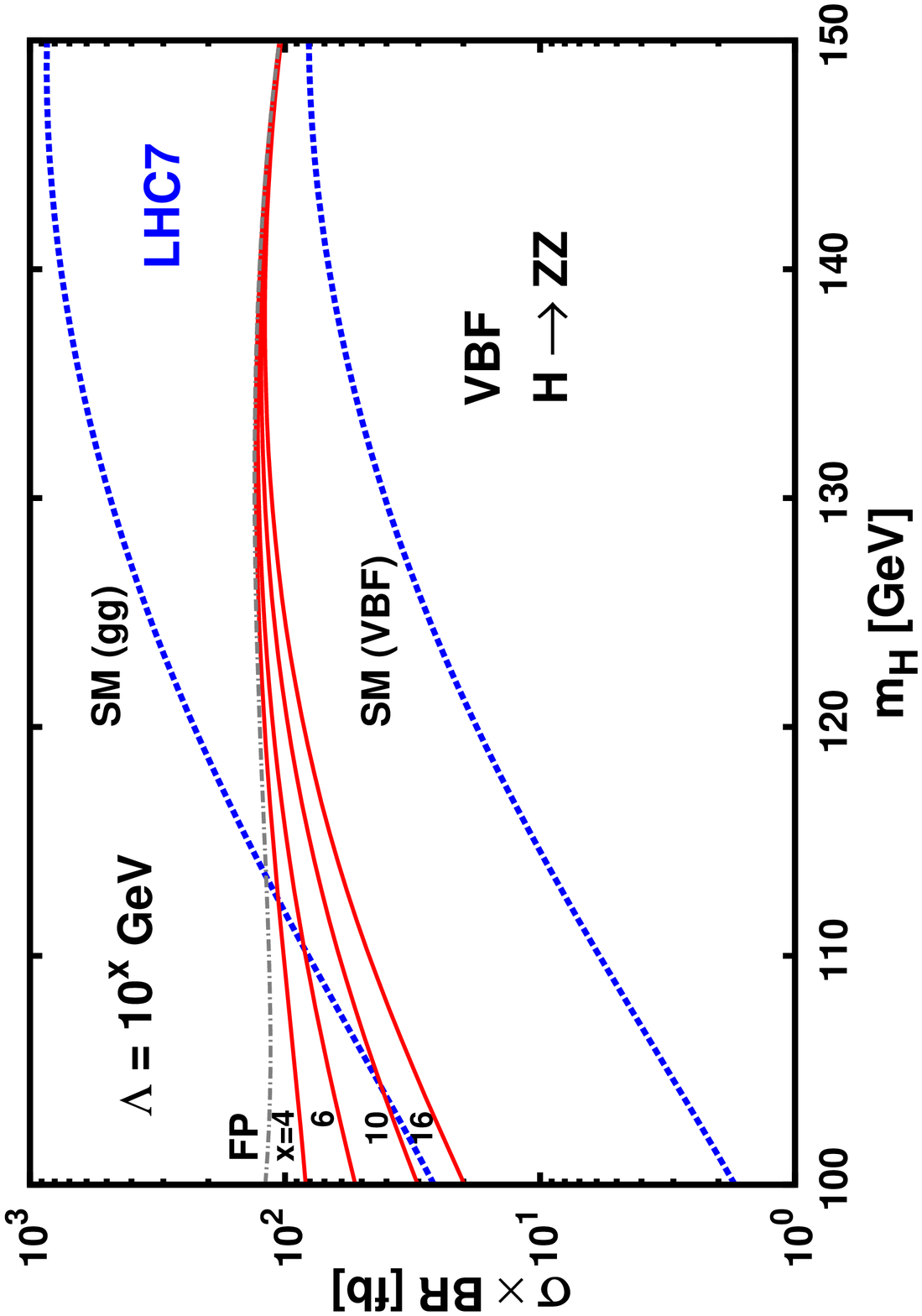}{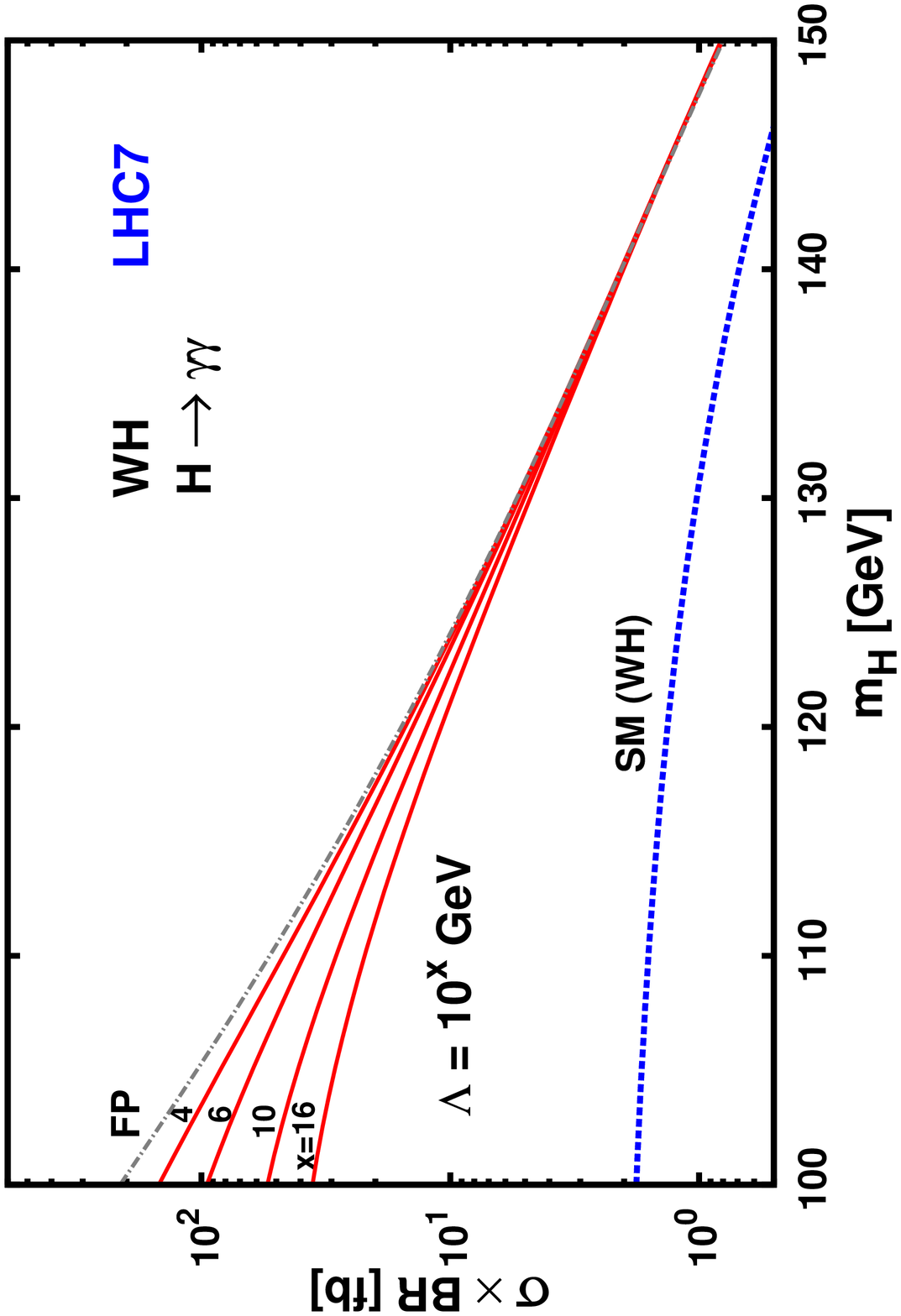}{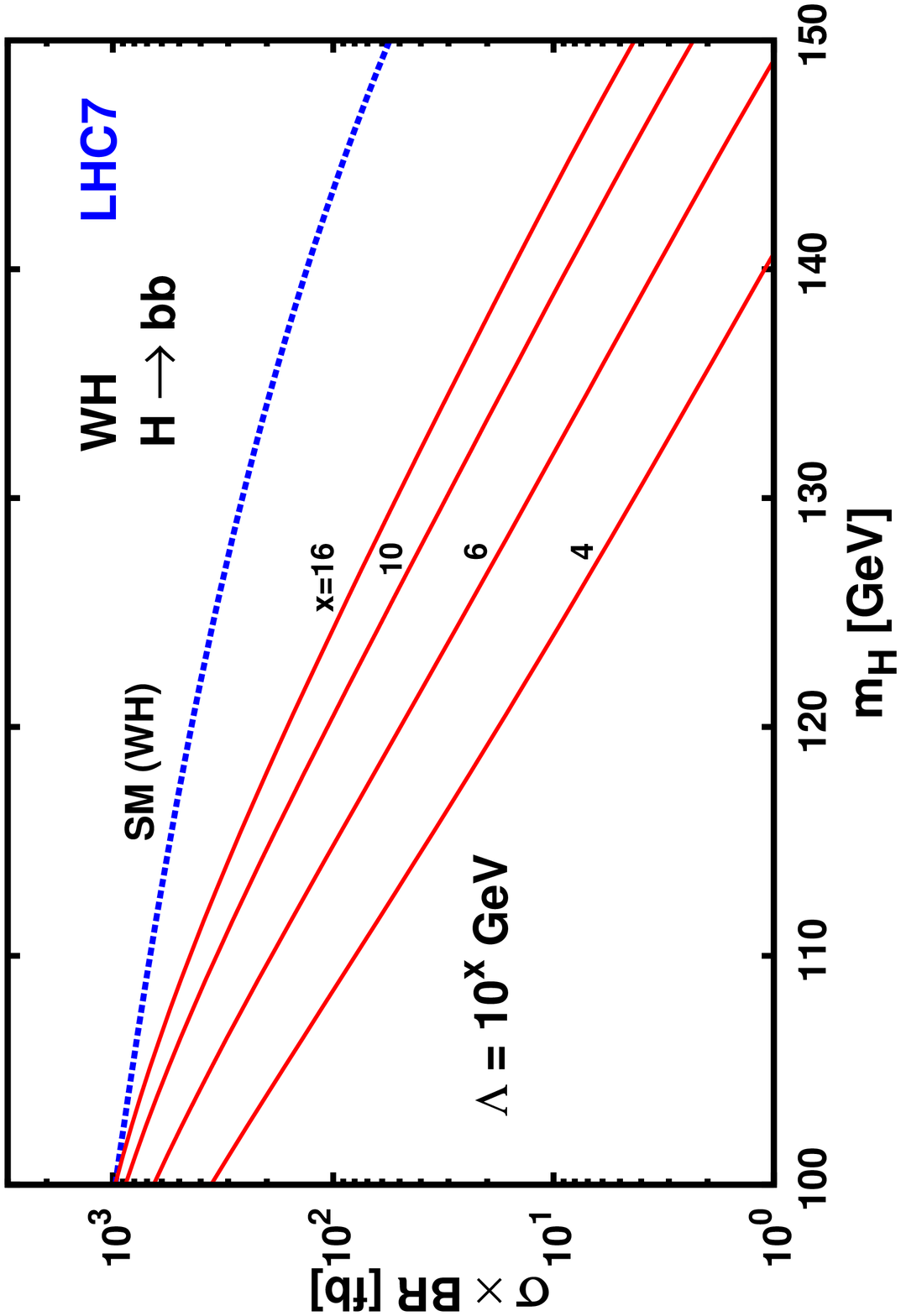}
\end{center}
\caption{\small 
Total cross sections times Higgs branching ratios, for $pp$ collisions at the LHC
with c.m. energy $\sqrt{S}=7$ TeV, for the Higgs decays 
$H\to \gamma \gamma$, $Z\gamma$, $WW$, $ZZ$, $b\bar{b}$ versus the Higgs 
mass. Continuous (red) lines
correspond to the VBF and $WH$ predictions in the effective Yukawas model, for 
$\Lambda=10^{4,6,10,16}$ GeV. The dashed (blue) lines and 
dot-dashed (grey) lines correspond to the SM (mediated by either $gg$ fusion, 
VBF or $WH$) and fermiophobic Higgs scenario (FP), respectively.
}
\label{LHC7}
\end{figure}

In Fig.\ref{LHC7}, the
total cross sections times the Higgs branching ratios are presented  for LHC collisions at 
 $\sqrt{S}=7$ TeV.  Four plots, corresponding to the Higgs decays
$H\to  \gamma \gamma$, $Z\gamma$, $WW$, $ZZ$ in  VBF production, plus two plots, corresponding to the Higgs decays
$H\to  \gamma \gamma$, $b\bar{b}$ in associated $WH$ production, are shown.
\\
Also here, we  neglect the top-loop gluon fusion 
production in the effective-Yukawa scheme\footnote{In Fig.\ref{LHC7}, total cross sections  corresponding
to the SM gluon-fusion  and VBF productions 
are  at NNLL+NNLO \cite{Hxsect} and at NNLO  \cite{VBFNNLO}, respectively. Total cross sections for the $WH$ associated production are taken from \cite{LHCHiggsCS}.}.
All total cross sections  drop by about a factor 3, for $\sqrt{S}$ falling from 14 TeV down to 7 TeV. Then, 
a study directed to pinpoint an anomalous behavior of the Higgs Yukawa couplings  seems feasible at low $m_H$ values, for the  amount of integrated luminosity of a few
fb's that is presently foreseen for the 7 TeV run of the LHC.

\section{Conclusions}
In this paper, we studied the phenomenological consequences
of a  theoretical framework where a Higgs boson gives masses to EW gauge bosons as in the SM, but it is not responsible for the generation of fermion masses. By setting the Higgs Yukawa couplings
 to zero at a scale $\Lambda$ (connected to the new mechanism for fermion mass generation), nonvanishing Yukawa couplings arise  radiatively, as an effect of  chiral symmetry breaking by explicit fermion mass terms. We computed these effects by RG equation techniques.  
 A nontrivial pattern for Higgs BR's in different channels ensues in the range 
 100 GeV$\lsim m_H\lsim 150$ GeV, with enhanced 
 $H\to  \gamma\gamma$, $WW$, $ZZ$, $Z\gamma\,$ decays,  and non negligible decay rates 
 into heavy-fermion pairs.
  VBF replaces  gluon fusion as the main Higgs boson production channel at the LHC, with quite considerable sensitivity to the $\Lambda$ scale in different decay signatures, for $m_H\lsim(130-140)$ GeV.
 
Present data from LEP \cite{LEPFP} and the Tevatron \cite{D0,CDF} can constrain this scenario today .
A dedicated analysis is needed, combining the effects of the simultaneous enhancement of the $H\to \gamma \gamma$ decay  and the nontrivial depletion of the $H\to b\bar b$ decay, in  order to find $m_H$ bounds versus $\Lambda$ (that, we stress, is the only new free parameter of the model).
The final potential of the Tevatron will depend on the integrated luminosity collected.
However, on the basis of present analysis carried out for the fermiophobic Higgs scenario \cite{D0,CDF}, it seems unlikely that Tevatron can probe the effective Yukawa scenario in the mass range
$m_H\gsim110$ GeV, with an integrated luminosity of about 10 fb$^{-1}$.

The potential of the LHC at $\sqrt{S}=7$ TeV and 14 TeV in further probing the effective Yukawa scenario 
looks excellent, deserving an accurate and dedicated analysis.
On the one hand, the enhanced VBF production for different Higgs signatures gives rise to total cross sections that are in general larger than, or comparable with, the SM ones for  $m_H\lsim120$ GeV, and will benefit, even for larger $m_H$, from the better 
signal-to-background ratio that VBF production enjoys with respect to the gluon fusion mechanism.
On the other hand, the excellent theoretical accuracy in the prediction of VBF processes \cite{VBFNNLO} could help to probe   
the cross section sensitivity to the scale $\Lambda$, even beyond  
$m_H\sim130$ GeV. Furthermore,
a remarkably promising role will be also played 
by the inclusive $WH/ZH$ associated production, when $H\to \gamma \gamma$.

Much better performances in precision measurements of the Higgs couplings are expected of course  from a possible $e^+e^-$ collider program \cite{ILC}. That could eventually also allow a quite accurate $\Lambda$ determination for $m_H>130$ GeV and  to directly  test  the radiatively induced Yukawa couplings.

\section*{\bf Acknowledgments}
We would like to thank Gian Giudice for many enlightening discussions. We also benefitted from conversations  with Giuseppe Degrassi, Gino Isidori, Mauro Moretti, Fulvio Piccinini, and Alessandro Strumia.
We thank Fabio Maltoni and Marco Zaro for providing us  LHC Higgs production cross sections at $\sqrt{S}=7$ TeV.
B.M. was partially supported by the RTN European 
Programme Contract No. MRTN-CT-2006-035505 (HEPTOOLS, Tools and Precision Calculations 
for Physics Discoveries at Colliders).

\renewcommand{\theequation}{A-\arabic{equation}}
\setcounter{equation}{0}  
\section*{Appendix}  

We provide here the main formula used for the calculation of the partial widths
in the present model.
The  decays of the Higgs boson with mass $m_H< 150$ GeV can be classified as tree-level and loop-induced channels. The corresponding
decay widths are given by
\begin{itemize}
\item {\bf Tree-level} \cite{Hfftree,HdecayggA,HVVtree,HVV4f}
\bea
\Gamma(H\to f\bar{f})&=& \frac{N_c\, m_H\, \left|{\rm Y}_f(m_H)\right|^2}{16\pi}
\left(1-4 \frac{m_f^2}{m_H^2}\right)^{3/2}
\label{Gammaff}
\\
\Gamma(H\to V^*V^*)&=& 
\frac{1}{\pi^2}\int^{m_H^2}_0\!\!\!\!\!
\frac{d\, \mu^2_1\, M_V \Gamma_V }{(\mu_1^2-M_V^2)^2+M_V^2 \Gamma_V^2}
\int^{(m_H-\mu_1)^2}_0\!\!\!\!\!\!\!\!\!
\frac{d\, \mu^2_2\, M_V \Gamma_V}{(\mu_2^2-M_V^2)^2+M_V^2 \Gamma_V^2}\,
\Gamma_0~~~~~~~~
\label{GammaVV}
\eea
where ${\rm Y}_f(m_H)$ is the effective Yukawa coupling of the fermion $f$
evaluated at the scale $m_H$, $x_i=m_i^2/m_H^2$, and 
the squared matrix element $\Gamma_0$ for the decay $H\to V^*V^*$, is given by
\bea
\Gamma_0=\frac{G_{F} m_H^3}{16\sqrt{2}\pi}\delta_V
\sqrt{\lambda(\mu_1^2,\mu_2^2,m_H^2)}
\left(\lambda(\mu_1^2,\mu_2^2,m_H^2)+\frac{12 \mu_1^2\mu_2^2}{m_H^4}\right)
\eea
with $\lambda(x,y,z)=(1-x/z-y/z)^2-4xy/z^2$ and $\delta_V=2(1)$ for $V=W(Z)$.
Equation (\ref{GammaVV}) includes 
the 2-, 3-, and 4-bodies decays.
\item {\bf One-loop-induced} \cite{Hgg,HdecayggA,HdecayggB,HZg,Hglgl}
\bea
\Gamma(H\to \gamma\gamma)&=& 
\frac{G_F\, \alpha^2\, m_H^3}{128\sqrt{2}\pi^3}
\left|\frac{4}{9}\,N_c\, A_t(y_t)\, {\bf \xi_t} + A_W(y_W)\right|^2
\label{Gammagg}\\
\Gamma(H\to g\, g)&=& \frac{G_F\, \alpha_S^2 \, m_H^3}
{36 \sqrt{2}\pi^3}\, \left|\frac{3}{4}\, A_t(y_t)\, {\bf \xi_t}\right|^2
\label{Gammaglgl}\\
\Gamma(H\to Z\, \gamma)&=& \frac{G_F^2\, M_W^2\, \alpha \, m_H^3}{64 \pi^4}
\left(1-\frac{M_Z^2}{m_H^2}\right)^3 \left|B_t(\tau_t,\lambda_t)\, {\bf \xi_t} +
c_W B_W(\tau_W,\lambda_W)   \right|^2~~~~~~~~
\label{GammaZg}
\eea
where 
 $y_t=m_H^2/(4m_t^2)$, $y_W=m_H^2/(4M_W^2)$, $\tau_i=1/y_i$,
$\lambda_t=4 m_t^2/M_Z^2$, $\lambda_W=4 M_W^2/M_Z^2$, and
\bea
A_W(x)&=& -\frac{2x^2+3x+3\left(2 x-1\right)F(x)}{x^2}
\nonumber\\
A_t(x)&=& \frac{2\left(x+\left(x-1\right) F(x)\, \right)}{x^2}
\nonumber\\
B_t(x,y)&=&\frac{2 N_c}{3 c_W}\left(1-\frac{8}{3}s_W^2\right)
\left(I_1(x,y)-I_2(x,y)\right)
\nonumber\\
B_W(x,y)&=&4\left(3-\frac{s^2_W}{c_W^2}\right) I_2(x,y)
        + \left[\left(1+\frac{2}{x}\right)\frac{s_W^2}{c_W^2}
        -\left(5+\frac{2}{x}\right)\right]I_1(x,y)\, ,
\label{loop}
\eea
\end{itemize}
where $c_W=\cos{\theta_W}$ and $s_W=\sin{\theta_W}$, with $\theta_W$
the Weinberg angle.
The functions $I_{1,2}(x,y)$ are given by
\bea
I_1(x,y)&=&\frac{xy}{2(x-y)}
+\frac{x^2 y^2}{2(x-y)^2}\left(F(\bar{x})-F(\bar{y})\right)+
    \frac{x^2y}{(x-y)^2}\left(G(\bar{x})-G(\bar{y})\right)
\nonumber\\
I_2(x,y)&=&-\frac{xy}{2(x-y)}\left(F(\bar{x})-F(\bar{y})\right)
\eea
with $F(x)=\left(\arcsin{\sqrt{x}}\right)^2$, 
$G(x)=\sqrt{\frac{1-x}{x}}\arcsin{\sqrt{x}}$, and $\bar{x}=1/x$, $\bar{y}=1/y$.
In eqs.(\ref{Gammagg})-(\ref{GammaZg}), ${\xi_f}=1$ and
${\xi_t}={\rm Y}_t(m_H)/{\rm Y}^{\rm SM}_t$ for the SM and 
effective Yukawa model, respectively, with 
${\rm Y}_t(m_H)$ 
evaluated at the scale $m_H$. The electromagnetic coupling 
constant $\alpha$, appearing in eqs.
(\ref{Gammagg}) and (\ref{GammaZg}), is taken at the scale 
$q^2=0$, namely $\alpha(0)$, since the final state photons 
in the Higgs decays $H\to \gamma \gamma$ and $H\to Z \gamma$ are on shell.
For the strong coupling $\alpha_S$ appearing in eq.(\ref{Gammaglgl}), we assume 
 $\alpha_S=\alpha_S(M_Z)$.

All the SM input parameters  in the numerical analysis  are given 
in table \ref{tabSM} \cite{Amsler:2008zzb}.
\begin{table} \begin{center}    
\begin{tabular}{|c||c|c|c|c|c|c|}
\hline 
$G_F({\rm GeV}^{-2})$ 
& $M_W({\rm GeV})$ 
& $M_Z({\rm GeV})$
& $\Gamma_W({\rm GeV})$
& $\Gamma_Z({\rm GeV})$
& $m_t({\rm GeV})$
& $m_b({\rm GeV})$
\\ \hline 
$1.16637\cdot 10^{-5}$
& 80.398
& 91.1875
& 2.141
& 2.4952
& 171.3
& 4.88
\\ \hline \hline 
$m_c({\rm GeV})$
& $m_s({\rm MeV})$
& $m_{\tau}({\rm GeV})$
& $m_{\mu}({\rm MeV})$
& $\alpha^{-1}(M_Z)$
& $\alpha^{-1}(0)$
& $\alpha_S(M_Z)$
\\ \hline
1.64
& 105
& 1.77684
& 105.658
& 128.9
& 137.036
& 0.1172
\\ \hline \end{tabular} 

\caption[]{
Values for the SM input parameters used for the numerical results.
The quark masses correspond to their pole masses.}
\label{tabSM}
\end{center} \end{table}

\end{document}